\documentclass[fleqn,usenatbib]{mnras}

\usepackage{newtxtext,newtxmath}

\usepackage[T1]{fontenc}

\DeclareRobustCommand{\VAN}[3]{#2}
\let\VANthebibliography\thebibliography
\def\thebibliography{\DeclareRobustCommand{\VAN}[3]{##3}\VANthebibliography}

\usepackage{graphicx}	
\usepackage{amsmath}	
\usepackage{color}
\usepackage{threeparttable}
\usepackage{hyperref}

\def\lambar{\lambda\llap {--}}

\title[Galactic ULPM characteristics]{Evidence for an abundant old population of Galactic ultra long period magnetars and implications for fast radio bursts}

\author[]{
P. Beniamini$^{*1,2}$
Z. Wadiasingh $^{*3,4,5}$
J. Hare$^{4,5}$ 
K. M. Rajwade$^{6}$
G. Younes$^{4,7}$
A. J. van der Horst$^{7}$
\\
$^{1}$ Department of Natural Sciences, The Open University of Israel, P.O Box 808, Ra'anana 4353701, Israel\\
$^{2}$ Astrophysics Research Center of the Open University (ARCO), The Open University of Israel, P.O Box 808, Ra'anana 4353701, Israel \\
$^{3}$ Department of Astronomy, University of Maryland, College Park, Maryland 20742, USA \\
$^{4}$ Astrophysics Science Division, NASA Goddard Space Flight Center, Mail Code 661, Greenbelt, Maryland 20771, USA \\
$^{5}$ Center for Research and Exploration in Space Science and Technology, NASA/GSFC, Greenbelt, Maryland 20771, USA \\
$^{6}$ ASTRON, the Netherlands Institute for Radio Astronomy, Oude Hoogeveensedijk 4, 7991 PD Dwingeloo, The Netherlands \\
$^{7}$ Department of Physics, George Washington University, 725 21st St NW, Washington, DC 20052, USA
}

\begin{document}
\label{firstpage}
\pagerange{\pageref{firstpage}--\pageref{lastpage}}
\maketitle

\begin{abstract}
Two recent discoveries, namely PSR J0901–4046 and GLEAM-X J162759.5–523504.3 (hereafter GLEAM-X J1627), have corroborated an extant population of radio-loud periodic sources with long periods (76 s and 1091 s respectively) whose emission can hardly be explained by rotation losses. We argue that GLEAM-X J1627 is a highly-magnetized object consistent with a magnetar (an ultra long period magnetar - ULPM), and demonstrate it is unlikely to be either a magnetically or a rotationally-powered white dwarf.
By studying these sources together with previously detected objects, we find there are at least a handful of promising candidates for Galactic ULPMs.
The detections of these objects imply a substantial number, $N \gtrsim 13000$ and $N \gtrsim 500$ for PSR J0901--4046 like and GLEAM-X J1627 like objects, respectively, within our Galaxy. These source densities, as well as cooling age limits from non-detection of thermal X-rays, Galactic offsets, timing stability and dipole spindown limits, all imply the ULPM candidates are substantially older than confirmed Galactic magnetars and that their formation channel is a common one. Their existence implies widespread survival of magnetar-like fields for several Myr, distinct from the inferred behaviour in confirmed Galactic magnetars. ULPMs may also constitute a second class of FRB progenitors which could naturally exhibit very long periodic activity windows. Finally, we show that existing radio campaigns are biased against detecting objects like these and discuss strategies for future radio and X-ray surveys to identify more such objects. We estimate that ${\cal O}(100)$ more such objects should be detected with SKA-MID and DSA-2000.
\end{abstract}




\def\thefootnote{*}\footnotetext{These authors contributed equally to this work}

\section{Introduction}
Since the first radio detection of an astronomical object by Karl Jansky, about 90 years ago, radio astronomy has become a major pillar in astrophysics. It has revealed classes of objects and types of radiation that were previously unknown such as pulsars, quasars, the cosmic microwave background, and fast radio bursts \citep[FRBs:][]{Linscott1980,Lorimer+07}. Still, much of the parameter space for radio detections remains as of yet unexplored \citep{2003ApJ...596.1142C,2015MNRAS.446.3687P}, particularly for rare or intermittent transients that are currently being discovered \citep[e.g.,][]{2021ApJ...920...45W,2022arXiv220902352W}. This in large part owes to the many trade-offs that must be made when designing a radio experiment or survey, e.g. selecting the bands to be observed, compromising between area of coverage and limiting flux or compromising between temporal and spectral resolution, sensitivity and specificity of triggering algorithms, and computational or storage constraints, just to name a few.

At the faint end, modern day radio experiments routinely measure lower luminosity emission, i.e. intrinsically faint sources, than probed in any other region of the electromagnetic spectrum. As an example, we can consider the electromagnetic observations of the afterglow of the first GW detected NS merger, GRB 170817A. In that case, the VLA radio limits reached as deep as $\!\sim\! 3\!\times\ 10^{-19}\mbox{erg s}^{-1}\mbox{cm}^{-2}$ \citep{Alexander2020} while even $\sim\! 100$ks Chandra $X$-ray limits are more than $10^3$ times brighter \citep{Troja2020}. Similarly, while many fast radio burst (FRB) models predict inherently bright multi-wavelength counterparts (often expected to be orders of magnitude more luminous than the radio emission), so far (barring the somewhat special case of the Galactic FRB \citep{Mereghetti+20,Tavani2020,Ridania2020,Li2020,MBSM2020,LKZ2020,2022ApJ...931...56L}, only upper limits exist in other wavelengths \citep[e.g.,][]{2021ApJ...921L...3M,2021A&A...656L..15P,2022ApJ...929..173L,2022ApJ...930..172L}, and these are typically far above the expected sources' output. 

In light of this, it is perhaps to be expected that many new and exciting astrophysical objects could be
found in current and future radio surveys. In the recent months, two particularly interesting such objects were detected, GLEAM-X J162759.5--523504.3 (henceforth GLEAM-X J1627 for short) detected by the MWA \citep{Hurley-Walker2022} and PSR~J0901--4046 discovered by the MeerTRAP program~\citep{caleb2022}. Both objects exhibit periodicity on long timescales (1091 s and 76 s respectively) and exhibit some magnetar-like phenomenology\footnote{Although neither would presently be classified as AXPs or SGRs, the original two sources classes which fit the historical high-energy operational definition of and motivation for ``magnetars" \citep{1984Ap&SS.107..191U,Paczynski1992,1992ApJ...392L...9D,1995MNRAS.275..255T,1996ApJ...473..322T,Kouveliotou98,2003ApJ...588L..93K} where the present luminosity is manifestly greater than spin-down power.}.  Indeed, there have been a variety of previously detected sources with some similar features as will be discussed in this work. However, the association of these objects with magnetars is far from obvious. Galactic magnetars, as historically defined, are observed to have periods $1\,\rm{s}\!\lesssim\!\,P\,\!\lesssim\!12\,$s with one notable exception~(\S\ref{RCW103}). The dearth of longer period conventional magnetars is understood to stem from the freezing of the spindown (SD) process as their (crust-anchored) dipole field $B_{\rm dip}$ decays on a timescale of $\tau_{d,{\rm dip}}\sim 10^3-10^4$\,yrs \citep{Kouveliotou98,Colpi2000,Dall'Osso2012D,Vigano2013,Beniamini2019,Jawor2022}. Clearly if the new class of observed objects are indeed highly magnetized, they require a different dominant spindown mechanism to drive their evolution, potentially related to a retardation of field decay in their crusts (with constraints on core field expulsion). In this work we adopt a broad definition of what a magnetar is -- a highly magnetized neutron star whose present observed luminosity does not necessarily exceed its spin-down power (but likely did exceed it sometime in its past).

Various lines of evidence \citep[e.g.,][]{Spitler+16,2017ApJ...838L..13L,Wadiasingh2019,Ravi2019b}, and perhaps most remarkably the association with a known Galactic magnetar SGR~1935+2154 \citep{STARE2020,CHIME2020}, connect FRBs with magnetar sources. Therefore, the discovery that some FRBs have periodic activity windows \citep{CHIMEperiodicity,Pastor-Marazuela2020,Rajwade2020,2021MNRAS.500..448C}, lasting tens of days and more, was unexpected. \cite{Beniamini+20} explored the idea that the observed periods reflect the rotational periods of magnetar-like sources. These authors have shown that giant flare (GF) kicks, charged particle winds or fallback accretion can all lead to enhanced spindown and that as a result, a small fraction of magnetars (that they dubbed ULPMs, for ultra long period magnetars) may develop extremely long rotation periods. It is still actively debated what is the source of the observed periodicity seen in FRBs (other suggestions include binary periodicity and free precession, \citealt{Lyutikov+20,IokaZhang20,Levin+20,Zanazzi&Lai20}). However,  FRB 20180916's chromatic windowing, lack of strong dispersion or rotation measure, random flux variation with activity phase, and absence of a low frequency cutoff \citep{Pastor-Marazuela2020,Pleunis2021,2022arXiv220509221M,2022arXiv220713669B} disfavors simple binary shrouding models. There are deep limits on any underlying shorter periodicity less than 1 ks for FRB 20121102 \citep{2021Natur.598..267L,2022MNRAS.515.3577H} and other prolific repeaters \citep[e.g., FRB 20180916,][]{CHIMEperiodicity}.

A possible connection to a Galactic (and nearby) class of objects could therefore hold an important key towards deciphering the origin of FRB periodicity. Even if FRB periodicity is unrelated to their rotational periods, the class of Galactic objects with magnetar-like features and long periods can be considered as ULPM-candidates and the enhanced spindown mechanisms discussed in the FRB context to create ULPMs should be revisited in this new context.
There is also tentative evidence that FRB sources may encompass two or more populations \citep{2021ApJ...923....1P}. ULPMs would naturally fit as a second, perhaps older population, compared to FRBs from relatively young SGR 1935+2154-like (or younger) magnetars. Evidence of offsets from a star forming region for the location of FRB 20180916 \citep{2021ApJ...908L..12T} also are compatible with an older neutron star (NS), disfavoring precession scenarios for the periodic windowing activity. Lastly, we note the recent discovery of a repeating FRB 20200120E in a globular cluster of M81 \citep{2021ApJ...910L..18B,2022Natur.602..585K,2021ApJ...919L...6M} that exhibits `burst storms' much like Galactic magnetars \citep{2022arXiv220603759N} and for which short-term periodicity signatures can be ruled out in the burst arrival time data (in the range $1\mbox{ ms}\!\lesssim\! P\lesssim \!25\!\mbox{ s}$). As pulsars in globular clusters are at least few Myr old \citep{1993MNRAS.265..449M,1996ApJ...460L..41L,2005AJ....129.1993M,2011ApJ...742...51B,2012ApJ...756...78L} and star formation is almost nonexistent in such environments, an object much older than conventional magnetars produced in a low-kick process is favored, possibly produced by electron capture supernovae, accretion-induced collapse or white dwarf (WD) collisions rather than a conventional core collapse supernovae (CCSN) \citep{LBK2021,Kremer2021}. 

In this paper we present the emerging class of Galactic ULPM candidates (\S \ref{sec:ULPMcand}). We then discuss empirical evidence supporting episodes of enhanced spindown in known Galactic magnetars in \S \ref{sec:enhancedSD}. We then explore the source densities of ULPMs inferred from observations (\S \ref{sec:numbers}) use this and other arguments to estimate the ages of the ULPM candidates, \S \ref{sec:Age}, and consider the implications regarding the temporal evolution of their magnetic field (\S \ref{sec:fieldevolution}). In \S \ref{sec:PeriodicFRB} we turn to show that the population of Galactic ULPM candidates can only be related to FRB periodicities if the latter are extremely strongly beamed and / or only a small fraction of ULPMs are efficient FRB producers. Finally, in \S \ref{sec:observestrat} we discuss prospects and strategies for future detections of ULPMS in radio and X-ray campaigns. Our results are summarized in \S 
\ref{sec:conclusion}.

\begin{table*}
\footnotesize
\setlength{\tabcolsep}{3pt}
\begin{center}
\begin{tabular}{|l|ccccccccccc|}
\hline \hline \hline
 \textbf{Magnetar} & {\textbf{P}} & \textbf{$\dot{P}$} & {\textbf{$B_{\rm SD}$}} &  {\textbf{SD age}} & \textbf{Dist.} & {\textbf{$F_X$}} & $k T_{\rm surface}$ &  {\textbf{Gal. b}} & \textbf{Duty Cycle} & \textbf{Refs.} \\
   \textbf{candidate} & (s) & (s s$^{-1})$ & (G) &  {(yr)} & (pc) & {(erg s$^{-1}$ cm$^{-2}$)} & (keV) &  &  (radio) & \\
\hline \hline \hline
GLEAM-X J1627 & 1091 & $<10^{-9}$ & $\lesssim 7 \times 10^{16}$ &  --  & $1300 $ & $<5\times10^{-14}$ & $<0.072$ & $-2.6^\circ$ & $2-5 \%$ & [1] \\
PSR~J0901--4046 & 76 & $\sim 2.21\times 10^{-13}$ & $2.6\times10^{14}$ & $5.3\times10^6$  & 327-450 & $<2\times10^{-13}$ & $<0.064$ & $+3.7^\circ$  & $\sim 0.7 \%$ & [2]\\
PSR J0250+5854 & 23.54 & $2.72 \times 10^{-14}$ & $2.56\times 10^{13}$& $1.37\times10^7$ & 1558 & $<2\times10^{-15}$ & $<0.053$ & $-0.5^\circ$ & $0.3-0.4\%$ & [3] \\
1E 161348--5055 & 24030.42 & $<1.6\times10^{-9}$ & $<10^{17}$ & -- & 3100-4700 & few$\times10^{-11}$& $0.58$ & $-0.4^\circ$ & -- & [4] \\
\hline 
FRB 20180916B (R3)& $1.4\times 10^6$ & -- & -- & -- & $4\times 10^7$ & -- & --  & -- & $\sim 40\%$ & [5] \\
\hline
IGR J16358--4726 & 5970 & -- & -- & -- &  >7000? & -- & -- & $\sim 0.0^\circ$ & -- & [6] \\
4U 2206+54$^\dagger$ & 5750 & $6\times 10^{-7}$ &  -- & -- &  3700 & 3$\times10^{-11}$ & $\sim 1.5$ & $-1.1^\circ$ & -- &  [7]\\
SGR 0755-2933 & 308 & -- &  -- & -- &  3500 & -- & -- & $-0.6^\circ$ & -- & [8] \\
4U 1954+319 & 20500 & -- &  -- & -- & 3300 & -- & -- & $+1.9^\circ$ & -- & [9] \\
4U 0114+65 & 9350 & -- &  -- & -- & 7000 & -- & -- & $+2.6^\circ$& -- & [10] \\
AX J1910.7+0917 & 36000 & -- & -- & -- & 16000 & -- & -- & $-0.02^\circ$ & -- & [11] \\
SXP 1062$^\dagger$ & 1070 & $3\times10^{-6}$ & -- & -- & SMC & -- & -- & -- & -- & [12] \\
\hline
GCRT J1745--3009 & 4630 & -- & -- & -- & 8500? & -- & -- & $-0.54^\circ$ & $\sim 13\%$ & [13]  \\
\hline
PSR J2251--3711$^{*}$ & 12.12 & $1.3 \times 10^{-14}$ & $2.5 \times 10^{13}$& $1.47\times10^7$ &  540-1344 & $\lesssim 10^{-14}$ & -- & $-62.9^\circ$ & $0.6-0.7\%$ & [14] \\
\hline \hline
RX J0420--5022 & 3.45 & $2.76\times 10^{-14}$ & $9.9 \times 10^{12}$ & $1.98 \times 10^6$ & 345 & $4.4\times10^{-14}$ & 0.046 & $-44.4^\circ$ & -- & [A] \\
PSR J0726--2612 & 3.44 & $2.93\times 10^{-13}$ & $3.2 \times 10^{13}$ & $1.86 \times 10^5$ & 100-1000 (<3200) & $1.6\times10^{-12}$ & 0.074 & $-4.64^\circ$ & $2-3\%$ & [B] \\
RX J0720--3125 & 8.39 & $7\times 10^{-14}$ & $2.45 \times 10^{13}$ & $1.9 \times 10^6$ & 286 & $3.2\times10^{-11}$ & 0.082 & $-8.17^\circ$ & -- & [A] \\
RX J0806--4123 & 11.37 & $5.6\times 10^{-14}$ & $2.54 \times 10^{13}$ & $3.24 \times 10^6$ & 250 &$2.1\times10^{-12}$ & 0.058 & $-4.98^\circ$ & -- & [A] \\
RX J1308+2127 & 10.31 & $1.12\times 10^{-13}$ & $3.44 \times 10^{13}$ & $1.46 \times 10^6$ & 500 & $8.4\times10^{-12}$ & 0.069 & $+83.1^\circ$ & -- & [A] \\
RX J1605+3249 & 6.88 & -- & --& -- & 390 & $1.1\times10^{-11}$ & 0.065 & $+48.0^\circ$ & -- & [A] \\
RX J1856--3754 & 7.06 & $2.98\times 10^{-14}$ & $1.47 \times 10^{13}$ & $3.76 \times 10^6$ & 156 & $2.7\times10^{-11}$ & 0.062 & $-17.2^\circ$ & -- & [A] \\
RX J2143+0654 & 9.43 & $4.1\times 10^{-14}$ & $2 \times 10^{13}$ & $3.65 \times 10^6$ & 150-430 & $1.3\times10^{-11}$ & 0.055 & $-33.1^\circ$ & -- & [A]\\ 
\hline \hline \hline
\end{tabular}
\end{center}
\caption{Candidates and potentially related Objects which demand more observational scrutiny. Data collected from the literature for ULPM candidates and (below the double solid line) another class of unusual NSs, XDINS. \\
$^\dagger$ The $\dot{P}$ is likely dominated by accretion torques.\quad $^*$ XDIN candidate.
\\ References: 
[1] \citet{Hurley-Walker2022} 
[2] \citet{caleb2022} 
[3] \citet{2018ApJ...866...54T} 
[4] \citet{DeLuca2006,2008ApJ...682.1185D,2011MNRAS.418..170E,2016MNRAS.463.2394D,2019MNRAS.489.4444B}[5] \citet{CHIMEperiodicity,Pastor-Marazuela2020,2022arXiv220713669B} 
[6] \citet{2003IAUC.8097....2R,2003IAUC.8109....2K,2004ApJ...602L..45P,2007ApJ...657..994P} 
[7] \citet{2018MNRAS.479.3366T} [8] \citet{2016ATel.8831....1B,2017AAS...22943104H} 
[9] \citet{2014ApJ...786..127E,2020ApJ...904..143H,2022MNRAS.510.4645B} 
[10] \citet{1999ApJ...513L..45L,2017A&A...606A.145S} 
[11] \citet{2011A&A...526A.122P,2013A&A...555A.115R,2017MNRAS.469.3056S}
[12] \citet{2012A&A...537L...1H}
[13] \citet{2005Natur.434...50H,2007ApJ...660L.121H,2009A&A...502..549S,2008ApJ...687..262K} [14] \citet{2020MNRAS.493.1165M} 
[A] \citet{2009ApJ...705..798K,2011MNRAS.417..617T,2019PASJ...71...17Y,2020MNRAS.496.5052P} [B] \citet{2006MNRAS.368..283B,2011ApJ...743..183S,2019A&A...627A..69R}  \\
}
\label{tab:sources}
\end{table*}

\section{Galactic ULPM Candidates}
\label{sec:ULPMcand}
We list below several Galactic objects which we consider as tenable ULPM candidates, and briefly summarize the evidence supporting such an association. Additional, less secure, associations are listed in Appendix \ref{sec:speccand}. These are not part of the analysis presented in the paper and are provided for completeness.
Table~\ref{tab:sources} summarizes some of the main observations of all the candidates as well as those of XDINS which share some of the properties with ULPM candidates (nearby objects, low inferred surface temperatures and magnetar-like fields).

\subsection{PSR~J0901--4046}
PSR~J0901--4046 is a prominent $P\approx 76$ s pulsar discovered by MeerTRAP~\citep{2020SPIE11447E..0JR} with a $\dot{P} \sim 2.21 \times 10^{-13}$ s s$^{-1}$ implying a spindown polar dipolar magnetic field component of $B_{\rm SD} \sim 2.6 \times 10^{14}$ G~(\citealt{caleb2022}; see \S \ref{sec:magPSRJ0901} below for a more general limit on $B$ that does not assume dipole SD). A faint radio shell appears to be spatially coincident with PSR~J0901--4046 \citep{caleb2022}, although, if associated, it is much too small to be consistent with a supernova remnant. PSR~J0901--4046 displays many of the radio characteristics seen in common radio pulsars, such as polarization position angle swings and variability within single pulses, yet with relatively wide duty cycle for its long period (see Appendix \ref{sec:radiodutycycle}).
It also exhibits unusual features, such as harmonically-spaced QPOs (see discussion in Appendix \ref{sec:crust}). Moreover, it is remarkably stable in timing, with timing residuals of $5.7$ ms over 7.4 months -- a relative stability of $\sim 10^{-6}$, and no timing instabilities or glitches in contrast to what is common within the young magnetar population. 

Some single pulses in PSR~J0901--4046 also exhibit partial nulling or spikes within the pulse that occurs on timescales $\ll 1$ ms, less than the temporal sampling resolution of MeerKAT 1.4~GHz observations. This, along with a characteristic radio luminosity of sub-components of $\gtrsim 1$ mJy implies a brightness temperature $T_b > 10^{16}$~K (assuming a distance 328 pc from the measured DM), i.e. favoring coherent pulsar-like emission and a NS origin. The discussion regarding PSR~J0901--4046 hosting a thick crust and strong core field are given in Appendix \ref{sec:crust} and \S\ref{sec:corefield}, respectively.

A potential optical counterpart of PSR~J0901--4046 is detected by Gaia, having a G-band magnitude of 17.5, and is offset from the source by about 3.3$''$. The measured Gaia parallax, places the optical source at a distance of 2.4 kpc \citep{2021AJ....161..147B}, which is far beyond the inferred dispersion measure distance for PSR~J0901--4046. Additionally, \cite{caleb2022} suggest the optical source is a reddened A-type star, based on optical spectroscopic and photometric observations of the source. For these reasons, it appears that PSR J0901--4046 does not have any plausible optical counterpart in Gaia.

\subsubsection{The magnetic field of PSR~J0901--4046}
\label{sec:magPSRJ0901}
The spin down value of $2\times 10^{-13}$ s s$^{-1}$ for PSR~J0901--4046 results in the inferred magnetar-like dipolar magnetic field estimate, which exceeds $10^{14}$ G and argues for its magnetar nature. Yet, it may be argued that this is an overestimate due to multipolar fields or particle winds enhancing spin down. The former argument can be largely discarded, as the influence of higher-order multipoles on spindown is negligible given its large light cylinder at $\sim10^4$ stellar radii (unlike the case for millisecond pulsars). Particle winds, and enhanced spindown, on the other hand may lower the magnetization estimate but even this is disfavored as described below. In order for the wind to dominate the spindown over dipole EM spindown, it must be powered magnetically. This demands the wind luminosity is limited by $L_{\rm p}\lesssim E_{\rm B}/\tau\propto B^2$. At the same time, the particle wind luminosity decreases with $B$ according to $L_{\rm p}\approx 6 c^3 I^2 \dot{P}^2/(B^2 P^2 R^6) \propto B^{-2}$  \citep{Harding1999}. Here $B$ is the dipole field at the pole. Combining these expressions, we find a lower limit on the required $B$ to sustain a sufficiently strong particle wind
\begin{equation}
\label{eq:Blimit}
    B\gtrsim \left( \frac{6 c^3 I^2 \dot{P}^2\tau}{P^2 R^9 f^2}\right)^{1/4} \approx 10^{13}f^{-1/2}\tau_{\rm Myr}^{1/4}\quad \mbox{G}
\end{equation}
where $\tau$ is the time over which this luminosity is sustained (normalized here by the age estimate of the source, $\sim 1$\,Myr) and $f\geq 1$ is the ratio between the internal and dipole field in the star. 
Eq.~(\ref{eq:Blimit}) is strictly a lower limit on $B$, as it assumes $100\%$ efficiency in converting the magnetic energy to a wind luminosity. As we can see a strong field is still required in this case.
This $B$ limit, in turn, corresponds to an upper limit on the particle wind luminosity, $L_{\rm p}$,
\begin{equation}
    L_{\rm p}\lesssim \left( \frac{6 c^3 I^2 \dot{P}^2f^2}{P^2 R^3\tau}\right)^{1/2}\approx 6\times 10^{30}f\tau_{\rm Myr}^{-1/2}\quad \mbox{erg s}^{-1}.
\end{equation}
An additional argument disfavors the existence of any dynamically significant wind (and in turn increases the lower limit on $B$): PSR~J0901--4046 is extraordinarily stable in its observed timing.

\begin{figure*}
\centering
\includegraphics[width=0.7\textwidth]{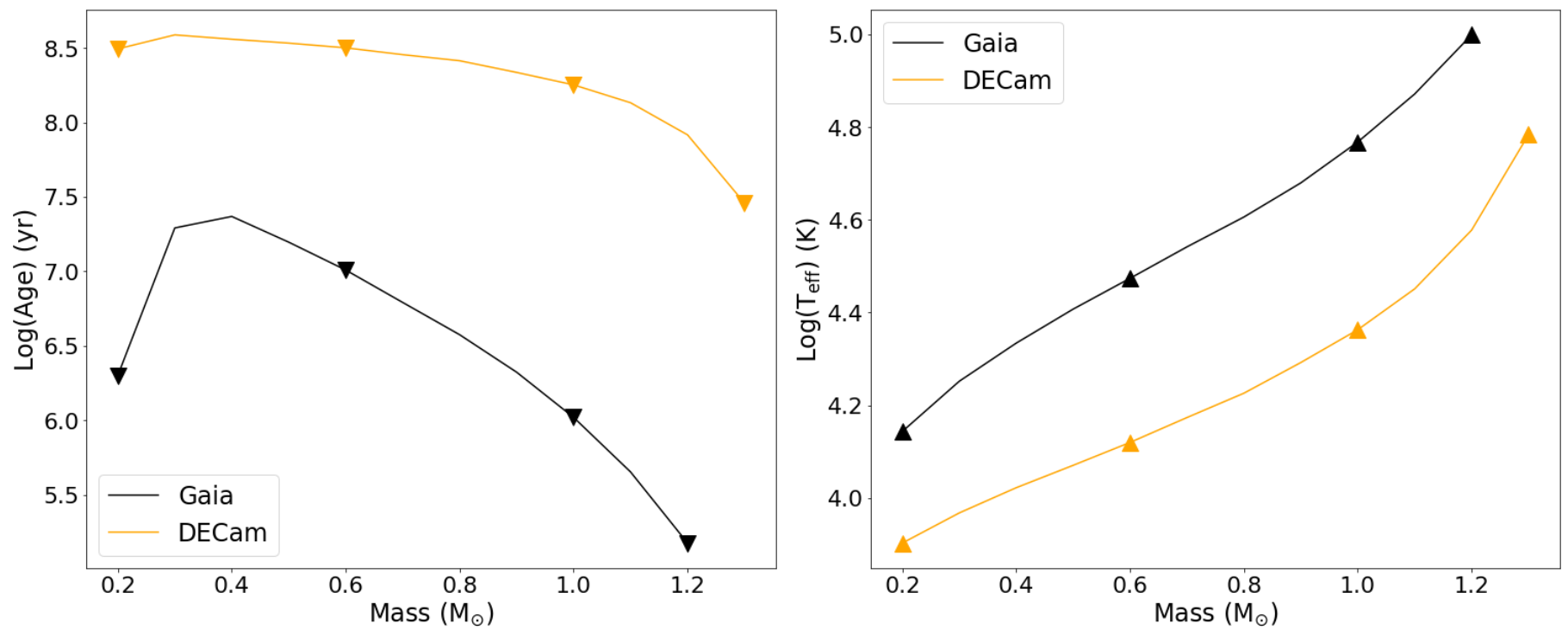}
\caption{{\bf Left:} Detectability,  with Gaia G band (black) and DECam r-band (orange), of hot WDs as a function of mass and age. WDs having ages younger than the respective lines for a given mass would be detectable. {\bf Right:} Similar to the left plot, but for effective WD temperature as a function of mass. WDs hotter than the respective lines for a given mass would be detectable. The WD evolutionary tracks of \citealt{2020ApJ...901...93B} are adopted here. }
\label{fig:wd_age_temp}
\end{figure*}

\subsection{GLEAM-X J1627}
\label{sec:WD}
GLEAM-X J1627 is a bright $20-50$ Jy periodic radio transient discovered at low ($\sim 70-230$\,MHz) frequencies by the MWA \citep{Hurley-Walker2022}. It exhibits pulsations at $P \approx 1091$ s which are nearly $100\%$ linearly polarized, as well as unresolved bright bursts that imply a brightness temperature $\gg 10^{12}$ K for its DM-inferred distance of $1.3\pm0.5$ kpc, possibly with $\gg100$ kJy bursts for the unresolved spikes if they are intrinsically ${\cal O}(10 \rm \, ms)$ in duration (i.e. similar in duration to extragalatic FRBs at low frequencies after accounting for scattering). At this distance, its persistent radio luminosity exceeded $10^{31}\mbox{ erg s}^{-1}$ in the limited bandwidth of MWA, and is possibly higher given its relatively flat radio spectrum. Owing to its long period, high persistent radio luminosity, and spindown upper limit ($\dot{P} \lesssim 10^{-9}$ s s$^{-1}$), the source is unlikely to be a rotation-powered object with moment of inertia compatible with neutron or quark stars\footnote{Of course, these conclusions depend on the beaming of the radiation as well as the efficiency of conversion between spindown luminosity and radio emission (see \citealt{2022MNRAS.514L..41E}). 
}. It's radio duty cycle is also much too large to be compatible with its physically small polar cap size or known empirical relations for rotation-powered radio pulsars (see Appendix \ref{sec:radiodutycycle}). This suggests it could be of a magnetar origin. If the observed period is due to rotation, this would imply that this object is an ULPM. Alternatively, \cite{2022arXiv220205160E} have suggested that GLEAM-X J1627 is a $\!\sim \! 0.1$\,s period magnetar that is undergoing precession with the $\!\sim \!1$\,ks observed period duration. A major difficulty with such a scenario is that the polarization would be variable and likely depolarized on MWA's 0.5 second temporal sampling. There could also be a beat period with the temporal sampling and the putative short period. In addition, the large toroidal magnetic field $(\!\sim \! 10^{16}$\,G) required for sustaining the deformation needed for precession, and a hot core required for the inhibition of superfluidity \citep[e.g.,][]{1977ApJ...214..251S} generally means the object must be extremely young which is inconsistent with a large number of observational constraints (see \S \ref{sec:Age} for details).
Other authors \citep[][]{2022RNAAS...6...27L,2022arXiv220308112K} have suggested that GLEAM-X J1627 may instead be a rotation or magnetic powered WD. We outline below some challenges that this interpretation encounters.

\subsubsection{Optical constraints on a WD or stellar association to GLEAM-X J1627}
\label{optgleammain}

We do not find a plausible counterpart to GLEAM-X J1627 in either Gaia or DECam archival data. Assuming no counterpart to GLEAM-X J1627 is detected by either Gaia or DECam (see Appendix \ref{sec:gleam_opt_count}), we can place constraints on the mass/age of a putative WD at 1.3 kpc. Using the evolutionary tracks of \cite{2020ApJ...901...93B}, we estimate at what age (or temperature) a WD of a given mass becomes undetectable\footnote{The actual spectra for the evolutionary tracks calculated by \cite{2020ApJ...901...93B} are not provided, so we use a black-body model for the WDs.}. For Gaia, we use the G band magnitude and a limiting magnitude of 21, while for DECam, we use the r-band to minimize the effects of absorption, and a limiting magnitude of 22.8. We account for the extinction in the direction of GLEAM-X J1627 at a distance of 1.3 kpc using the {\tt mwdust} package \citep{2016ApJ...818..130B}. These estimates are shown in Figure \ref{fig:wd_age_temp}. For a large range of masses ($0.2\!<\!M_{\rm WD}\!<\!1.2$ $M_{\odot}$) a WD requires an age $t_{\rm WD}\gtrsim (1\!-\!3)\times10^{8}$ yrs to be undetected by DECam. Such an old age is a severe constraint to WD models of GLEAM-X J1627 (see below).

\subsubsection{Why GLEAM-X J1627 is likely a magnetized neutron or quark star and not a WD}
\label{sec:WDdiscuss}

\noindent{\bf Magnetically powered WDs}

Below we consider some limitations on a magnetically active WD as the source of GLEAM-X J1627. The energetic radio output of GLEAM-X J1627 in its two-month active episode was $\sim (10^{37}-10^{39}) f_{\Omega}$~erg assuming a typical flux density of $10-30$ Jy and flat spectral index over bandwidth of $\sim 100$ MHz (at a distance of 1.3 kpc). \cite{Hurley-Walker2022} estimate $L_R \sim$~few~$\times 10^{31}$~erg~s$^{-1}$ (isotropic-equivalent) based on the spectral index and extrapolation to higher frequencies. The $f_{\Omega}<1$ above is the intrinsic (unknown) collimation of emission, which may be conservatively taken as $f_\Omega \sim 10^{-2}$, the observed duty cycle of pulsations.
This disfavors magnetically-powered activity from a highly magnetized WD similar to Ar Sco \citep{2016Natur.537..374M,2017NatAs...1E..29B,2018AJ....156..150S,2022MNRAS.516.5052P}, as the total field (and ephemeral free energy, the magnetic field helicity) reservoir corresponds to $E_B\!\sim \!B_8^2 R_{8.5}^3/6\!\sim\!5\times 10^{40}$\,erg. Here $R \! \sim\! 3\times 10^8$\,cm is a typical WD radius.
The efficiency of conversion (from magnetic free energy) to coherent radio emission needs to be $\eta\!\sim \!2\!\times\! 10^{-4}-2\!\times\! 10^{-2}$ for the energetics to be justifiable. Yet, the non-thermal radio fluence of the outburst episode likely severely underestimates the true energetics (e.g. source Poynting flux and particle kinetic power, transient or persistent), as in known natural coherent radio emitters. For bright broadband coherent radio emitters such as rotation-powered pulsars $\eta \ll 1$ \citep[e.g., Figure 4 in][]{Wadiasingh2019} except the oldest, most charge-starved\footnote{For WDs, as pair production cannot take place without extreme values of magnetization, and a relativistic plasma is required for coherent radio emission, significant kinetic power must go to baryons implying $\eta \ll 1$.} ones in the class \citep[e.g.,][]{2002ApJ...568..289A}. Likewise, the bright radio burst from SGR 1935+2154 was only $\sim 10^{-5}$ the fluence of the associated hard X-ray burst \citep{MBSM2020}. Then, given the radio luminosity of GLEAM-X J1627, it can maintain this level of activity up to at most
\begin{equation}
\label{eq:tauB}
\tau_B\lesssim \frac{\eta E_B}{f_{\Omega} L_R}\sim 50 B_8^2R_{8.5}^3\eta f_{\Omega}^{-1} L_{R, 31.5}^{-1} \quad \mbox{yr,}
\end{equation}
seven orders of magnitude shorter in age than the lower limits corresponding to the optical limits from WD cooling (\S\ref{optgleammain}). One may imagine a situation in which the active lifetime of the sources is shorter than their true ages, however it would be highly unlikely to detect such an object if its ``active duty cycle" (i.e. active lifetime compared to true age), $\eta_{\rm act}$ were $\ll1$. In particular the number of inferred Galactic objects of this type (see \S \ref{sec:simplenum}) would be increased by $\eta_{\rm act}^{-1}$.
Independent of $\eta_{\rm act}$, the low value of $\tau_{\rm B}$ leads to unrealistic minimum requirements on the formation rate of high $B$ WDs as detailed next. As shown in \S \ref{sec:simplenum}, the distance to GLEAM-X J1627 requires that the number of similar active objects in the Galaxy be $N_{\rm act}\gtrsim 2500$ (where we have considered the more conservative limit, that is independent of object type and we have considered a 3D distribution of the putative objects, as appropriate for WDs). This corresponds to a formation rate $\dot{r}\gtrsim 2500 \tau^{-1}\mbox{ yr}^{-1}$ where $\tau \leq \tau_{\rm B}$ is the active lifetime of the sources. Combining with Eq.~(\ref{eq:tauB}), and relating to the Galactic WD formation rate, $\dot{r}_{\rm WD}\approx 0.25 \mbox{ yr}^{-1}$ \citep{Hills1978}, we find
\begin{equation}
    \frac{\dot{r}}{\dot{r}_{\rm WD}}\gtrsim 200 B_8^{-2}R_{8.5}^{-3}f_{\Omega}\eta^{-1} L_{R, 31.5}.
\end{equation}
We see that for $B\lesssim 2\times 10^9\mbox{ G}$, the required formation rate is greater than that of all WDs, which is clearly ruled out. Even for $B\approx 10^{10}$~G, one needs the formation rate of such extremely magnetic WDs to be more than 2$\%$ of the total WD formation rate. This is inconsistent with the observed population \citep{2015SSRv..191..111F,2020AdSpR..66.1025F}---much weaker magnetized WDs with $B\approx 10^8\mbox{ G}$ WDs are already limited to at most a few percent of the total population.

\vspace{0.1in}
\noindent{\bf Rotation powered WDs}

A typical WD with $B\sim 10^6$\,G, has a dipole spin power that is much too low to account for the observed radio luminosity. Thus, even though the rotational energy is $\sim 10^{45}M_{1M_{\odot}}R_{8.5}^2\mbox{ erg}\gg E_B$, spindown cannot power the radio luminosity through dipole spindown, unless the magnetic field is substantially larger.

An outlandishly high-field magnetized WD (with moment of inertia $I \sim 10^{50}$~g~cm$^2$, corresponding to one expected from WDs) rotationally powering the radio emission is also disfavored. Such a putative WD would require a global dipolar field $B\gtrsim 10^{11}$ G to power the observed luminosity (assuming $\eta /f_{\Omega} \sim 1$), a regime in which the total stellar magnetization is only an order of magnitude less than the gravitational binding energy (i.e. the Chandrasekhar hydromagnetic limit for stability), and two to three orders of magnitude larger than Ar Sco or the most highly magnetized WDs known \citep{2015SSRv..191..111F,2020AdSpR..66.1025F}. For a typical WD core conductivity, the resulting ohmic timescale for field decay is $1-3$ Gyr \citep{2002MNRAS.333..589C}. This implies a persistent ohmic magnetic dissipation luminosity of $L_{\rm md} \gtrsim 10^{30.5} B_{11}^2 R_{8.5}$ erg s$^{-1}$ and a minimum effective temperature of $T_{\rm eff} \gtrsim 1.4\times 10^4 \, B_{11}^{1/2} R_{8.5}^{-1/4}$~K\footnote{This estimate is strictly a lower limit, as this magnetic dissipation is in additional to any residual heat from formation that may be estimated from appropriate WD cooling models (for varying age and mass).}. At its distance and expected extinction, optical/UV observations with HST or Webb could readily rule out a WD origin, irrespective of high nebular emission expected from energetic WDs. Such temperatures are ruled out by DECam in Figure~\ref{fig:wd_age_temp} for masses up to $\sim0.8 M_\odot$.

A core difficulty with the WD origin of GLEAM-X J1627, is its small duty cycle, on the order of a few percent, the likes of which has not been previously observed in pulsating WDs. For instance, Ar Sco's pulse duty cycle exceeds $30\%$ \citep[][]{2017NatAs...1E..29B}. Observing highly collimated radiation (as needed for a small duty cycle) from a WD, presents also a theoretical difficulty. To attain a high degree of beaming, one typically needs a relativistic particle population. However, it is generally difficult to source and accelerate plasma near the surface of a WD to relativistic velocities as in NSs, considering that the escape velocity from the surface of the WD is $\ll c$ so large voltages are not necessary to launch plasma. Furthermore single-photon magnetic pair production requires high magnetic fields and small curvature radii. This then sets demands on any voltage and magnetic field an isolated WD must attain to source its relativistic plasma (in contrast to Ar Sco, which has a stellar companion).

\begin{figure}
\centering
\includegraphics[width = 0.45\textwidth]{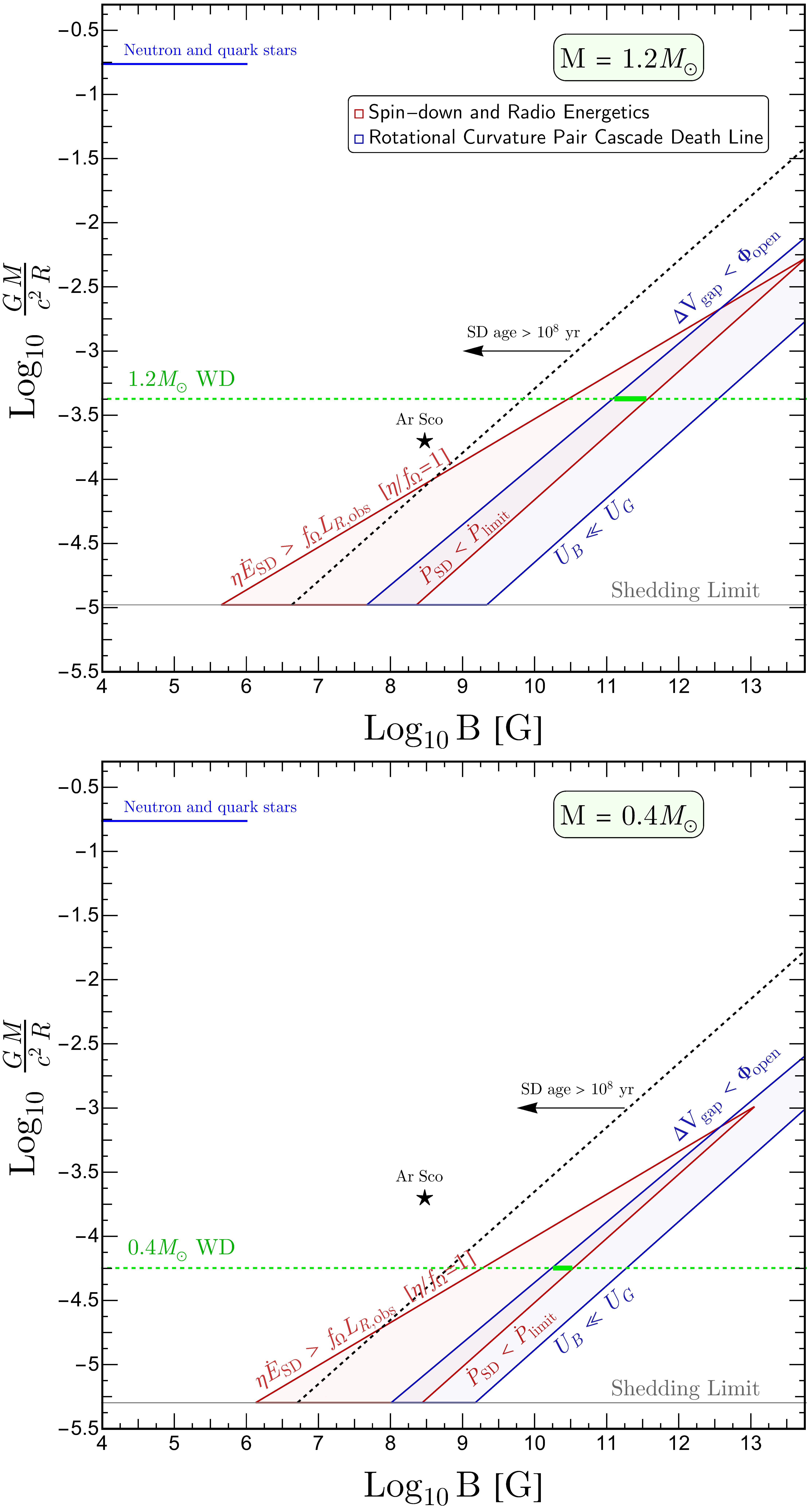}
\caption{Constraints on a rotation-powered origin for GLEAM-X J1627 ($P=1091$ s) expressed in terms of stellar compactness versus dipolar magnetic field at the surface, $B$. The red shaded region indicates the allowed region from spindown energetics $\eta \dot{E}_{\rm SD} > f_\Omega L_R \sim f_\Omega 10^{31.5}$ erg s$^{-1}$ and the limit $\dot{P}_{\rm SD} < P_{\rm limit} \approx 10^{-9}$ s s$^{-1}$ while the blue region is a curvature pair cascade death line bounded by the hydromagnetic stability limit; both regions are for mass $1.2 M_\odot$. Only the region within the blue {\it and} red shaded regions is allowed for rotation powered WDs. Likewise, the second lower panel indicates similar regions for $0.4 M_\odot$. The mass shedding or break up limit is indicated by the gray horizontal line, as well as relevant compactness for stable WD equation of state at the indicated masses. The thick solid green segment indicates the allowed region, requiring outlandishly high magnetization $B~>~10^{10}-10^{11}$~G for a putative WD. This allowed parameter region is narrow and largely untenable. Furthermore, even this narrow region is ruled out for optical limits corresponding to ages from standard WD cooling models (black dashed line, applicable to the range of standard WD compactness and composition).   }
\label{fig:sdconstraints}
\end{figure}

We may generically quantify the above arguments and constraints at the measured period of GLEAM-X J1627 for arbitrary stellar compactness $G M/(c^2 R)$ versus dipolar surface magnetic field $B$,  assuming the moment of inertia is approximately $2 M R^2/5$ (a good approximation for degenerate stars within a factor of two). The following restrictions largely rule out rotation-powered WDs for GLEAM-X J1627. Broadly, we require (i) a putative rotation-powered nature of radio emission $\eta \dot{E}/f_\Omega >  L_{R,\rm obs} \sim~10^{31.5}$~erg~s$^{-1}$ (ii) electromagnetic spindown rate less than the observed constraint\footnote{A more restrictive smaller value, e.g. in Extended Figure 2 in \cite{Hurley-Walker2022} yields no viable phase space for a rotation-powered WD scenario in Figure~\ref{fig:sdconstraints}.} $\dot{P} < P_{\rm limit} \sim 10^{-9}$ s s$^{-1}$, (iii) hydromagnetic stability $U_B \sim B^2 R^3 \ll G M^2/R \sim U_G$ (iv) compactness above the rotational shedding or breakup limit $\Omega^2 R \ll G M/R^2$ ($\Omega = 2\pi/P$) and (v) curvature photon magnetic pair cascades for high brightness temperature, highly polarized, pulsar-like emission in a vacuum gap-like condition. Propagation of such radio emission in the magnetosphere of the object also sets constraints on the hierarchy of plasma and cyclotron frequencies demanded \citep[if, for example, observed emission are ordinary eigemodes of the magnetized plasma, e.g.,][]{1986ApJ...302..120A}, but it turns out that the pair cascade constraint is most limiting for the scales of interest here. As shown in \S \ref{sec:voltagegap}, the gap constraint is essentially a radio death line (assuming curvature radius $\rho_c \sim 10 R$), derived using the condition that the curvature photon gap height is much smaller than the polar cap rim boundary $\ell_{\rm gap} < r_{\rm pc} \sim \sqrt{2 \pi R^3/(P c)}$. Equivalently the voltage drop in the gap is less than the open field line potential $\Delta V_{\rm gap} < \Phi_{\rm open}$, or the accelerated primary luminosity (which cascades into $e^+e^-$ pairs) $L_{e^\pm} \sim 2 \pi A \ell_{\rm gap}^2\rho_{\rm gap}^2 c < \dot{E}_{\rm SD}$. Here $A$ is the characteristic polar cap area $\pi r_{\rm pc}^2$, while $\rho_{\rm gap} \sim \rho_{\rm GJ} \sim B/(P c)$ is the Goldreich-Julian charge density. Finally, for a WD scenario, optical limits and standard WD cooling models (\S\ref{optgleammain}) imply that the true age and thus also the characteristic spindown age ($\propto P/\dot{P}$) is greater than about $10^8$~yr. This is a conservative estimate as epochs of enhanced spindown only lower the maximum magnetization allowed.

We plot these constraints for $M= \{0.4,1.2\} M_\odot$ in Figure~\ref{fig:sdconstraints}, along with a horizontal line indicating a typical compactness value from known WD mass-radius relations, as well as the mass and magnetization of Ar Sco \citep{2018AJ....156..150S,2019ApJ...887...44D,2022MNRAS.516.5052P}. There are several important features of the constraints in Figure~\ref{fig:sdconstraints} to note. First, the observed limit $\dot{P}_{\rm limit} \lesssim 10^{-9}$ s s$^{-1}$ is quite constraining to models with large magnetic fields, particularly objects with low compactness or large radii. At the same time, the gap constraints above and Eq.~(\ref{eq:minradius}) require a minimum compactness for a given $B$, yielding the rather restrictive allowed parameter space of a WD with outlandishly high surface dipole field $B> 10^{10}$ G and $\dot{E}_{\rm SD} \sim 10^{33}-10^{34}$ erg s$^{-1}$. Note that even the narrow $B$ range that is consistent with conditions (i)-(v) listed above, is ruled out for standard composition (and mass) cooling limits considering the optical upper limits (\S\ref{optgleammain}). Thus the arguments presented above largely rule out a rotation-powered WD nature for GLEAM-X J1627.

\vspace{0.1in}
\noindent{\bf Other WD models}

We close this discussion with a brief description of other models that have been suggested to power GLEAM-X J1627. \cite{2022RNAAS...6...27L} have suggested that GLEAM-X J1627 is a hot proto-WD or sub-dwarf. Due to the large ($\sim 0.3 R_{\odot}$) radius of the WD in this model (required to explain the observed period as corresponding to the break-up limit), the spindown luminosity in this model is extreme ($\sim 10^{38}\mbox{erg s}^{-1}$). The corresponding effective temperature at the surface would be $\sim 10^{4.5}-10^5\mbox{ K}$ \citep{Heber2009}. The existence of such an object is strongly ruled out by existing Gaia/DECam observations discussed above (see Figure \ref{fig:wd_age_temp} and Appendix \ref{sec:gleam_opt_count}). For instance, a hot proto-WD with $T\sim 10^5$\,K and other parameters as listed above would have a Gaia G band magnitude of $\approx12.5$ at the distance and corresponding absorption of GLEAM-X J1627, which is about 8.5 magnitudes brighter than Gaia's limiting magnitude. 
We conclude that an ultra-long period magnetar description for GLEAM-X J1627 is a more natural working hypothesis.

\subsection{1E 161348--5055}
\label{RCW103}
1E 161348--5055 is the pulsating ($P \sim 6.7$ hour) central compact object in SNR RCW 103, and its magnetar nature is well-established \citep{DeLuca2006,2008ApJ...682.1185D,2011MNRAS.418..170E,2016ApJ...833..265T,2016ApJ...828L..13R,2016MNRAS.463.2394D,2017ApJ...841...11T,2017MNRAS.464L..65H,2018MNRAS.478..741B}. From detailed modeling of the SNR, \cite{2019MNRAS.489.4444B} estimate the age of 1E 161348--5055 to be $880-4400$\,yrs.

\subsection{PSR J0250+5854}

Until recently, PSR~J0250+5854 was the longest period {\it{bona fide}} radio pulsar with $P\approx 23.5$\,s and $B \approx 3 \times 10^{13}$\,G \citep{2018ApJ...866...54T}. It is however consistent with being rotation-powered and has not yet exhibited any magnetar-like activity. That being said, it's spindown age of $14$ Myr is the largest in Table~\ref{tab:sources}. We note that in this case, the characteristic age is likely an unreliable estimate of the true age as slow can pulsars show a significant amount of timing noise leading to spurious measurements of age~\citep[see][for more details]{lower2020} and because of long-term magnetic field evolution. Thus its currently-measured spindown age can be regarded as an upper limit on the true age.

\subsection{GCRT J1745--3009}
GCRT J1745--3009 is the Galactic center ``Burper" discovered serendipitously by the VLA at 330 MHz \citep{2005Natur.434...50H}. It exhibited bright $\sim 1$ Jy 10-minute-wide ``pulses" at a cadence of $P \sim 77$ minutes, as well as sporadic on/off epochs in follow-up observations \citep{2007ApJ...660L.121H,2009A&A...502..549S}. The $\lesssim$ minute-timescale variability implies a brightness temperature $\gg 10^{12} (d/70 \rm pc)^{2}$~K \citep{2007ApJ...660L.121H,2009A&A...502..549S} consistent with a coherent emission mechanism, if the source is not associated with a nearby stellar object. Optical observations \citep{2008ApJ...687..262K} largely rule out any plausible nearby M/brown-dwarf counterpart within $\sim 0.1-1$ kpc, implying GCRT J1745--3009 is indeed a distant coherent emitter. The only known bright coherent emitters are NSs, and if the pulsations are ascribed to spin periodicity\footnote{\cite{2009A&A...502..549S} disfavors aperiodicity in the \cite{2005Natur.434...50H} data-set, suggesting the source is a relatively stable rotator.}, the source cannot be rotation-powered, suggesting a magnetar origin. In this regard, the source is analogous to GLEAM-X J1627 but whose full window of activity was missed owing to technological limitations of the time.

\section{Evidence for enhanced spindown}
\label{sec:enhancedSD}
As shown in \S \ref{sec:ULPMcand}, there are several ULPM candidates in the Galaxy. The existence of such systems requires a mechanism for substantially spinning down some magnetars to spin frequencies well below those that are realized by the confirmed Galactic magnetar population. Indeed it has been shown that in the presence of field decay, dipole spindown becomes increasingly inefficient, and as a result standard magnetars' spin periods tend to reach a maximum value of $\sim 12$\, s \citep{Dall'Osso2012D,Beniamini2019}. However, the presence of enhanced spindown mechanisms, operating for limited durations can be inferred from observed features in the confirmed magnetar population. Below we list some of this evidence, and discuss how it could result in extremely prolonged spin periods in a fraction of the magnetar population.

Several known magnetars exhibit enhanced spindown in association with high-energy activity. For example,  SGR\,1900+14, has  exhibited an enhanced 
spindown of $x_{\rm P}\equiv \Delta P /P\approx10^{-4}$ after its 1998 giant flare (GF). 
Since its prolific 2004 GF, SGR\,1806$-$20 has increased its $\dot{P}$ considerably relative to its pre-GF value (\citealt{Younes2015};  although $x_{\rm P}\lesssim 5\times 10^{-6}$ during the time of the main spike of the GF, \citealt{Woods2007}). By 2012, its $P$ increased by $\sim\!\,2\%$ compared to the extrapolated value from 1994. Kinematic age constraints for SGR\,1806$-$20 and SGR\,1900+14 \citep{2012ApJ...761...76T} that imply effective braking indices $n\ll3$ also demand enhanced spindown in the past history of these magnetars, either as a result of field decay or sporadic stretches with monopolar ($n\approx 1$) winds \citep{BT1998,Harding1999}. More generally, it has been observed that magnetars suffer `glitches' or `anti-glitches', sometimes close to an outburst onset, where their periods decrease or increase by $|x_P|\sim 10^{-9}-10^{-5}$ abruptly \citep{Archibald2013,dib14ApJ,younes20ApJ:2259,younes2022:1935}. Perhaps more important, is that these phenomena are often followed by a subsequent episode of enhanced spindown, the effect of which often dominates the temporal evolution of $P$.
An illuminating example is 1E 1048.1-5937, which experienced several glitches over the years, that were followed by enhanced spindown episodes on years-long timescale \citep{Archibald2020}. These episodes resulted in $x_P\sim 2\times 10^{-4}$, about two orders of magnitude greater than associated with the glitches themselves (and with opposite sign). Another interesting recent example is that of Swift J1818.0--1607 which after an outburst in 2020 has experienced both a candidate glitch and an anti-glitch, as well as a long-term highly fluctuating (on a timescale of days) spindown rate \citep{Hu2020}.

Such events of small fractional increase in $P$ can accumulate to produce a large total increase in $P$, if their number per magnetar lifetime is $N_{\rm P}\gtrsim\!x_{\rm P}^{-1}$. Indeed, if the value of $x_{\rm P}$ is independent of $P$, then $P$ can grow exponentially, $P_{\rm f}=P_{\rm 0}\exp(N_{\rm P} x_{\rm P})$ and $P_{\rm f}\gg P_{\rm 0}$ for $N_{\rm P}\gtrsim\!x_{\rm P}^{-1}$.
For the inferred GF energy, $E_{\rm GF}\approx 4\times 10^{44}\,$erg and $x_{\rm P}\approx10^{-4}$ from SGR\,1900+14, a significant increase of $P$ due to the cumulative effect of many GFs requires an initial magnetic energy reservoir of $E_{B,0}\approx E_{B,{\rm int},0}>E_{\rm GF}x_{\rm P}^{-1}\sim4\times10^{48}\,$erg that can be used to power GFs. This magnetic energy corresponds to a quite reasonable initial internal magnetic field strength $B_{{\rm int},0}~>~5\times10^{15}$~G (compare this to the current surface field strength of SGR\,1900+14 inferred from dipole spindown $B_{\rm d}=7\times 10^{14}$~G).
Although far from being a proof, this simple calculation shows the idea that some magnetars (that start with a sufficiently large energy reservoir) can spin down substantially due to the accumulated effect of outbursts, is plausible and merits further investigation. 

\section{Source densities of ULPMs}
\label{sec:numbers}
The local density of sources similar to a given object, can be estimated by the distance to that object, assuming it to be the closest member of the group. If due to observational selection effects (for example due to beaming), there are closer (and missed) similar objects, this analysis will overestimate the minimum distance, $d_{\rm min}$ and thus underestimate the true number of sources. The discussion below is divided into two parts. In the first (\S \ref{sec:simplenum}) we provide an approximate, but more straightforward estimate for the minimum source density, designed to give intuition to the reader. This analysis has the advantage of being applicable to a general category of underlying sources (not restricted to magnetars). In the second part (\S \ref{sec:popsynthnum}), we consider a more elaborate and realistic model that is however, dependent on the ULPM candidates being NSs. Moderate errors due to these types of analysis are unavoidable due to small-number Poisson statistics. However, at a 90\% confidence limit, we can still constrain the number of sources to better than an order of magnitude, which is constraining for the underlying physical picture.

\subsection{Simplified estimate - isotropic distributions}
\label{sec:simplenum}
We consider the underlying distribution to be approximately isotropic (in either 3 or 2D, see below for details) on a scale comparable to $d_{\rm min}$ (motivated by the proximity of PSR J0901--4046 and GLEAM-X J1627).
The method applied follows the analysis of \cite{LBK2021}. Given a source number density, $n$, the probability density of the volume enclosing the closest member, $V_{1}$ is (using Poisson statistics) $\mbox{dP}/\mbox{d}V_{1}|_n\!=\!n\exp{(-V_1n)}$. Using Bayes theorem with an uninformative prior, this can be converted to $\mbox{dP}/\mbox{d}n|_{V_1}\!=\!V_{1}\exp{(-V_{1}n)}$.
Under the assumptions, we have, by construction, $d_{\rm min}\!=\!(3V_1/4\pi)^{1/3}$ for a 3D distribution and $d_{\rm min}\!=\!(V_1/\pi)^{1/2}$ for the 2D case. We see that $d_{\rm min}$ gives an estimate of the local source density.

Using this approach we turn to estimate the total number of ULPM candidate sources in the Galaxy. This is attained by first calculating the fraction of Galactic systems that reside up to a distance $d$ from the Sun. Assuming an object population that tracks the stellar population of the Milky Way (we address below the possibility that ULPMs are confined to the disk only), this fraction is given by:
\begin{equation}
\label{eq:fd}
    f_d=\frac{4\pi\int_0^d dr'\int_0^{\pi}d\theta'\int_0^{2\pi}d\phi' n_*(r,z) r'^2}{2\pi \int_0^{\infty} dr \int_0^{\infty} dz n_*(r,z)r}
\end{equation}
where $r,z,\phi$ are polar coordinates measured from the Galactic center and $n_*(r,z)$ is the (unnormalized) stellar density in the Milky Way. As a first approximation, the latter can be described by $n_*(r,z)=\exp(-r/h_r)\exp(-|z|/h_z)$ with $h_r=3.5\mbox{ kpc}, h_z=0.25\mbox{ kpc}$ \citep{1980ApJS...44...73B} being the scale-lengths in $r,z$ (more sophisticated distributions are discussed in \S \ref{sec:popsynthnum}). Finally, $r',\theta',\phi'$ in Eq.~\ref{eq:fd} are spherical coordinates as measured from the Sun, i.e
\begin{eqnarray}
& r'=\left[(r\cos \phi -R_{\rm sun})^2+r^2\sin^2\phi+z^2\right]^{1/2} \nonumber \\
& \cos \theta' = \frac{z}{r'} \quad \& \quad  \tan \phi=\frac{r\sin \phi}{r\cos \phi-R_{\rm sun}}
\end{eqnarray}
where $R_{\rm sun}=8\mbox{ kpc}$.

With these definitions, the total number of objects in the Galaxy is on average $\langle N \rangle =\int dn \int dV n\frac{dP}{dn} 2\pi h_r^2 h_z n_*(r,z)=f_{\rm d}^{-1}$. 
Taking a distance of $0.4$kpc for PSR~J0901--4046 we estimate a lower limit of $N\!\approx\!3620^{+9000}_{-3000}$ similar objects in the entire Galaxy (where quoted uncertainties are 90\% confidence limits, given Poisson statistics, i.e. we estimate that for the quoted number,$X^{+Z}_{-Y}$, we have $P(N\!>\!X)\!=\!0.5$, $P(N\!>\!X\!-\!Y)\!=\!0.95$, $P(N\!>\!X\!+\!Z)\!=\!0.05$).
Similarly, with a distance estimate of $1.3$~kpc for GLEAM-X J1627 we estimate a lower limit on the number of similar objects to be $N\!\approx\!225^{+565}_{-190}$. If we take a further conservative assumption, that all similar objects reside in the Galactic plane \citep[as magnetars are observed to have small offsets from the plane,][]{KaspiBeloborodov2017,2017ApJS..231....8E}, the numbers are reduced to $N_{\rm 2D}\!\approx \!2500^{+5450}_{-2145}$ for PSR~J0901--4046 and $N_{\rm 2D}\!\approx\!220^{+490}_{-188}$ for GLEAM-X J1627. These lower limits are likely to be overly conservative as the observation of magnetars being confined to the Galactic plane could be an observational bias towards detecting X-ray active and therefore also much younger objects.
A similar analysis can be realized with the other ULPM candidates but becomes less constraining as the distances get larger. For example, for PSR J0250+5854, we get $N\!\approx\!150^{+450}_{-135}$ ($N_{\rm 2D}\!\approx\!140^{+440}_{-130}$). If some or all of these sources belong to the same category, then the estimate will be dominated by the closer object. 

\subsection{More realistic estimate - population synthesis approach}
\label{sec:popsynthnum}
A more realistic (but model dependent) estimate of the (lower limit on the) number of Galactic sources needed to explain a given observed $d_{\rm min}$ is achieved by accounting for the birth rates, locations, velocities and ages of a hypothetical population. Our analysis here builds upon the work by \cite{GiguereKaspi2006} done in the context of the general pulsar population. We perform a Monte Carlo calculation to derive the birth properties of the objects. We then propagate the motion of each object according to the Galactic potential.

The initial location of each system is assumed to follow the spiral arm structure of the Milky Way (see \citealt{GiguereKaspi2006} for more details) in either 2 (i.e. confined to the plane) or 3 (i.e. taking into account the disk scale height) dimensions. As ULPM candidates are most likely NSs (see discussion in \S \ref{sec:WD}), the magnitudes of the birth kicks (relative to the Galactic rotational velocity curve at their birth radius) are taken to follow the two-component Gaussian favoured by the analysis of pulsar proper motions \citep{2002ApJ...568..289A,2020MNRAS.494.3663I}. As we show below, the results are weakly dependent on this assumption. The directions of the kick velocities are arbitrary relative to the Galactic coordinates. The characteristic lifetime of ULPM candidates is left as a free parameter, $T_{\rm age}$. The age of a specific object, $t$, is uniformly distributed in the range $t\!\in\![0,T_{\rm Age}]$. Similarly, the formation rate of systems (systems formed in the Galaxy per unit time), $r$ is left as a free parameter. At a given time, the expected number of Galactic systems is $N_0\!=\!rT_{\rm age}$, and the actual number is given by a Poisson distribution with a mean of $N_0$.

\begin{figure}
\centering
\includegraphics[width = 0.48\textwidth]{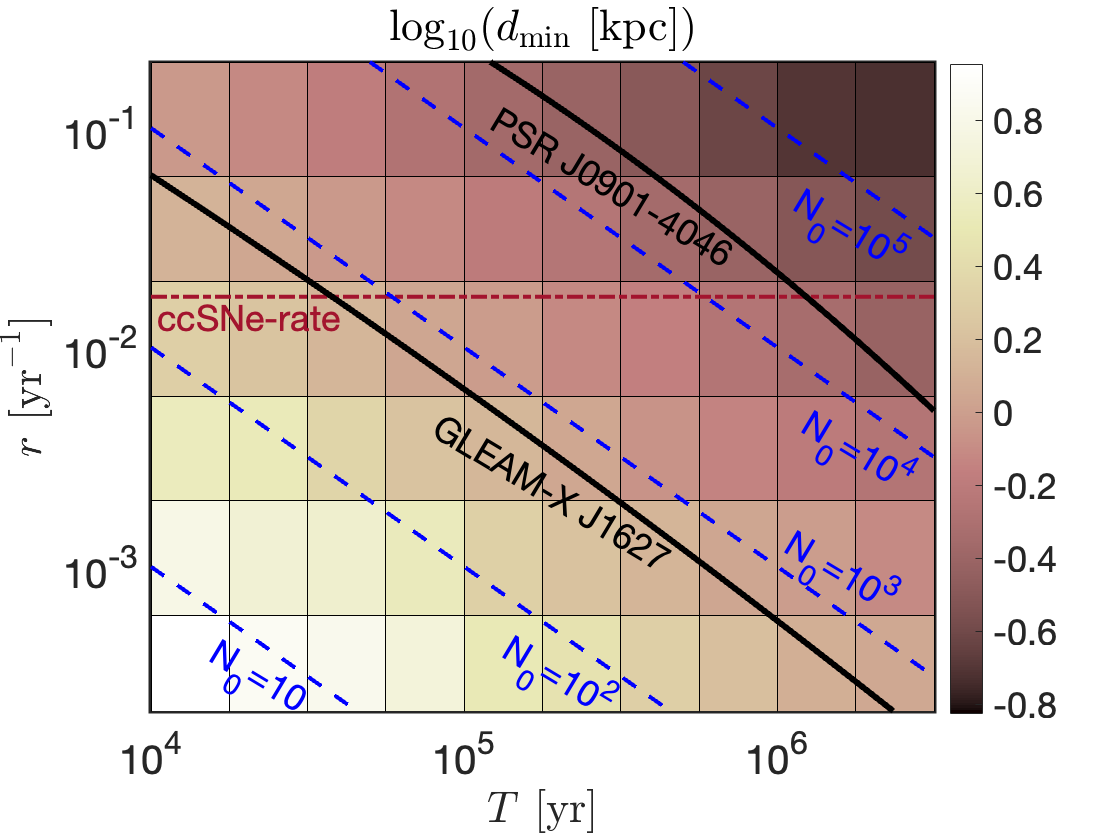}\\
\includegraphics[width = 0.48\textwidth]{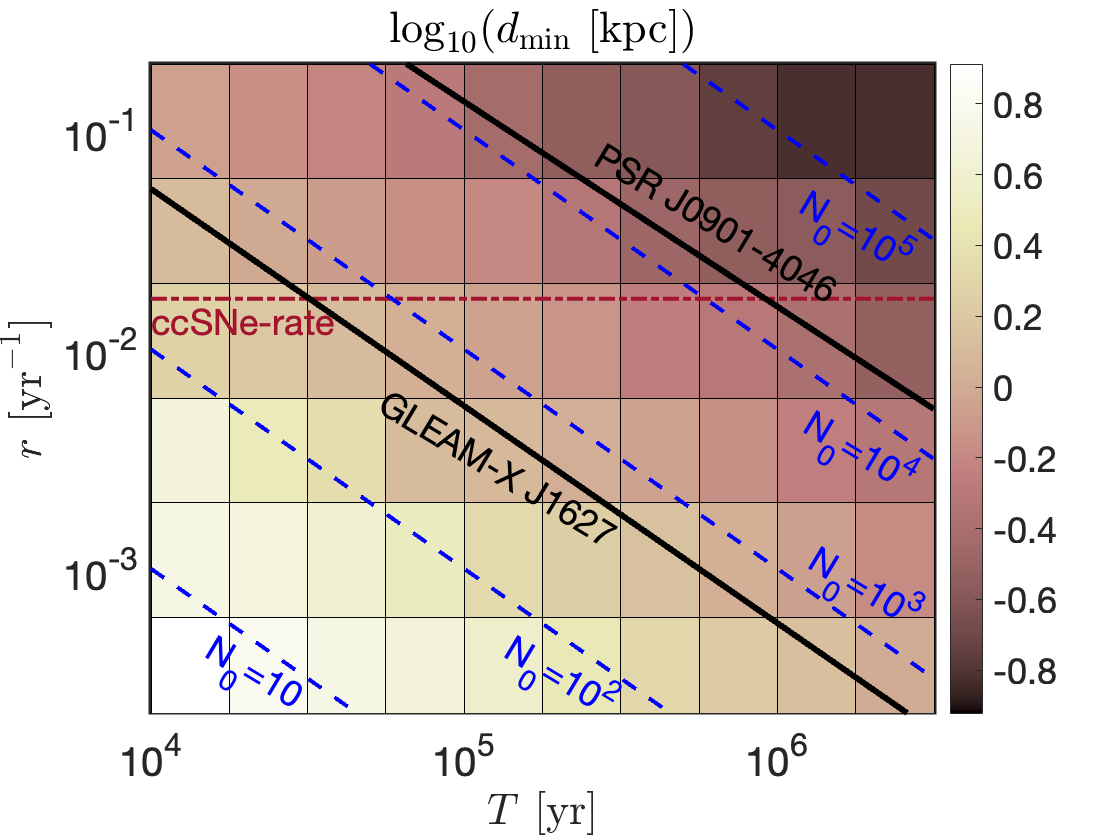}
\caption{Minimum distance between the Sun and ULPM candidates, assuming the later are born with rate $r$ and have a characteristic lifetime of $T_{\rm Age}$. Candidates' birth locations and velocities are randomized as described in \S \ref{sec:popsynthnum}. The numbers shown are median estimates of $d_{\rm min}$, averaged over many realizations with the same underlying properties. The top panel depicts results for sources initially distributed throughout the Galaxy and the lower panel for sources being born in the Galactic plane.}
\label{fig:popsynth}
\end{figure}

After initializing the birth properties, we propagate each system's Galactic motion for a duration $t$ with \texttt{galpy}\footnote{\url{http://github.com/jobovy/galpy}} \citep{Bovy2015} using the \texttt{MWPotential2014} Galactic potential. Finally, we compare the systems' locations to that of the Sun and calculate $d_{\rm min}$, between the Sun and any of the simulated systems. This process is repeated $10^3$ times for each $\{r,T_{\rm Age}\}$. The results are shown in Figure \ref{fig:popsynth}. As expected from \S \ref{sec:simplenum}, the results depend mostly on $N_0$, and are largely insensitive to other combinations of $r,T_{\rm Age}$.
We see that at a $90\%$ confidence level, $N\!=\!19200_{-13600}^{+40000}$ ($N\!=\!590_{-500}^{+1010}$) for PSR~J0901--4046 (GLEAM-X J1627 ) if the initial distribution of systems is taken in 3 dimensions. Confining the birth locations to the Galactic plane, only slightly changes these estimates: $N_{\rm 2D}\!=\!12800_{-10100}^{+19000}$ ($N_{\rm 2D}\!=\!510_{-420}^{+500}$) for PSR~J0901--4046 (GLEAM-X J1627).

Finally, we note that if systems are preferentially born closer to the Galactic center (e.g. within the Galactic bar), $d_{\rm min}$ increases dramatically for a given $\{r,T_{\rm Age}\}$. The implication is that the total required number of systems in the Galaxy will increase significantly compared to the estimates above under such an assumption.

\section{Age estimates}
\label{sec:Age}
We turn next to show that the Galactic ULPM candidates are likely to be old systems relative to SGRs or AXPs.
This is motivated by several independent considerations that, combined, allow us to estimate the ages of ULPM candidates. 

\subsection{Spindown age limits}

The period and period derivative of a NS provide us with the ``spindown age", $\tau_{\rm SD}=P/(2\dot{P})$. This is generally an upper limit on the true age of the NS (and this conclusion is strengthened in the presence of magnetic field decay, see \citealt{Dall'Osso2012D,Vigano2013,Beniamini2019}). For GLEAM-X J1627, the extant upper limit on $\dot{P}$ leads to $\tau_{\rm SD}>170\mbox{ kyr}$. For PSR~J0901--4046, $\dot{P}$ is measured, yielding $\tau<\tau_{\rm SD}=12\mbox{ Myr}$.

\begin{figure*}
\centering
\includegraphics[width=0.95\textwidth]{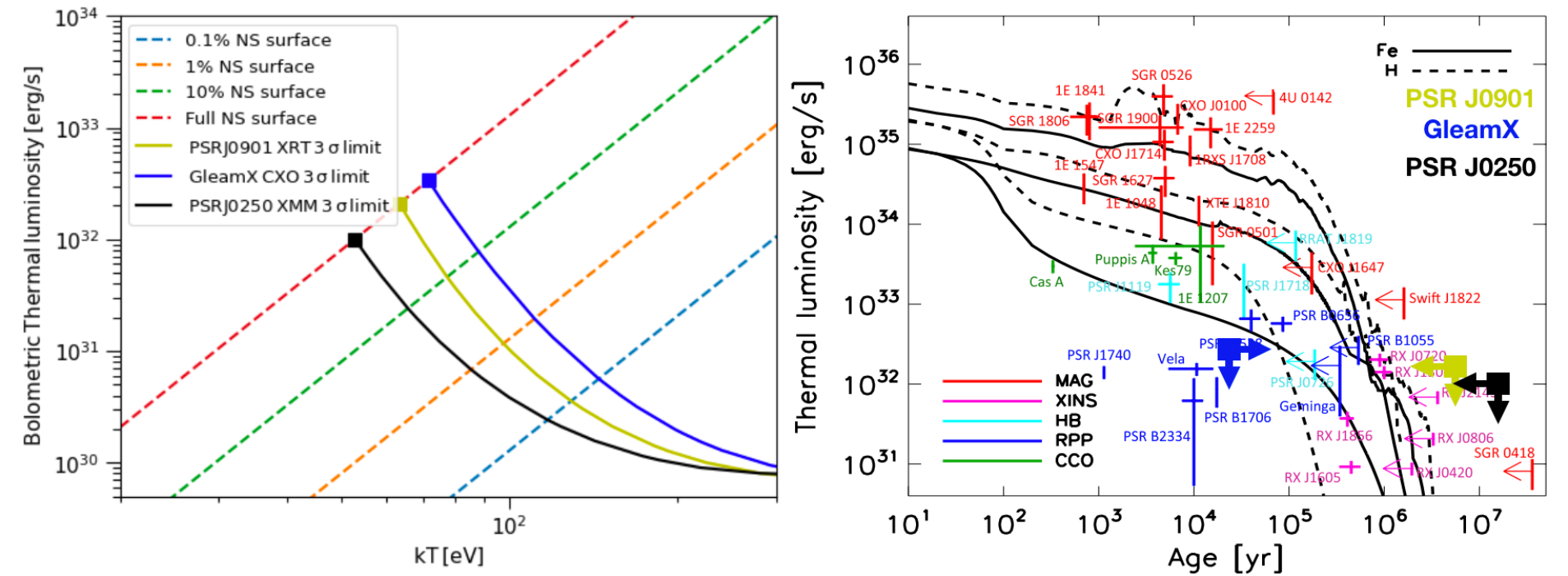}
\caption{ {\sl Left:} 3$\sigma$ upper-limit on the thermal X-ray luminosity as a function of temperature for PSR~J0901--4046, GLEAM-X J1627, and PSR J0250+5854. PSR J0901--4046 is a factor of $\!\sim\!3$ and 4 closer than GLEAM-X J1627 and PSR J0250+5854, respectively. {\sl Right:} Bolometric thermal luminosity versus age for different types of thermally emitting NSs (figure adapted from \citealt{Vigano2013}). The $3\sigma$ upper-limits for PSR~J0901--4046 (yellow),  GLEAM-X~J1627 (blue), and PSR J0250+5854 (black), assuming the entire NS surface is thermally emitting, have been placed on the plot for comparison (see Appendix \ref{xray_analysis} for details).  }
\label{fig:thermalage}
\end{figure*}

\begin{figure}
\centering
\includegraphics[width = 0.48\textwidth]{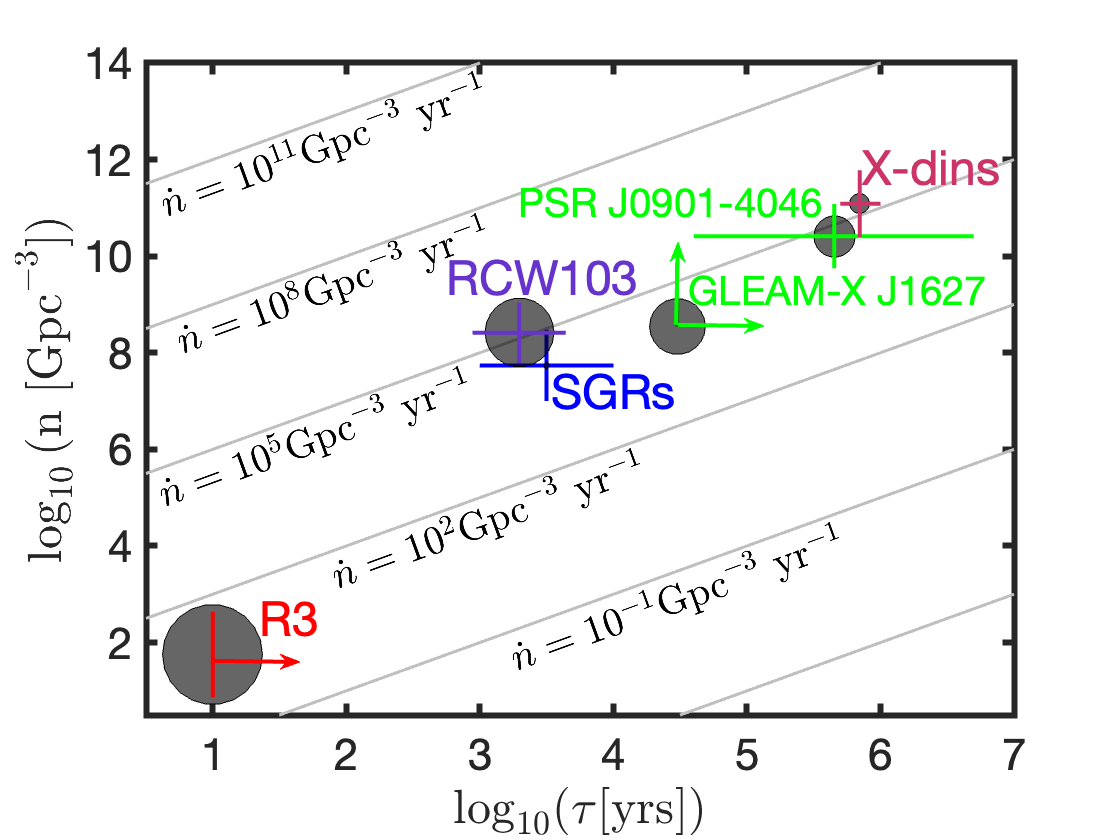}\\
\includegraphics[width = 0.23\textwidth]{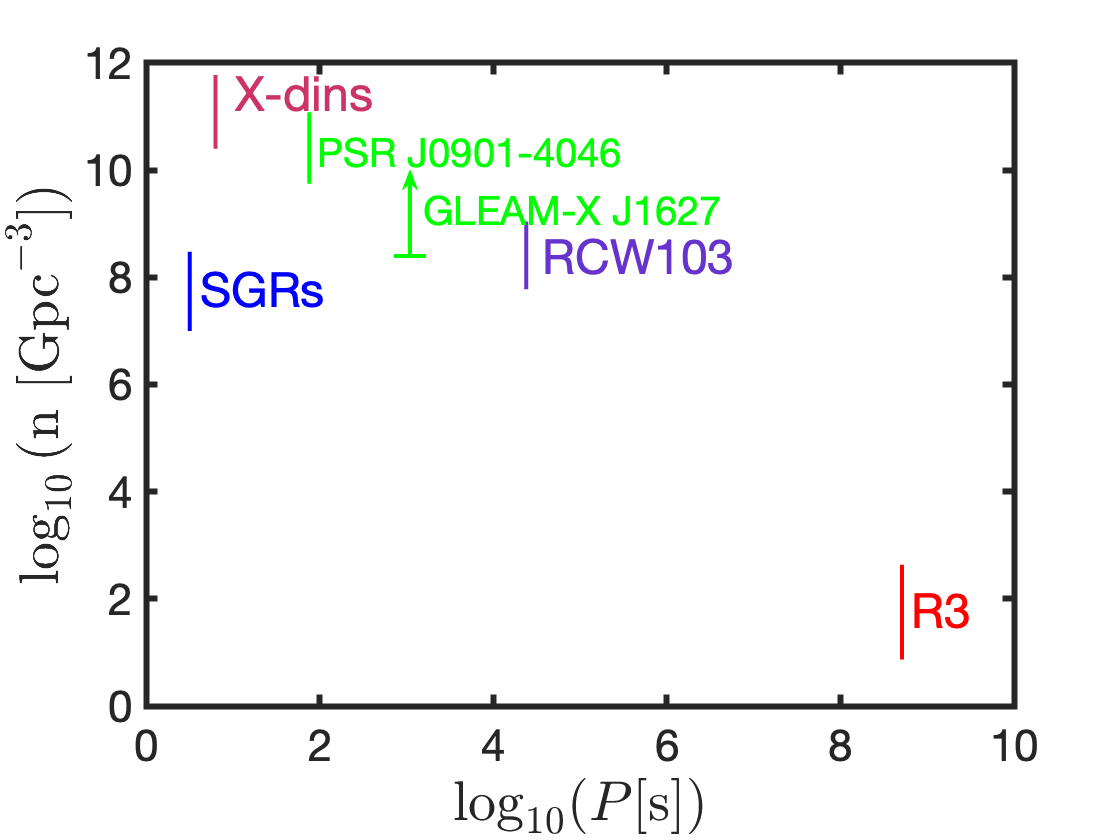}
\hspace{-0.3cm}
\includegraphics[width = 0.23\textwidth]{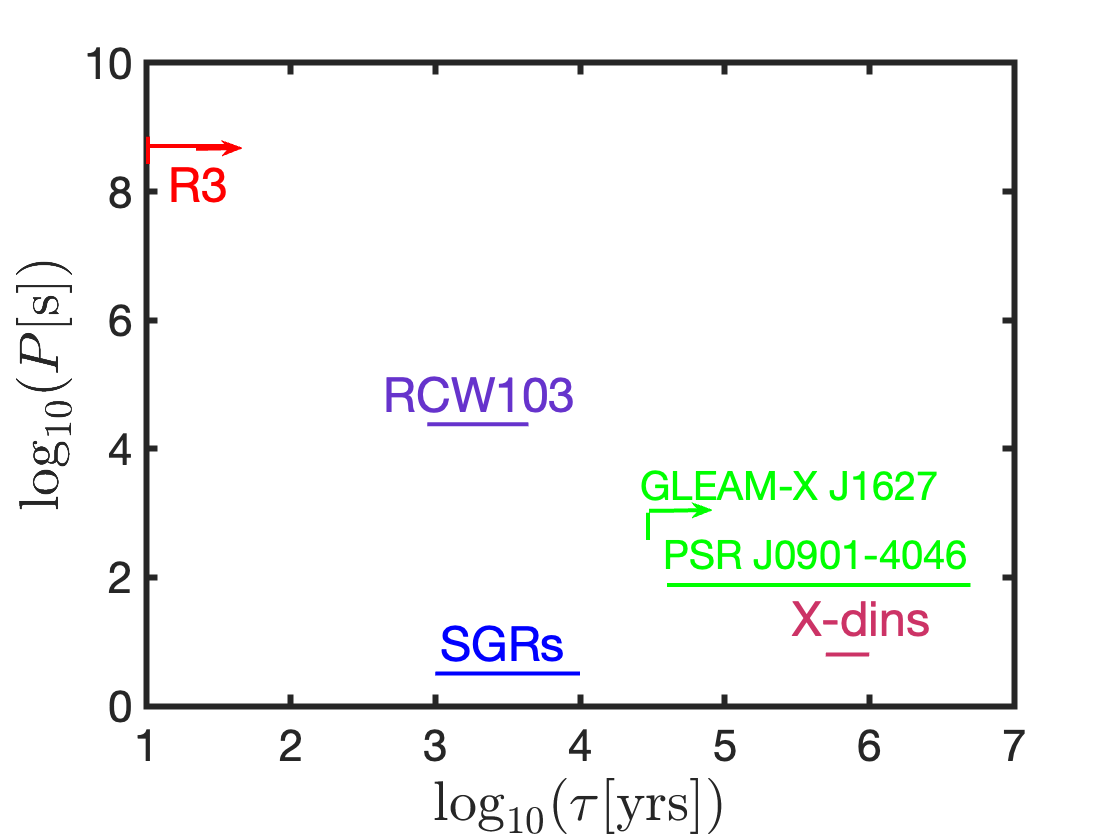}
\caption{Top: Inferred active lifetime $\tau$ and source density $n$ for FRB 20180916B (aka R3, the closest of the highly active cosmological FRBs), SGR 1935+2154 type FRBs, ULPM candidates discussed in this work and XDINS. $n$ is inferred from the distance to the nearest object of that class taking into account Poisson statistics and 90$\%$ confidence limits. For Galactic populations, we correct for the density of Milky Way-like galaxies taking $n_{\rm MW}=0.01\mbox{Mpc}^{-3}$. In addition, for SGRs, the density ranges consider whether it is only SGR 1935+2154 or all Galactic magnetars which are capable of producing an FRB. Age estimates for the SGR-type population are informed by Galactic magnetar evolution studies \citep{Dall'Osso2012D,Beniamini2019}, while for R3 we only take a conservative lower limit of 10 years, informed by its steady behaviour since it has first been observed. The figure demonstrates that the two types of FRBs correspond to vastly diverging source birth rates. The circle sizes corresponds to $\log_{10}(P [s])$ and shows that the different Galactic sources are unlikely to be part of a single evolutionary trajectory. Bottom panels depict the projections on the $n-P$ (left) and $P-\tau$ (right) planes.}
\label{fig:nVsP}
\end{figure}

\subsection{Timing stability}
PSR J0901--4046 exhibits remarkable timing stability for a magnetar, the stability is more commensurate with middle-aged pulsars. Over the span of 7 months, PSR~J0901--4046 has shown a significantly lower timing noise compared to other slow radio pulsars~\citep{lower2020}. The timing residuals show no signature of higher order frequency derivatives~\citep[see][for more details on the timing properties]{caleb2022}, suggesting an exceedingly stable magnetospheric state. This provides valuable information about the age of the NS. Assuming that increasing stability with age is universal, a relative timing stability (over 7 months) and glitch size constraint of $\sim \Delta \nu/\nu \lesssim 10^{-6}$ implies an age $\gtrsim 100$ kyr from the known glitch activity distribution of young rotation-powered pulsars \citep{2011MNRAS.414.1679E,2017A&A...608A.131F,2022MNRAS.510.4049B}.

\subsection{Source density age limits}
A conservative limit on the local Galactic formation rate is given by the CCSN rate, $\dot{N}<\dot{N}_{\rm ccSNe}\approx 17\pm 2\mbox{ kyr}^{-1}$. Given this birth rate and our estimates for $N$, the number of ULPM candidates in the Galaxy (\S \ref{sec:numbers}), we can estimate the typical lifetime of a given object as $\tau \approx N/\dot{N}\gtrsim 1.2_{-0.9}^{+0.8}\mbox{ Myr}$ for PSR J0901--4046 ($\tau \approx N/\dot{N}\gtrsim 38_{-28}^{+44}\mbox{ kyr}$ for GLEAM-X J1627). These numbers are slightly modified if the systems are only born in the plane, in which case we find $\tau \approx N/\dot{N}\gtrsim 0.8_{-0.6}^{+2.7}\mbox{ Myr}$ for PSR J0901--4046 ($\tau \approx N/\dot{N}\gtrsim 35_{-27}^{+33}\mbox{ kyr}$ for GLEAM-X J1627).

\subsection{Cooling age limits}
An independent limit on the age comes from the X-ray flux limits which can be translated to effective cooling luminosity and surface temperature upper limits (see Appendix \ref{xray_analysis} for details). 

Relating the luminosity and temperature to the age of the NS, depends on various uncertain physical inputs, such as the NS mass, equation of state, the existence or lack of superfluidity and superconducitivity, the composition of the core and the envelope and the magnetic field structure throughout the star. Nonetheless, a data-driven comparison can be made with other NSs with measured thermal luminosities and age estimates.
Such a comparison is shown in Figure \ref{fig:thermalage}. The left panel shows the derived X-ray upper-limits on the bolometric thermal luminosity of the three long period pulsars. The right panels shows how these pulsars' upper-limits on the bolometric thermal luminosities compare with other pulsars of various classes (e.g., magnetars, RPPs, XINs, CCOs). Based on the cooling curves (adapted from \citealt{Vigano2013}), we find that GLEAM-X J1627 is likely to have an age $\tau\!\gtrsim \!10^{5}$\,yr. The upper-limits on the thermal luminosities of PSR~J0901--4046 and PSR J0250+5854 are consistent with their inferred characteristic ages, but they could be as young as $\tau\!\approx\!1$ Myr and still remain undetected in X-rays.

\subsection{Proper motion}
The most nearby supernova remnants to GLEAM-X J1627 are RCW~103, Kes~32, and G332.0+00.2 \citep{2019JApA...40...36G}. The first hosts a known NS remnant (\S \ref{RCW103}). The other two have angular offsets of $2.72^\circ$ and $2.84^\circ$, respectively. Considering the distance to GLEAM-X J1627, this corresponds to a physical offset of $\!\sim \!60$\,pc. The Earth's proximity to GLEAM-X~J1627 implies that a nearby SNR would have remained detectable unless it was extremely old. 
Combining this with the fastest space velocities of observed pulsars ($\sim2000$~km~s$^{-1}$), we conservatively estimate a minimum kinematic age of $\gtrsim 3\times 10^4$~yr. Indeed $60$\,pc is also the approximate offset between GLEAM-X J1627 and the Galactic plane (see Table \ref{tab:sources}). Another possibility is that GLEAM-X J1627 formed in a young (i.e., $<10$ Myr) stellar cluster. The nearest cluster to GLEAM-X J1627 with a compatible distance and age is DBSB 154 \citep{2013A&A...558A..53K}, which has an estimated distance of $d\approx2.1$ kpc and an age of 5 Myr. GLEAM-X J1627 is offset from this cluster by a distance of $2.84^\circ$, leading to a similar lower-limit on its kinematic age\footnote{This presumes the cluster is at the GLEAM-X~J1627 distance, and not at 2.1 kpc. If instead GLEAM-X J1627 is at 2.1 kpc, the lower-limit on the kinematic age would be $\gtrsim 5\times 10^4$ yr.}. Therefore, requiring that the NS formed in the plane results in a commensurate lower limit on the age of the object. 

We have also carried out a similar study of the field surrounding PSR~J0901--4046. The closest SNR is G261.9+05.5 (2.14$^{\circ}$ offset) but it is at a distance of 2.9 kpc \citep{2019JApA...40...36G}, which is likely too far to be compatible with PSR J0901--4046's DM distance. The next closest SNRs (G266.2-01.2 and the Vela complex) all have known compact objects associated with them. G272.2-03.2 is the next closest SNR, which is 11.4$^{\circ}$ offset from PSR J0901--4046. This corresponds to a physical offset of about 80 pc and to a limit on the kinematic age of $\gtrsim40$ kyr. The closest stellar cluster with a compatible distance estimate is Trumpler~10, which lies at 420 pc and has an age of about 24 Myr \citep{2013A&A...558A..53K}. This cluster is $\approx$3.1$^{\circ}$ offset from PSR J0901--4046. This leads to a fairly unconstraining lower-limit on the kinematic age of $>11$ kyr.

\subsection{Combined age limits and implications}
These different age estimates suggest that systems like PSR~J0901--4046 and GLEAM-X J1627 are rather old, perhaps $\sim\!10^{5.5}- 10^6$~yr or so. Although there are uncertainties involved in each of the limits discussed above, they are independent from each other, lending overall credence to the old age of these systems compared to the known magnetar population. We consider the ULPM candidates' location on the $n\!-\!\tau$ diagram shown in the top panel of Fig.~\ref{fig:nVsP}, where we have also supplemented an additional class of objects, the XDINS, a nearby  ($\sim$\,150\,--\,500\,pc) class of cooling NSs whose origin is still debated but likely related to standard magnetars (relevant details of these sources are given in Table \ref{tab:sources}). The tentative alignment of the confirmed magnetars with the RCW103 source, GLEAM-X J1627, PSR J0901--4046 and the XDINS along a roughly constant value of the volumetric rate density, $\dot{n}$, suggests that these objects arise from similar formation channels, commensurate with common CCSN. At the same time, their vastly differing periods (changing non-monotonically along this supposed sequence), imply that these represent different evolutionary sequences, with similar occurrence rates. It is noteworthy that FRB 20180916B (R3) is a clear outlier from these objects (see discussion in \S \ref{sec:PeriodicFRB}) suggesting it probes a much rarer population altogether, involves fine-tuning in physical conditions necessary for FRB emission, or strong observational selection effects against R3-like objects. Finally, we note that loosely $n \propto P^{-1}$ (even ignoring R3) and that the SGR source density is possibly an underestimate, as the sample is likely not as complete as that of XDINS. This will become important in \S \ref{sec:PeriodicFRB}, when we discuss the possible connection with periodic FRBs.

\section{Magnetic field and spin evolution - Different decay to standard magnetars}
\label{sec:fieldevolution}

\subsection{Phenomenological constraints on field evolution}
\label{sec:phenomfieldevol}

\begin{figure}
\centering
\includegraphics[width = 0.34\textwidth]{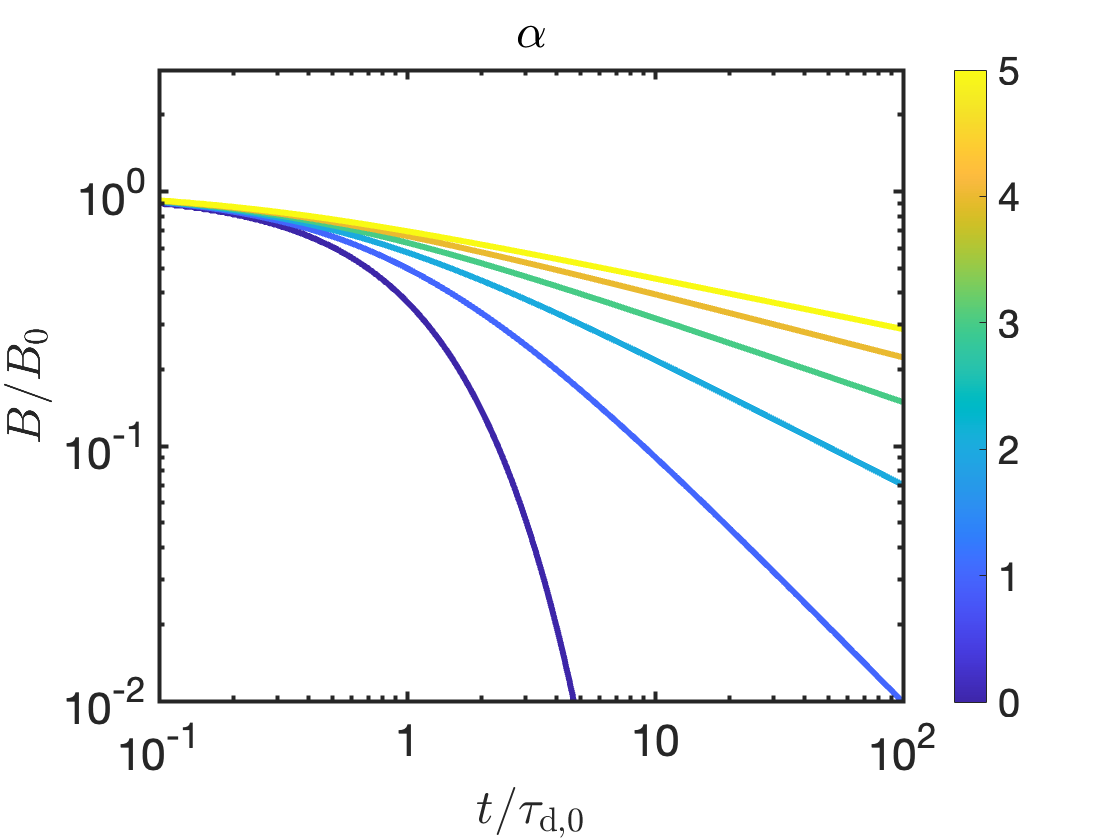}
\caption{Magnetar's $B$-field strength normalised by its initial value, $B/B_0$, versus its true age $t$ normalized by the $B$-field's initial decay time, $t/\tau_{d,0}$ (different curves correspond to $\alpha=5,4,3,2,1,0$ from top to bottom).}
\label{fig:alpha}
\end{figure}

The dipole surface field evolution of confirmed Galactic magnetars has been characterized through $\dot{B}_{\rm d}\!\propto\! B_{\rm d}^{1+\alpha}$ \citep{Colpi2000}. It follows that a magnetar's age $t$ is related to the fractional decrease in the dipole magnetic field, $B_{\rm d}/B_{\rm d,0}$ as
\begin{equation}
\label{eq:alphaconst}
    \frac{t}{\tau_{\rm d,0}}=\frac{1}{\alpha}\left(\left(\frac{B_{\rm d}}{B_{\rm d,0}}\right)^{-\alpha}-1\right).
\end{equation}
where $\tau_{\rm d,0}$ is a characteristic time on which the dipole field decays by a significant amount. We see that a lower limit on $B_{\rm d}/B_{\rm d,0}$ and $t/\tau_{\rm d,0}$ translates to a lower limit on $\alpha$ (see also Fig. \ref{fig:alpha}). Comparison to Galactic magnetar properties suggests typical initial decay times of the dipole field of $\tau_{\rm d,0}\sim 10^3-10^4\mbox{ yr}$ for an initial surface dipole field of $B_{\rm d,0}~\lesssim~10^{15}\mbox{ G}$. Motivated by our results in \S \ref{sec:Age}, we consider the age of PSR J0901--4046 to be approximately $1$\,Myr. If, in addition, we take the surface dipole field as estimated from dipole spindown $B_{\rm d}\sim 2.6 \times 10^{14}$~G, Eq.~(\ref{eq:alphaconst}) leads to an effective $\alpha \gtrsim 4.5$ ($\alpha \gtrsim 6.5$) for $t/\tau_{\rm d,0}\gtrsim 100$ ($t/\tau_{\rm d,0}\gtrsim 1000$). Such values are inconsistent with the known Galactic magnetar population that generally imply $-1\lesssim \alpha \lesssim 1$ for $\tau_{\rm d,0}\sim 10^4\mbox{ yr}$ \citep{Beniamini2019} or $1\lesssim \alpha \lesssim 2$ for $\tau_{\rm d,0}\sim 10^3\mbox{ yr}$ \citep{Dall'Osso2012D}.
They are also inconsistent with certain theoretical predictions (e.g. $\alpha_{\rm int}=6/5$ for the solenoidal mode of ambipolar diffusion, \citealt{Dall'Osso2012D}). This, along with the arguments given in \S \ref{sec:Age}, suggests a different field decay evolution for the ULPM candidates than seen in the confirmed Galactic magnetar population. We note that the ULPM field decay mechanism may be episodic rather than continuous, in which case the effective $\alpha$ above describes only an averaged property of this mechanism on long timescales.

A different possibility is that ULPM candidates have similar $\alpha$ to confirmed magnetars, but begin their life with much larger values of $B_{\rm d,0}$. However, for either $\alpha<1, t/\tau_{\rm d,0}\gtrsim 100$ or $\alpha<2, t/\tau_{\rm d,0}\gtrsim 1000$, one requires $B_{\rm d,0}\gtrsim 10^{16}\mbox{ G}$. Such large dipole fields (and likely even larger internal fields that accompany them) are difficult to obtain from standard channels. This problem becomes more severe considering the large implied formation rates of ULPM candidates (see Figure \ref{fig:nVsP}), comparable to those of confirmed magnetars.

Finally, if the current $\dot{P}$ of the ULPM candidates is not dominated by dipole spindown, then their surface field may be overestimated by $\{P,\dot{P}\}$. As demonstrated in \S \ref{sec:magPSRJ0901}, even if spindown is dominated by a particle wind, we still demand $B_{\rm d}\gtrsim 10^{13}\mbox{ G}$. Such a reduced field strength leads to $\alpha>1$ ($\alpha >1.6$) for $t/\tau_{\rm d,0}\gtrsim 100$ ($t/\tau_{\rm d,0}\gtrsim 1000$) which is marginally consistent with the confirmed magnetar population, but requires a distinct spin evolution between ULPM candidates and confirmed magnetars. Furthermore, the possibility that the current rate of change in $P$ is dominated by a wind could be tested with future observations by constraining the braking index of the ULPM candidates, which should approach in this case $n=1$ rather than the dipole value, $n=3$. We conclude that the ULPM candidates require a distinct magnetic and / or spin decay mechanisms to standard Galactic magnetars.

\subsection{Physical origins of long-lived strong fields in ULPMs and powering transient radio emission}
\label{sec:corefield}

The large phenomenological effective values of $\alpha$ above, and the high source density of PSR J0901--4036 like objects implies the physics of field evolution in the crust and core, substantially different from standard magnetars, ought to be generically realized. Apparently magnetar-like fields, with dipolar components of $\gtrsim 10^{13}-10^{14}$~G, are commonly produced and survive for Myr or longer with limited observational consequences beyond the radio. 

For young magnetars, their field evolution is compatible with nonlinear Hall evolution (in a resistive electron MHD treatment) in their crusts \citep[e.g.,][]{1992ApJ...395..250G,2004ApJ...609..999C,2013MNRAS.434.2480G,Vigano2013,2019LRCA....5....3P,2022Symm...14..130G} on a timescale on about $10^4$~yr (with significant dependence on crust thickness), about 2--3 orders of magnitude shorter than Ohmic evolution from electron scattering. The tentative evidence in favor of a crust in PSR~J0901--4046 (Appendix \ref{sec:crust}) and the high source density disfavors nonstandard crust scenarios (e.g. quark stars, or very light or massive NSs) for this class of object. Crust-threading fields (which may possess significant toroidal components) are thought to power the energetic high-energy activity of conventional magnetars \citep[e.g.,][]{2022arXiv220908598L}. The existence of non-dipolar components is exemplified by the ``low-field" magnetar SGR 0418+5729 (with $P=9.1$ s) whose dipolar field is only $B_p \!\sim\! \!1.2\!\times 10^{13}$~G  \citep{2013ApJ...770...65R}; yet evidence points to much stronger local field strengths up to $10^{14}\!-\!10^{15}$ G \citep{2013Natur.500..312T} as expected from 3D crustal Hall evolution \citep[e.g.,][]{2015PhRvL.114s1101W,2016PNAS..113.3944G}. For electromagnetic spin evolution in quiescence, especially on long timescales, the dipolar component is most germane, as that is most influential at the light cylinder. 

The X-ray limits on PSR J0901--4046 and GLEAM-X J1627 (see Figure~\ref{fig:thermalage}) suggest low magnetic dissipation within the crust, commensurate with the crustal Ohmic timescale \citep{2004ApJ...609..999C} or even lower. A rough guide on the relevant magnetic dissipation timescale in the current epoch is $B_{14}^2 R_6^3/\tau_{\rm decay}\! \sim \! L_X \ll 10^{32}$ erg~s$^{-1}$ implying $\tau_{\rm decay}\!\gg\!1$\,Myr. This is generally beyond the standard Hall evolution timescale for typical densities of the inner crust, and suggests a different physical mechanism currently reigning for field evolution. The crustal nonlinear Hall evolution may be substantially retarded with ``Hall attractor" solutions where the Hall evolution saturates to a level commensurate with Ohmic evolution \citep{2014PhRvL.112q1101G,2014MNRAS.438.1618G,2015PhRvL.114s1101W,2016PNAS..113.3944G}. Such attractor solutions are predominantly independent of initial conditions. The dipolar component in this scenario may actually increases with time, implying unconventional tracks in the $P-\dot{P}$ diagram \citep[see also][]{2017JPhCS.932a2048P}. However, a major caveat is that these models assume a crust-dominant field, compatible with Meissner type-I superconductor for the outer core where the field is entirely expelled. This core boundary condition is a strong assumption and unrealistic for common CCSN formation channels (see below). Another assumption in these models is a time-independent specification for the crust conductivity, rather than coupled magnetic and thermal evolution. The assumption of a crust dominant field is also likely inapplicable for old systems which manifest strong global dipolar components exceeding $10^{13}$~G (see \ref{sec:magPSRJ0901}). This perhaps points towards strong core-bound fields in PSR J0901--4046, GLEAM-X J1627 and similar objects. Note that coupled core-crust evolution, with simplifying assumptions and model choices of toroidal/poloidal field evolution, can also obtain Hall attractor solutions \citep{2018MNRAS.473.2771B}, including longer evolutionary timescales that are required for ULPMs.

Proton superconductivity is expected to set in very soon after the formation of NSs by CCSN \citep{1969Natur.224..673B,2004ARA&A..42..169Y}, as the core temperature drops from neutrino cooling \citep[by either modified URCA or direct URCA processes,][]{1983bhwd.book.....S} below the critical proton superconducting temperature. In general, two approximate critical fields exist, $H_{c1} \sim 10^{15}$~G and $H_{c2}\!\sim \!10^{16}$~G (which depend on depth and local properties) where for the magnetic field $H\!<\!H_{c1}$ the protons are in a type-I Meissner state, while for $H\!>\!H_{c2}$ superconductivity is destroyed. Numerous works have considered the evolution of core fields in NS interiors in the context of neutron superfluidity and proton superconductivity \citep[e.g.,][]{1985Ap&SS.115...43M,2006MNRAS.365..339J,2011MNRAS.410..805G,2011MNRAS.413.2021G,2013MNRAS.431.2986H,2013PhRvL.110g1101L,2014MNRAS.437..424L,2015MNRAS.452.3246P,2015MNRAS.453..671G,2015PhRvC..91c5805S,2016MNRAS.456.4461E,2016PhRvD..94h3006G,2016PhRvD..93f4033G,2017MNRAS.465.3416P,2017MNRAS.469.4979P,2017PhRvC..96f5801H,2017MNRAS.467L.115D,2017AN....338.1090G,2018ASSL..457..401H,2021PASA...38...43S,2022Univ....8..228W} following the pioneering work of \citet{1969Natur.224..673B,1969Natur.224..674B}.  The physics in this multicomponent and multiscale system is a poorly understood domain of NSs, and still debated. The initial configuration of the field at formation also influences its evolution at later times, both in the crust and the core, and there can be strong differences based on the equation of state, constituents in the core, and mass of the NS. Among complications are how superfluid vortices interact with quantized flux tubes (or ``fluxoids"), vortex pinning \citep[e.g.,][]{2012MNRAS.421.2682L,2012MNRAS.422.1640L} and entrainment, formation of magnetic domains, and the rate of diffusion of flux tubes out of the core into the crust. For a recent discussion of $B\!\neq \!H$ in the core, see \cite{2021MNRAS.506.4632R}. Yet, the timescales for core field evolution are much longer than the crust \citep[e.g.,][]{2011MNRAS.413.2021G,2015MNRAS.453..671G} -- strong fields may be retained in the core for $\gg$Myr in superconducting cores. Diffusion of fields out of the core into the crust is a slow process. Proton superconductivity in the core is also expected in all magnetars soon after formation \citep[e.g.,][]{2012MNRAS.422.2632H,2017PhRvC..96f5801H}. Here, the typical diffusive timescale near the outer core is $\sim 0.1-0.3$ km Myr$^{-1}$ \citep[e.g.,][]{2015MNRAS.453..671G,2016MNRAS.456.4461E,2017PhRvC..96f5801H}, or equivalently, $\tau_{\rm sc} \gtrsim 10 \, \ell_{\rm km}^2$ Myr where $\ell$ is a characteristic lengthscale for evolution. As the NS cools, increasingly deeper regions of the core become superconducting with the outer core type-II proton superconductor. If the core field is $H\!<\! H_{c1}$, a metastable arrangement (where fluxoids and fields are frozen within interspersed zones of superconducting and normal matter) is likely as the growth rate of superconducting zones (from neutrino cooling) is much faster than the flux expulsion timescale \citep[e.g.,][]{2017PhRvC..96f5801H}. Deeper in the core, where the Ginsburg-Landau parameter is small, a type-I Meissner state is likely realized.

The field diffusion out of the outer core into the crust, and its timescale, may eventually power transient radio emission by stressing by the crust \citep[possibly beyond the breaking strain for high fields, e.g.,][]{2015MNRAS.449.2047L,2022ApJ...938...91K}, whose stored energy may be released abruptly (as FRBs) or gradually with low field external twists (both scenarios requiring currents, particle accelerate, pair production and concomitant radio emission). Indeed, much of the dissipation can occur at the core-crust interface \citep{2014MNRAS.437..424L,2016MNRAS.456.4461E} where current sheets may develop. At late times, i.e., $\gtrsim 10^5-10^6$ yr, weak toroidal components may also develop \citep[e.g.,][]{2016MNRAS.456.4461E,2018MNRAS.473.2771B} that may stress the crust and twist external fields. 

The metastable superconducting arrangement is stable insofar as the neutrino cooling timescale is shorter than the flux expulsion timescale. Yet, at around a Myr, the cooling timescale begins to become commensurate with the flux expulsion timescale \citep{2017PhRvC..96f5801H}. The timescale associated with flux migration from vortex-fluxoid pinning is commensurate with the spin-down timescale \citep{1992ApJ...399..213C}, i.e. also Myr for PSR 0901--4046, and similar in value to $\tau_{\rm sc}$. This late-time diffusion and evolution is likely most germane for radio activity from some Galactic ULPMs or the M81 FRB (FRB 20200120E). Flux expulsion and vortex pinning in NS cores has actually previously been invoked in deprecated models of GRBs, e.g. \cite{1991ApJ...382..576R,1991ApJ...382..587R} that may have renewed qualitative applicability to late-time radio activity from ULPMs and FRBs. 

In all of these physical scenarios, where the fields are in the crust only or both core and crust, the field evolution generally results in strongly dominant poloidal fields after $\sim 100$ kyr. For the external magnetosphere, this is a low-twist state with low persistent charge density -- and as such is highly susceptible to single photon pair cascades even from weak crust disturbances. Such magnetospheric states are generically suited for generation and escape of transient coherent radio emission and FRBs \citep{Wadiasingh2019}.

\section{Are periodic FRBs related to Galactic ULPMs?}
\label{sec:PeriodicFRB}

Long activity window periods were reported in the two most prolific repeaters, FRB 20180916B and FRB 20121102A (with periods of $\sim16.3$ and $\sim160$ days, respectively)~\citep{CHIMEperiodicity, Rajwade2020}. \cite{Beniamini+20}  have suggested that these could be a manifestation of the extremely slow spin periods achieved by a rare sub-class of magnetars. These authors have shown that the natural predictions of the two main competing mechanisms -- binary companion or precession -- are both disfavoured by the data. In particular, both mechanisms would require an underlying shorter period associated with the magnetar spin. However, a continuous $\sim\!5\,$hr observation of 20121102A revealed no such periodicity in burst arrival time, suggesting either a wide beaming cone (relevant only to the binary model, a precession model necessarily implies beaming) or else a long spin period of $P\!\gtrsim\!5\,$hr. Likewise, FAST observations of 1652 bursts from FRB 20121102A spanning over a month \citep{2021Natur.598..267L} find no periodicity up to at least $10^3$~s. \cite{2021ApJ...908L..12T} also have ruled out any stellar companion brighter than O6V for FRB 20180916B, and favor an older ${\cal O}$(Myr) age if its birth site is the nearest observed clump of star formation. Other evidence disfavouring a binary companion include the observed lack of an active phase fraction that increases with period (\citealt{Pleunis2021}; this is expected since absorption of radio waves decreases with increasing frequency) and the lack of change in DM between bursts in the beginning and the center of the active phase. Moreover, all observed Galactic magnetars have neither a confirmed binary companion nor do they show robust signatures of precession on similar timescales.

As FRB 20180916B is much closer to us than FRB 20121102A, it is natural to expect that it represents a significantly more common type of object, which, at this stage, dominates the source density estimates of periodic FRBs \citep{LBK2021}. The source densities estimated for ULPM candidates are at least one order of magnitude greater than estimated for the Galactic FRB (see \S \ref{sec:numbers}), two orders of magnitude greater than estimated for the M81 FRB and seven(!) orders of magnitude greater than estimated for FRB 20180916B \citep{LBK2021}. This source density discrepancy may be at least partially resolved if ULPM sources very rarely produce any FRBs (such as due to peculiar necessary physical conditions or triggers) and / or the beaming of their FRB emission is much narrower than that of the radio pulses of the Galactic ULPM candidates.
It is interesting to note that FRB 20200120E (associated with a globular cluster in M81) represents a class of FRB sources with significantly larger source densities than either FRB 20180916B or FRB 20121102A. Remarkably, recent observations of FRB 20200120E reveal a lack of underlying periodicity with values comparable to those seen in confirmed magnetars or pulsars. At the same time, while the existence (or lack of) of a much longer active periodic window (like in FRB 20180916B or FRB 20121102A) cannot yet be established, there is a hint of periodic activity window with a period of 12.5 days \citep{2022arXiv220603759N}. If this connection is eventually established, it suggests that the discrepancy between the ULPM candidate and periodic FRB source densities are much reduced.

\begin{figure}
\centering
\includegraphics[width = 0.2\textwidth]{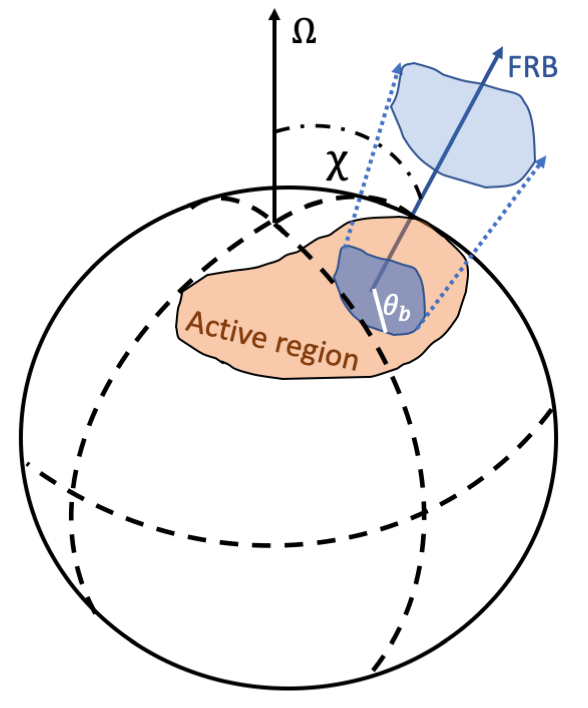}
\caption{Schematic of the generic beaming model discussed in \S \ref{sec:PeriodicFRB}. The region from which bursts can be observed is separated by $\chi$ from the principal rotation axis $\boldsymbol{\Omega}$ and has a typical angular size of the same order (to account for $\sim 0.25-0.5$ active phase observed in periodic FRBs). The beaming of each particular burst has an angle $\theta_{\rm b}$, which may be much smaller than $\chi$.}
\label{fig:sketch}
\end{figure}

To explore the possibility that FRB beaming accounts for the vast source density gap, we consider a simplified but generic beaming model. Our basic assumption is that the periodicity of FRBs is associated with their spin periods. To allow for dark periods (where no bursts are seen from FRB 20180916B), the direction of burst emission (denoted by $\chi$) must be removed from the rotation axis by at least $\theta_{\rm b}$ (where $\theta_{\rm b}$ is the beaming or collimation angle). i.e. $\chi>\theta_{\rm b}$. Considering that the `active phase' covers $\sim \!1/4\!-\!1/2$ of the burst period, the region of the sky from which bursts can be produced, can be very large compared to the emission collimation.
The large active fraction suggests that the size of the region is itself is of the same order as $\chi$. These considerations are schematically depicted in Figure \ref{fig:sketch}.
We denote the time averaged isotropic equivalent luminosity of bursts from FRB 20180916B by $L_{\rm iso}$. If bursts are beamed, the true energy of each burst is lower by $\sim \theta_{\rm b}^2/4$. At the same time, a fraction $\pi \chi/\theta_{\rm b}$ of all bursts produced by the active region are beamed away from us and not observed. Putting all this together, the energy output of FRB 20180916B is
\begin{equation}
\label{eq:energyR3}
    E\sim L_{\rm iso} \tau \epsilon_{\rm r}^{-1} \theta_{\rm b} \chi \approx 10^{45}\tau_{10\rm yr}\epsilon_{\rm r, -3}^{-1}\theta_{\rm b}\chi \mbox{ erg}
\end{equation}
where $\tau$ is the active FRB producing lifetime and $\epsilon_{\rm r}$ is the intrinsic radio efficiency. We see that a smaller beaming cone relaxes the energetic requirements on FRB~20180916B. Therefore, energetic requirements alone do not pose a problem for the $\chi,\theta_{\rm b}\ll1$ scenario.

Increasing the source density of FRB 20180916B like sources due to beaming, would require $\chi\!\ll\!1$. This would imply that for each FRB 20180916B-like source observed, there are $\sim \chi^{-1}$ sources whose emitting regions are always beamed away from us and are therefore missed\footnote{Using Eq.~(\ref{eq:energyR3}), we see that the total energy output from FRB 20180916B-like sources is independent of $\chi$.}. That is, the true source density of FRB 20180916B-like sources could be larger by $\chi^{-1}$ compared to the one mentioned above.

The beaming model discussed above requires in particular $\chi\!>\!\theta_{\rm b}$. In FRB models where the emission takes place far from the NS surface (outside the light-cylinder), the geometric model proposed above, works only if both $\theta_b\!\ll\! 1$ and the Lorentz factor of the emitting material is $\Gamma\!\geq \!\theta_{\rm b}^{-1}$. While some beaming can be accounted for in this way, these models generically find much larger values of $\theta_{\rm b}$ and smaller values of $\Gamma$ than needed to explain the Galactic ULPM candidate and periodic FRB source density discrepancy.

An alternative set of FRB models, consist of radiation from regions close to the NS surface. In this case, beaming can be the result of radio waves having to propagate along field lines\footnote{This is because in a homogeneous plasma, the low-altitude waves are in fact the normal modes \citep[e.g.,][]{1979AuJPh..32...61M,1986ApJ...302..120A,2019JPlPh..85c9011R} of the ambient upstream plasma (these are necessarily superluminal O-modes to avoid Landau damping or nonlinear wave-particle interactions) and are ducted or refracted along field lines analogous to an optical fiber, decoupling at higher altitudes transforming into electromagnetic vacuum modes.}.
One may invoke the extremely small polar cap region to explain the small active region size. Indeed, for a period of 16 days, and dipole spin down, the polar cap opening angle is tiny $\theta_{\rm c}\sim 10^{-5}$. Empirically, the radio in known pulsars appears to decouple around $10-100$ NS radii (see Appendix~\ref{sec:radiodutycycle}), which suggests an effective beaming angle that is at least $3-10\theta_{\rm c}$ (due to field line flaring with a dipolar structure). Furthermore, for such long periods as discussed above, the open field line region is very easily increased much beyond this value due to outflows \citep{Beniamini+20} or due to any small perturbance to the dipole field line structure. In addition, intrinsically nonneglibile surface area is expected to participate in FRBs and magnetar oscillations \citep[e.g.,][]{Wadiasingh2020,2020ApJ...903L..38W}, particularly quasi-polar closed zones as observed in SGR 1935+2154 \citep{2020ApJ...904L..21Y,2021NatAs...5..408Y}. The field line flaring in closed zones away from the pole can be more extreme compared to the flaring in the near polar cap region (small angle approximation), substantially increasing the area of the sky where the radio could decouple from the field and escape.
Finally, if we were to associate observed FRB periodicities with $\chi\ll1$, it would require a physical mechanism that causes small $\chi$ to be associated with a higher activity rate, while at the same time preventing $\chi=0$, which would wipe out the observed periodicity in the FRB signal. This circumstance is perhaps unnatural and requires fine-tuning, unless the mechanism which removes angular momentum to produce a ULPM also leads to small but non-vanishing magnetic obliquity.

In the Timokhin-Arons model for pulsar radio emission \citep{2010MNRAS.408.2092T,2013MNRAS.429...20T}, a crude estimate of $\theta_{\rm b}$ resulting from an infinitesimally narrow active region, is the angle of the pair formation front. This quantity couples the characteristic gap and field radius of curvature length scales. This field curvature is essential in the Timokhin-Arons model for the generated waves to be electromagnetic (rather than electrostatic) in character \citep{2020PhRvL.124x5101P,2022ApJ...933L..37T}. The collimation may be much wider than this estimate if large surface areas (involved in global magnetar oscillations) is tied to flux tubes where pair cascades occur, or the field lines flare significantly where the waves decouple. Thus, the pair formation front angle estimate constitutes a conservative lower limit on collimation. For at-threshold pair cascades, the angle is approximately $\vartheta \sim h_{\rm gap}/\rho_c$ where  $h_{\rm gap}$ is defined in Eq.~(30) of \cite{Wadiasingh2020}. This yields,
\begin{equation}
\vartheta \sim 10^{-3} \,  \rho_{c,7}^{-5/7} B_{14}^{-3/7} \nu_2^{-3/7} \xi_2^{-3/7} \lambda_5^{3/7} 
\end{equation}
which may be significantly wider for smaller amplitudes $\xi$ viable for FRBs in charge-starved ULPMs \citep{Beniamini+20}. Thus, we conclude that $\theta_b \gg \vartheta$, while we generically require $\chi\geq \theta_{\rm b}$ in our toy model. Thus beaming is \textit{likely insufficient} to explain the much lower source density of FRB 20180916B over ULPMs, and could suggest the FRB mechanism may require special or contrived conditions \citep[e.g., particular magnetosphere states or excitation of specific global oscillation modes,][]{Wadiasingh2019} to transpire frequently in a rare subset of ULPMs. Such peculiarity is already suggested by SGR 1935+2154, where only one of its thousands of X-ray short bursts in April 2020 resulted in a contemporaneous bright radio burst \citep{2020ApJ...904L..21Y,2021NatAs...5..408Y}.

Finally, the observed period of GLEAM-X J1627 is three orders of magnitude smaller than that of FRB 20180916B. It is plausible that the number of sources decreases rapidly enough with period to accommodate their respective source densities, see bottom left panel of Figure~\ref{fig:nVsP}. This scenario will be constrained with future discoveries of objects with long periods (\S\ref{sec:observestrat}).

\section{Observing Strategy}
\label{sec:observestrat}
\begin{figure*}
\centering
\includegraphics[scale = 0.34]{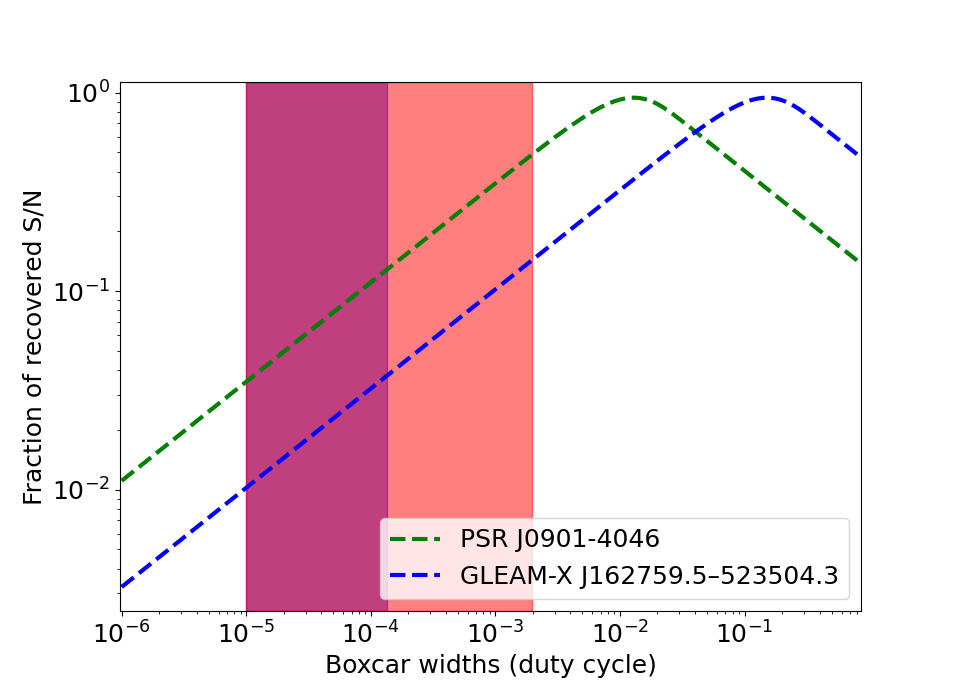}
\includegraphics[scale = 0.34]{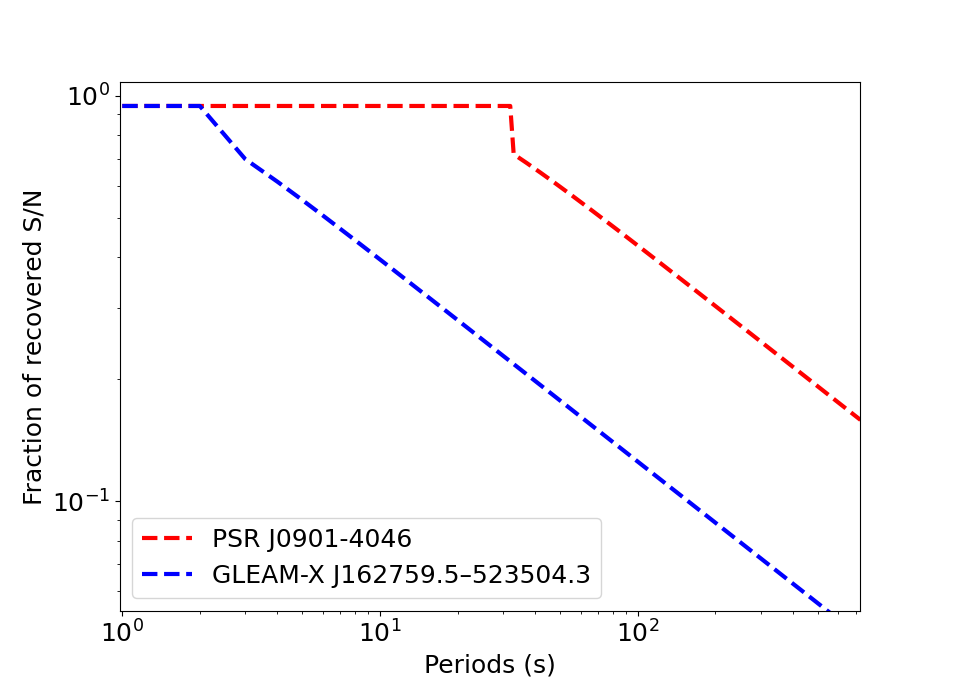}
\caption{\textbf{Left Panel}: Fraction of S/N recovered after boxcar convolution during a single pulse search as a function of boxcar-trails for the measured duty cycles and periods of PSR~J0901--4046 and GLEAM-X~J1627. The pink and purple shaded regions depict the space covered by the boxcar trials used in current surveys for PSR~J0901--4046 and GLEAM-X J1627, respectively. \textbf{Right Panel}: Fraction of recovered S/N as a function of spin period for fixed pulse duty cycles (taken to be equal to those of PSR~J0901--4046 and GLEAM-X~J1627).}
\label{fig:eff_boxcar}
\end{figure*}

\subsection{Radio}

The recent discovery of ULPM candidates like GLEAM-X J1627 and PSR~J0901--4046 has demonstrated that there might be large population of ultra-long period radio-loud NSs that has remained hidden to date. This dearth of radio-loud ULPMs can mostly be attributed to a selection bias of all current time-domain searches for radio transients. Typically, a time-domain survey for pulsars and other transients spends 10--20 minutes on any given part of the sky~\citep{manchester2001}. This limited amount of integration time immediately precludes detection of long-term periodic sources similar to GLEAM-X J1627. Furthermore, current real-time single pulse search pipelines at various facilities are strongly biased against single pulses with large widths ($>$few hundred ms). Single radio pulses from ultra-long period NSs can be fairly wide depending on the duty cycle. For example, a NS with a period of 5 minutes and 1$\%$ duty cycle can have a pulse as wide as 3~seconds, which would be completely missed by single pulse search pipelines. To demonstrate this, we estimate the reduction in the recovered signal to noise ratio (S/N) of a single pulse after the convolution with a series of boxcars with varying widths. Following the analytical treatment from \citet{2022MNRAS.510.1393M}, for a Gaussian pulse with a width $\sigma$ (we ignore any sub-pulse structure in the pulse to simplify our computations) and a boxcar width $w$, both measured in units of pulse phase (going from 0 to 1), the efficiency is
\begin{equation}
    \eta = \left(\frac{4 \pi \sigma^{2}}{w^{2}} \right)^{1/4}~{\rm erf}\left( \frac{w}{2\sigma \sqrt{2}}\right),
\end{equation}
where $\rm erf(x)$ is the error function.
We compute the efficiency for various values of boxcar width trials for the duty cycle of PSR~J0901--4046 and GLEAM-X J1627, and show this in the left panel of Figure~\ref{fig:eff_boxcar}. For a pulse with a width of 500~ms or more, the maximum boxcar trial will only cover $\lesssim 30\%$ of the total width. Thus, the recovered S/N in the best case scenario will be only as high as 40--50$\%$ of the true S/N of the pulse for any wide single pulses from ULPMs. Furthermore, the right panel of Figure~\ref{fig:eff_boxcar} shows the recovered S/N as a function of period for the duty cycles of PSR~J0901--4046 and GLEAM-X J1627, clearly indicating that the efficiency of current surveys reduces very rapidly for longer periods and that they are insensitive to the majority of the ULPM population in the Galaxy.

\begin{figure}
\centering
\includegraphics[scale=0.25]{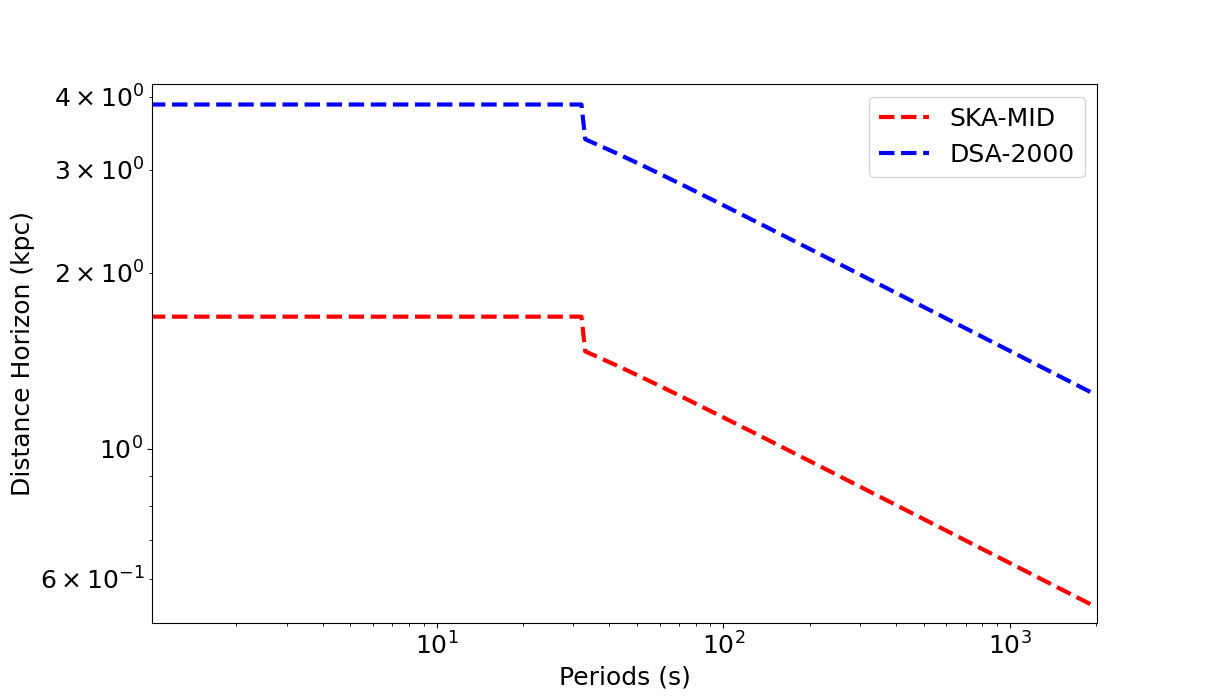}
\caption{Distance horizon for putative ULPMs with the same radio luminosity as PSR~J0901--4046 for SKA-MID and DSA-2000 time domain surveys as a function of spin period.}
\label{fig:horizon}
\end{figure}

Therefore, it is important to modify observing strategies or develop new techniques in order to search for these sources in the future. One way of overcoming the bias is the use of Phased Array Feeds (PAF) on single dish radio telescopes. PAFs can increase the effective field of view of the telescope manifold which can enable larger dwell times for surveys, thus becoming more sensitive to long period sources. On the other hand, performing searches in radio images for these sources is an effective way to combat the selection effects. Image domain surveys have large dwell times (few hours) with large integration times (a few seconds). Since the width of pulses from ULPMs is expected to be large, radio images can be used effectively to search for radio-loud ULPMs over a large parameter space without hindering the sensitivity of the telescope due to the large integration times. 

This provides a framework for predicting a horizon for radio-loud ULPMs assuming they have a radio luminosity comparable to PSR~J0901--4046. To illustrate this, we assume a time-domain radio survey with the SKA-MID~\citep{dewdney2017} and the DSA-2000 (deep synoptic array; \citealt{Hallinan2019}) array. We assume an integration time of 3600 seconds per pointing and use the parameters for the telescope (for e.g. gain, system temperature, bandwidth) as provided in the literature~\citep[][Lina Levin, priv. comm.]{keane2015, Hallinan2019}. Using the radiometer equation, for a signal to noise ratio threshold of 8, we obtain a sensitivity of 5.7~$\mu$Jy for the duty cycle of PSR~J0901--4046. We use this sensitivity to compute the distance horizon for detecting ULPMs with properties similar to PSR~J0901--4046 as a function of spin period. Figure~\ref{fig:horizon} shows the result of such a simulation. We can in turn use these estimates to put a limit on the detection rate of ULPMs similar to PSR~J0901--4046. Using the estimates in \S \ref{sec:popsynthnum} for the number of ULPMs of this type in the entire Galaxy, we estimate that $N_{\rm SKA}=21_{-5}^{+4}$ similar objects will be detectable by all-sky surveys with SKA-MID and $N_{\rm DSA}=130_{-13}^{+11}$ with DSA-2000. These numbers, are only slightly decreased in case the initial birth locations of ULPMs are confined to the Galactic plane: $N_{\rm SKA2}=16_{-4}^{+4}, N_{\rm DSA2}=89_{-10}^{+10}$.

\subsection{X-ray}

The detection of a persistent high energy counterpart to any ULPM candidate would provide a crucial step forward to understanding their origin and evolutionary path. As discussed in Section~\ref{sec:Age}, older ULPMs such as PSR J0901--4046 are expected to be $\gtrsim \!10^5$~years old. Under any B-field decay scenario, their surface temperature is not expected to exceed the $100$~eV \citep[e.g.,][]{2020MNRAS.496.5052P} with thermal bolometric luminosities comparable to those of the XDIN population, $L_X \lesssim 10^{31}$~erg~s$^{-1}$ \citep{2008AIPC..983..331K,2009ApJ...705..798K}. The faintness of these sources, coupled with their cool surface temperatures poses a challenge for their detection. For instance, an absorption column density $N_{\rm H}$ larger than $10^{22}$~cm$^{-2}$ will severely attenuate the observed flux below $\sim2$~keV. From \citet{2013ApJ...768...64H}, this $N_{\rm H}$ limit is reached at around 4 kpc, implying that the current fleet of X-ray telescopes are practically insensitive to such population beyond this horizon. Even in the absence of strong absorption, deep exposures are required to detect these sources beyond 1~kpc, where their expected X-ray fluxes are $\lesssim8.0\times10^{-14}$~erg~s$^{-1}$~cm$^{-2}$. Moreover, it remains complex to discern a ULPM from other soft X-ray sources without the detection of a pulsed signal, increasingly restricting the parameter space to which we are currently sensitive. These expectations are consistent with the results of \citet{israel16MNRAS} and \citet{deluca21AA} which found no tenable ULPM candidate in their systematic pulsation searches in the Chandra and XMM archives. From the current population of ULPM candidates, PSR~J0901--4046 is the prime source for deep X-ray counterpart searches due to its DM-implied distance of 330~pc and $N_{\rm H} \lesssim 10^{21}$~cm$^{-2}$; both are the smallest among ULPMs and comparable to the XDIN population. The latter is detectable with currently operating X-ray telescopes, e.g., XMM and NICER.

Prospects of independently discovering ULPM candidates in X-rays must rely on large field-of-view imaging soft X-ray instruments, such as eROSITA. From extensive simulations of the Galactic NS population and the eROSITA all sky survey scans, \cite{2017AN....338..213P} derived a detection estimate of about 95 XDINs candidates within a distance of 1.7~kpc. Given our source density estimate of ULPMs and XDINs (Figure~\ref{fig:nVsP}), we surmise that $\!\sim \!10$ ULPMs candidates may hide within the total eROSITA XDIN population. While eROSITA survey observations will not be suited for pulsation searches due to the $\lesssim2$~ks exposures in most of the Galactic plane regions \citep{erosita21AA}, any candidate ULPM could be targeted for deeper exposures with {\sl XMM-Newton}, NICER, or pointed eROSITA observations to search for slow X-ray pulsations to confirm their nature. Another potential route to uncover ULPMs is through relatively deep (reaching, e.g., $\!<\!10^{-15}$~erg~s~cm$^{-2}$ observed flux in the 0.2-2~keV band) observations of large swatch of the Galactic plane. This is achievable with sensitive, large FOV ($\!>\!1$~degree$^2$) soft X-ray telescopes, e.g., STAR-X \citep{2022HEAD...1910845Z} with larger grasp than any current mission. In such cases, search is more sensitive to farther distances, balancing the smaller FOV of such instruments compared to eROSITA.

\section{conclusions}
\label{sec:conclusion}
The recent discovery of several intriguing Galactic radio sources with long periods compared to the pulsar population reveals a potentially large and unexplored class of objects. We have considered the observations of these objects and scrutinized the two latest additions to this list: PSR J0901--4046 with $P\approx 76$\, s and GLEAM-X J1627 with $P\approx 1091$\, s. We have shown that the periodicity in these objects is highly unlikely to be associated with a binary companion and that in both cases, the observations are hardly explained by a WD nature. Instead we find that PSR J0901--4046 and GLEAM-X J1627 are naturally explained as ultra-long period highly-magnetized NSs (ULPMs). Together with 1E 161348-5055 ($P\!\approx \!2.4\times 10^4$\,s), PSR~J0250+5854 ($P\!\approx\! 23.5$\,s) and GCRT J1745-3009 ($P\!\approx \!4620$\,s), they form our list of ULPM candidates.

Evidence based on the confirmed magnetar population suggests they undergo episodes of enhanced spindown associated with their outbursts. The accumulated effect of outburst associated spindowns is sufficient to drive, at least some magnetars, to ultra-long periods.

We have shown that the nearby distances of the most recent ULPM candidates require that there are thousands to tens of thousands such objects in the Galactic field. This in turn constrains the formation rate and typical ages of these objects. Indeed the ages of ULPM candidates, are further constrained by several other considerations, including dipole spindown ages, high timing stability of the radio pulsations, upper limits on the X-ray fluxes (requiring the objects to have cooled down sufficiently) and their offsets from SNR and stellar clusters. The conclusion is that GLEAM-X J1627 and PSR J0901--4046 should have ages of $\sim 10^5-10^6$\,yr, much older than the ages of the confirmed Galactic magnetar population. These results suggest that ULPMs represent a distinct evolutionary sequence from confirmed magnetars and have undergone a distinct evolution of their poloidal fields. Future 3D mapping of the locations of ULPMs in the Galaxy (and particularly the Galactic halo) could be used as an independent test for the ages of these systems (see e.g. \citealt{Rajwade2018}).

The long magnetic field decay timescales discussed above (which are associated with a common magnetar sub-population) are a sizeable fraction of the time to merger in the rapid channel associated with a large fraction of binary neutron star systems \citep{BP2019}. This suggests that magnetar strength-fields might still be tenable in binary neutron star systems, just before their merger and might lead to detectable precursor emission from those systems \citep[e.g.,][]{1996A&A...312..937L,Lyutikov2019,2022arXiv221017205C}.

Ultra-long period magnetars have also been suggested as an explanation for many days long periodic activity windows seen in some repeating FRBs. We explored the possible connection between the Galactic ULPM population and these periodic FRBs. We found that the source density of the latter is between five and seven orders of magnitude lower than that of the former. It is on the one hand fortunate that we have more than enough long period objects than required by FRB observations. On the other hand, the large source density discrepancy is puzzling and needs to be understood if the connection between these two phenomena is a real one.
Only a fraction of this discrepancy could be ascribed to beaming.
Other factors which could be important are (i) the $\sim 3$ order of magnitude gap between the period of GLEAM-X J1627 and that of periodic FRBs (considering the apparent reduction in the density of sources with increasing periods, Figure \ref{fig:nVsP}) and (ii) some selection process causing the longest period magnetars to be preferential producers of bursts (see e.g. \citealt{Beniamini+20}). It therefore remains to be determined whether periodic FRBs are directly connected to the Galactic ULPM population.

Finally, the recent observations of Galactic ULPM candidates serves as a vivid demonstration of the large amount of discovery parameter space that remains unexplored beyond the standard curvature pair cascade pulsar death line. We have shown that current time-domain searches for radio transients are heavily biased against finding long period systems, and many more exciting and nearby long-period objects therefore likely remain undetected. We advocate the need to explore different observational strategies in order to truly open up the window to the detection of this potentially large population of objects hidden in our Galactic backyard.

\section*{acknowledgements}
We especially thank Kostas Gourgouliatos, Sam Lander and Wynn Ho for useful discussions regarding models for field evolution in the crust and core. We also thank Jonathan Granot, Alice Harding, Oleg Kargaltsev, Demos Kazanas, Tom Maccarone, Cole Miller, Sk Minhajur Rahaman, and Andrey Timokhin for interesting discussions. PB's research was supported by a grant (no. 2020747) from the United States-Israel Binational Science Foundation (BSF), Jerusalem, Israel. The material is also based on work supported by NASA under award number 80GSFC21M0002. This research has made use of NASA’s Astrophysics Data System. KMR acknowledges support from the Vici research programme ``ARGO'' with project number 639.043.815, financed by the Dutch Research Council (NWO). JH acknowledges support from an appointment to the NASA Postdoctoral Program at the Goddard Space Flight Center, administered by ORAU. JH thanks Brad Cenko for helpful discussions related to the access and use of DECam data.

\section*{Data Availability}
Data produced in this study will be shared upon reasonable request.





\appendix

\section{Magnetar Candidates in Binaries}
\label{sec:speccand}

For completeness, we here consider magnetar candidates in binaries listed in Table~\ref{tab:sources} \citep[also see][for a recent review]{2022arXiv220107507P} excluding pulsating ULXs\footnote{ In the authors' assessment, evidence currently favors \citep[e.g,][]{2002astro.ph..2488P,2008AIPC.1010..303P,2012ApJ...752...90P,2018ApJ...857L...3W,2018NatAs...2..312B,2018NatAs...2..282B,2022ApJ...937..125B} pulsating ULXs as harboring magnetars, over highly-beamed emission from NSs of lower magnetization.} \citep{2014Natur.514..202B,2016ApJ...831L..14F,2017Sci...355..817I,2017MNRAS.466L..48I}. The magnetar nature of the compact objects in these binary sources (detailed in Table~\ref{tab:sources}) are less secure, and chiefly based on indirect arguments based on source luminosity, and spin equilibrium in the quasi-spherical wind fed propeller/settling regime, and their position in the Corbet diagram \citep{1984A&A...141...91C,1986MNRAS.220.1047C} in catalogs of high mass X-ray binaries \citep{2006A&A...455.1165L,2016A&A...586A..81H,2019NewAR..8601546K,2020ApJ...896...90M}. By construction, these sources exhibit long spin periods. Also included in Table~\ref{tab:sources} but not detailed below are two exceptional HMXBs AX~J1910.7+0917 at $P\approx 10$ hr and SXP 1062. We have also omitted long period sources \citep[e.g.,][]{2018MNRAS.475..220V} that are compatible with standard NS magnetizations via indirect arguments without fine tuning but could harbor higher magnetization objects.
The rarity (i.e. low source density) of all such HMXB systems (including pulsating ULXs) implies they cannot be the dominant channel for production of ULPMs. The number of magnetars residing in binaries is also likely subdominant to the isolated population \citep{2022MNRAS.513.3550C}.

\subsection{IGR 16358--4726}

IGR 16358--4726 \citep{2003IAUC.8097....2R} is an unusual periodic ($P\sim 6000$ s) source originally detected as a nonthermal hard-spectrum {\it{INTEGRAL}} transient that exhibited a K$\alpha$ iron florescence line characteristic of accreting compact objects \citep{2003IAUC.8109....2K,2004ApJ...602L..45P}. The hard spectrum disfavor it being a compact low-mass X-ray binary (LMXB), and thus orbital modulation origin for the reported periodicity. A subsequent large spin-up $\Delta \nu/\nu \!\sim \!10^{-2}$ over a baseline of $10^6$ s reported by \cite{2007ApJ...657..994P} also disfavor an orbital origin for the periodicity. Interpreting such a large spin-up as orbital contraction that would imply a huge transient companion mass loss (that also does not result in significant spectral or flux changes) of $\dot{m}\! \gtrsim \!10^{-8} m_c\! \sim \!10^{25}$ g/s ($m_c \!\sim \! 0.1-1 M_\odot$ is the companion mass in a putative compact LMXB). Thus a NS origin of the compact object and association of periodicity with spin is preferred. A disk accretion scenario for the spin-up implies a large magnetic moment for the NS for a large lever arm (the Alfv\'{e}n radius). A standard calculation implies $B_* \!\sim\! 10^{14} d^{-6}_{7}$ G for a distance of 7 kpc. A lower limit to distance of $\!\sim \!7$\,kpc is also inferred if the source flux is not in the propeller regime. These results suggest for any plausible distance within the Galaxy, IGR 16358--4726 possesses a strong magnetic field $B_* \!\gtrsim \!10^{13}$ G.

\subsection{SGR 0755-2933}

SGR 0755-2933 is a candidate magnetar discovered by magnetar-like short bursts by Swift/BAT \citep{2016ATel.8831....1B}. The error region of BAT contains a HMXB harboring a NS with spin period $P \!\approx\!308$ s \citep{2017AAS...22943104H}. Although \cite{2021A&A...647A.165D} suggest the candidate HMXB may be unassociated with the BAT burst, we will argue against this possibility in a future dedicated work. The parallax distance to the counterpart is $3.5$ kpc.

\subsection{4U 1954+319}

 4U 1954+319 is a peculiar  HMXB hosting a 12-50 Myr M supergiant \citep{2020ApJ...904..143H}. It was suggested by \cite{2014ApJ...786..127E} to host a magnetar with spin period $P \!\approx\! 5.7$\,hr. Since \cite{2014ApJ...786..127E}, Gaia has attained distance of $3.3 \pm 1 $\,kpc \citep{2018AJ....156...58B,2020ApJ...904..143H} for the counterpart. Stellar evolution and lack of a supernova remnant constrains the age of the NS to $\!<\!43$ Myr. Recently, \cite{2022MNRAS.510.4645B} argue a highly-magnetized (or magnetar-like) magnetic field is demanded given the recent Gaia association, in either a wind accretion or spin-equilibrium propeller scenario.

\subsection{4U2206+54}

First suggested by \cite{2010Ap.....53..237I,2013ARep...57..287I} as a possible wind-fed magnetar, 4U2206+54 is a HMXB with NS spin period $P \!\approx \! 5750$ s \citep{2018MNRAS.479.3366T}. A high spin frequency derivative of $\dot{\nu}\! \approx\! -1.8 \time 10^{-14}$ Hz s$^{-1}$ implies a relatively large level arm and magnetization. Gaia parallax implies a distance of $3.7 \pm 0.4$ kpc for the optical counterpart. \cite{2018MNRAS.479.3366T} suggest that a wind settling accretion scenario requires a NS with magnetar-like field.

\subsection{4U0114+65}

Suggested by \cite{1999ApJ...513L..45L} as a potential magnetar, 4U0114+65 has a spin period of  $P\approx  9350$ s and positive frequency derivative $\dot{\nu} \approx 10^{-14}$ \citep{2017A&A...606A.145S}. The companion distance is estimated $7 \pm 3$ kpc.

\section{GLEAM-X J1627 and PSR J0901--4046, and the Radio Duty Cycles of Rotation-Powered Pulsars}
\label{sec:radiodutycycle}

The radio duty cycles of GLEAM-X J1627 and PSR J0901--4046 are generally inconsistent empirical relations of period dependence of pulsar beam width and frequency in young rotation-powered pulsars. This is independent of standard (rotational) theoretical death lines for pair formation tenability, for which both GLEAM-X J1627 and PSR J0901--4046 are beyond without significant fine-tuning. This suggests the radio emission is not rotation-powered in both objects, and that they are entirely (GLEAM-X J1627) or perhaps partially (PSR J0901--4046) magnetically powered.

More specifically, in a fixed band, rotation-powered pulsar beam full widths (empirically) scale as $\Delta \phi \sim C/\sqrt{P}$ for a ``core" component aligned with the magnetic axis, and hollow ``conal" components. This is the expected scaling for the corotating magnetospheric open zone and polar caps where single photon pair production occurs. The absolute value of the beam width, however, depends on the decoupling height of the radio emission and other variables such as the pulsar magnetic obliquity. Empirically, $C \!\sim\! 6^{\circ}$ s$^{1/2}$ for ``conal" components and $\!\sim \!1-1.5^{\circ}$ s$^{1/2}$ for ``core" Gaussian components. Such core-cone components \citep{1983ApJ...274..333R,1990ApJ...352..247R,1993ApJ...405..285R} are expected as the pair formation front angle and gap physics varies between the conducting boundary of the polar cap rim and changing curvature photon angle with respect to the magnetic field. These empirical relations constitute the maximum observed pulse width expected from a rotation-powered pulsar, since the observer may have a glancing impact with the radio beam. The core component angular size scales with the square root of decoupling height \citep{1998MNRAS.299..855K}, $\rho_{\rm core} \!\approx\! 1.2^\circ \sqrt{h/(h_{\rm kg,0} P)}$, where the height is normalized to $h_{\rm kg,0}\!\sim\!40 R_6 \dot{P}^{0.07}_{-15} P^{0.3} \nu^{-0.26}_{\rm GHz}$ \citep{2003A&A...397..969K}. This empirical relation is an explicit radius-to-frequency mapping \citep{1978ApJ...222.1006C}.
For a putative ULPM magnetar with $P\!\sim \!10^2$\, s such as PSR~J0901--4046, naive extrapolation of these rotation-powered relations yields $\rho_{\rm core, ulpm}\! \sim \!0.4^\circ $ or a duty cycle of $0.1\%$, about $20-50$ times smaller than that observed in the L-band. For GLEAM-X J1627, the situation is even more at odds. Thus, ULPM emission is unlikely to be a mere extrapolation of the behaviour of regular pulsars, and different physical effects are involved. The core component is where most of the flux density resides in rotation-powered pulsars, and the presence of a conal components are revealed by a double-peaked pulse profiles. However, note conal components begin to dominate over core components at larger periods \citep{2002ApJ...568..289A}, up to the death line where single photon pair cascades become untenable. Yet, neither GLEAM-X J1627 or PSR~J0901--4046 exhibit features (pulse profile or polarization) typically associated with conal components.

\section{Radio QPOs and The presence of a crust in PSR~J0901--4046}
\label{sec:crust}

PSR~J0901--4046 is a ``pulsar inside a pulsar" exhibiting many QPOs modulating emission within the envelope of its 76-second period radio pulsations, including $\!\sim \!50$ ms period QPOs consistent with low-order nonmagnetic $l=2$ or $l=3$ ($n=0$) crustal shear modes (i.e. torsional modes). The QPOs in PSR~J0901--4046 appear to persist for several hundred milliseconds or longer and appear in about 30\% of pulse envelopes \citep{caleb2022}. That is, the covering fraction is $\dot{q} {\cal{T}}\!  \sim \!1/3$ where ${\cal{T}}$ is the mode damping time and $\dot{q}$ is the excitation rate. If continually but randomly excited this implies $\dot{q}\!<\! 1/(3 \Delta t)$ ($\Delta t \!\sim \!0.5$ seconds the typical pulse width) for $ {\cal{T}}\!>\! \Delta t $. This rate of QPO excitations is much larger than rare observed QPOs in high-energy magnetar bursts \citep[e.g.,][]{2022ApJ...931...56L}, which may be detectable only if the burst is bright enough and the corresponding amplitude of oscillations is large. Thus the radio observations in PSR~J0901--4046 are likely sampling a population of lower in amplitude (but much more numerous) crustal vibrations. This suggests a hierarchy in magnitude of vibrations versus commonality, much like the logN-logS power-law event size distribution of magnetar bursts or earthquakes \citep{1996Natur.382..518C}. Note that if the effective power (total elastic energy divided by  ${\cal{T}}\!\sim\! 1$ s) in these magneto-elastic toroidal oscillations (which appear significantly modulate the radio emission) are comparable to or greater than the radio luminosity, then their required amplitude is larger than only about $\xi\!\gtrsim\! 10^{-4}$ cm. The required screening charge density for such small amplitude oscillations is much smaller than the persistent one from corotation ($\propto B/(c P)$), and ought not to lead to dramatic modification of the pair cascades and radio luminosity as in FRBs, but rather apparently only the wave/plasma dynamics and inhomogeneity along open field lines \citep[see also,][]{2000MNRAS.316..734T,2010MNRAS.408..490M,Wadiasingh2019}.
 
If these are indeed crustal magneto-elastic modes, they may damp by coupling to a continuum of Alfv\'{e}nic modes associated with the magnetized core of the NS \citep{2006MNRAS.368L..35L}. The QPOs appear inconsistent with bouncing magnetospheric Alfv\'{e}n modes along field lines, as the pulse profile is commensurate with polar cap radio emission and has a characteristic ``S" shape curved in its polarization position angle. This would also imply field lines involved in radio emission attain very high altitudes (or are open, sampling a characteristic scale of the light cylinder at few $\!\sim \!10^{11}$ cm) and thus the observed harmonic frequencies (and potential damping) timescales are inconsistent with low-amplitude Alfv\'{e}nic modes in a bounded cavity as sometimes invoked for QPOs in magnetar bursts and giant flares \citep[e.g.,][]{2021Natur.600..621C}. 
s
The presence of crustal modes rules out PSR~J0901--4046 as a strange (quark) star \citep{1984PhRvD..30..272W,1986ApJ...310..261A,2005PrPNP..54..193W,2008RvMP...80.1455A}, as strange quark matter phase transitions and formation scenarios generally preclude the existence of a thick ionic crust bound by Coulomb interactions \citep[e.g.,][]{2004PhRvD..70f7301U,2006ApJ...646L..17N}. Although nonbare or hybrid star models exist, their predicted quasinormal frequencies are much higher into the kHz regime \citep[e.g.,][]{2015ApJ...815...81M}.

It is presently unknown if the QPOs of PSR~J0901--4046 are long lived (i.e. if ${\cal{T}}\!\gg \!\Delta t$). However, it appears that $P\!>\! {\cal{T}}$ since QPOs do not usually persist across consecutive single pulses, nor are they coherent across pulses. The expected damping time of crustal shear modes is quite model and magnetic field configuration dependent. Configurations with (MHD-mode) spectral gaps, tangled magnetic fields, or fields confined largely to the crust can significantly enhance the longevity of modes \citep[e.g.,][]{2009MNRAS.395.1163S,2011MNRAS.414.3014C,2012MNRAS.423..811C,2021MNRAS.504.5880B}. In the simplest models with a core-bound global dipolar field \citep{2012MNRAS.421.2054G} that is not entirely expelled by type II proton superconductivity, the damping (i.e. e-folding) time is $\eta_{\rm damp} \!\sim\! 0.1$ Alfv\'{e}n crossing times and yields an upper bound on a core-tied dipolar magnetic field: $B_{\rm core} \overset{?}{\lesssim} 2\times 10^{13} \, (\eta_{\rm damp}/0.1) \,  {\cal{T}}_{-0.5}^{-1} R_6 \sqrt{\rho_{14}}$ G. Here ${\cal{T}}\! \sim\! 0.3-0.5$ s is the minimum damping time, $\rho \!\sim \!10^{14}$ g cm$^{-3}$ the core density at the crust boundary and $R \!\sim \!10^6$ cm the characteristic spatial scale. This together with the constraints on $B_{\rm d}$ found in \S\ref{sec:magPSRJ0901} leaves a relatively narrow parameter space for the simplest picture of the mode structure. The implication could be that a more complicated scenario is required, such as the core field structure is significantly modified by proton superconductivity, and that there are spectral gaps in the MHD continuum of the core field.

\section{Constraints on optical counterparts to GLEAM}
\label{sec:gleam_opt_count}

The positional uncertainty of GLEAM-X J1627 was reported as 2$''$ \citep{Hurley-Walker2022}. A search in the Gaia EDR3 catalog returns one optical source within this positional uncertainty, which is offset from the radio position by 0.94$''$, making  it a potential optical counterpart to the GLEAM-X J1627. The Gaia source has an estimated distance of 5.4~kpc, suggesting the source is not related to GLEAM-X J1627 which has an estimated distance of 1.3~kpc (\citealt{Hurley-Walker2022}; \citealt{2021AJ....161..147B}). However, the parallax ($\bar{\omega}=0.63\pm0.68$ mas) is not well measured, making this distance estimate unreliable and warranting additional investigation. The Gaia source does have well measured proper motion, suggesting that this optical source is Galactic.

\begin{figure}
\centering
\includegraphics[width = 0.35\textwidth]{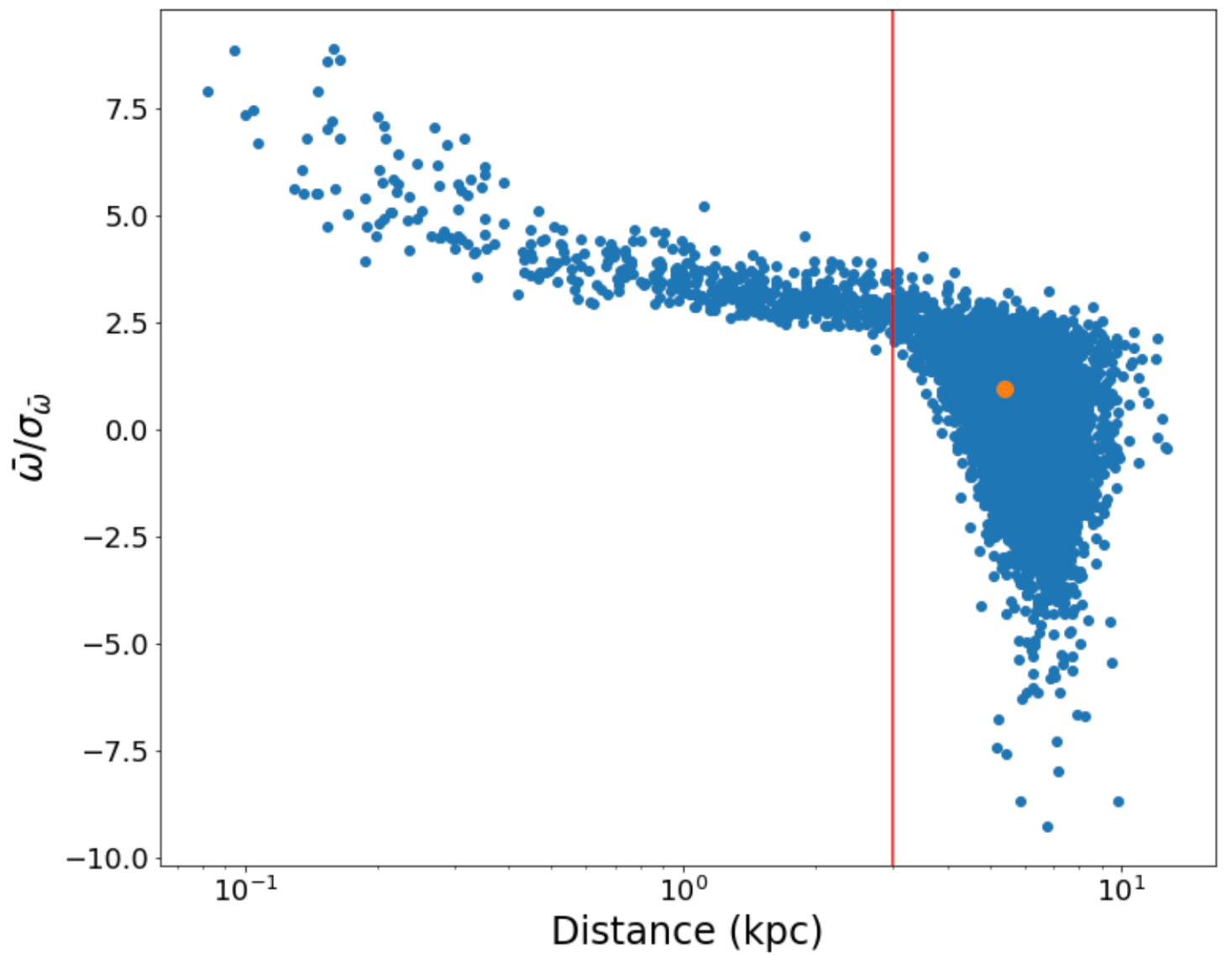}
\caption{Parallax inferred distances from the catalog of \citealt{2021AJ....161..147B} plotted versus the measured parallax divided by the parallax error (i.e., $\bar{\omega}/\sigma_{\bar{\omega}}$) for all Gaia sources within 10$'$ of the potential GLEAM-X J1627 optical counterpart and having a similar G band magnitude (i.e., $>20.0$). The orange point shows the location of the potential Gaia counterpart in this plot. The red line shows roughly where sources begin to have poorly measured parallaxes ($\bar{\omega}/\sigma_{\bar{\omega}}\lesssim2$), corresponding to a distance of $\sim3$ kpc. }
\label{fig:gaia_dist_ll}
\end{figure}

To place a lower-limit on the distance to this Gaia source, the following procedure was undertaken: First, we extracted all Gaia sources within 10$'$ of GLEAM-X J1627's position. Then we removed all sources with G-band magnitudes $<20.0$, to keep only those sources with a comparable brightness to this potential optical counterpart (i.e., $G=20.25$). Next, we cross-matched these Gaia sources with the parallax distance catalog of \cite{2021AJ....161..147B}. We then plotted the derived distances from \cite{2021AJ....161..147B} (i.e., {\tt rgeo}) versus the measured parallax divided by the parallax error (i.e., $\bar{\omega}/\sigma_{\bar{\omega}}$). This allows us to determine out to roughly what distance sources of a similar brightness to the potential GLEAM-X J1627 counterpart will have meaningful parallax measurements (i.e., not due to statistical noise), thus allowing us to place a lower-limit on the distance to the potential Gaia counterpart. The results of this procedure are shown in Figure \ref{fig:gaia_dist_ll}. It appears that sources still have well measured parallaxes (i.e., $\bar{\omega}/\sigma_{\bar{\omega}}\gtrsim2$) out to distances of about 3 kpc. Therefore, it is likely that this Gaia source lies far beyond GLEAM-X J1627 and is an unrelated background source.

There also exists more sensitive optical observations of this field from the Dark Energy Camera Plane Survey (DECaps; \citealt{2018ApJS..234...39S}). The DECaps observed in five bands, $grizY$, and goes much deeper than Gaia. The Gaia G-band limiting magnitude is about 21, while the DECaps images of this field go to 5$\sigma$ limiting magnitudes of 23.7, 22.8, 22.8, 22.4, and 21.5 in the five bands, respectively. There are four sources detected in the DECaps images in the 2$''$ error circle of GLEAM-X J1627. Two of these sources have less than three secure detections and have magnitudes very close to the limiting magnitude, suggesting that they may be spurious. Additionally, their magnitudes are not well constrained so we exclude these sources, leaving only two sources. The brightest of these sources is the source detected by Gaia, which we  already ruled out as a potential counterpart. The second source is 1.3$''$ offset from the GLEAM-X J1627 position and has magnitudes $g=<22.8$, $r=22.8$, $i=21.8$, $z=21.0$, $Y=20.7$. 

This leaves three potential options, the first is that the optical source is the counterpart to GLEAM-X J1627. The second is that GLEAM-X J1627 may be in a binary with the optical counterpart being the binary companion. Lastly, the source could simply be a background star, unrelated to the GLEAM-X J1627. To constrain the source properties in the first two scenarios, we assume the counterpart lies at a distance of 1.3 kpc,  and account for extinction using the {\tt mwdust} package \citep{2016ApJ...818..130B} with the \cite{1999PASP..111...63F} reddening law tabulated in the Dark Energy Camera bands by \cite{2011ApJ...737..103S}. We find that the $E(B-V)=0.21$ at 1.3 kpc in the direction of GLEAM-X J1627. After accounting for distance and extinction, we find the potential optical counterpart has an absolute magnitude $M_{r}=11.8$ and color $r-z=1.6$. 

To explore the first scenario, we use the WD atmospheric models of \cite{2020ApJ...901...93B} to compare the color and absolute magnitude of this potential counterpart. We find that the absolute magnitude of the potential optical counterpart is too large for a source with such a red color to be consistent with a WD, thus ruling out this scenario. If, however, the source is instead a binary companion to GLEAM-X J1627, it would be more consistent with a late-type star. Given the absolute magnitude and color, the source would be most consistent with an M4-6 type M-dwarf (see Figures 15 and 16 in \citealt{2019AJ....157..231K}). This spectral type is also consistent with this sources absolute $M_i\!=\!10.9$ and $M_z\!=\!10.2$ magnitudes and non-detection of the source by 2MASS as all NIR magnitudes are expected to be $\gtrsim18$. Further follow-up observations to obtain a source spectrum could be carried out to confirm or refute this possibility. 

The final possibility is that the source is a background star that is coincident with GLEAM-X J1627 by chance. The probability of having one or more DECaps sources in the $\delta r\!=\!2''$ positional uncertainty of GLEAM-X J1627 can be calculated as $P_{\rm DECaps}\!=\!1-e^{\rho_{\rm DECaps}\pi\delta r^2}$, where $\rho_{\rm DECaps}$ is the density of sources with three or more good detections in the DECaps survey and $\delta r$ is the positional uncertainty of GLEAM-X J1627 (see e.g., \citealt{2018ApJ...865...33H}). This calculation yields a large chance coincidence probability of $\!\approx\! 90\%$, suggesting that this scenario is the most likely. If this is the case, it is difficult to place any constraints on the spectral type as the source distance and extinction are unknown.

\section{Gap Voltage and the Curvature Pair Death Line for Arbitrary Stellar Radii}
\label{sec:voltagegap}
The three conditions ($\ell_{\rm gap} < r_{\rm pc}$, $\Delta V_{\rm gap} < \Phi_{\rm open}$, and  $L_{e^\pm} <  \dot{E}_{\rm SD}$) yield identical parameter dependence of the ``death line" on $B$ and compactness within factors of unity, verifying the self-consistency of the constraints. Requiring the pair luminosity is greater than the uncorrected radio luminosity $L_{e^\pm}\! >\! L_{R,\rm obs}$ applies only to very compact stars (e.g. NSs) in an untenable part of the parameter space and is not shown in Figure~\ref{fig:sdconstraints}. For these gross estimates, we adapt the curvature photon gap height $\ell_{\rm gap}\! \sim\! (B_{\rm cr}^4 \lambar^2 c^3/\pi^3)^{1/7} \rho_{c}^{2/7} P^{3/7} B^{-4/7}$ \citep{2015ApJ...810..144T} where $B_{\rm cr}$ is the quantum critical or Schwinger field, $\lambar$ the reduced Compton wavelength and $\rho_c$ is the characteristic field curvature radius near the polar cap. We omit weak order unity factors associated with the relative gap speed and pair attenuation exponential of the Erber formula \citep{1966RvMP...38..626E}. The open field line voltage is $\Phi_{\rm open} \sim \Omega^2 B R^3/c^2$ while for a vacuum gap $\Delta V_{\rm gap} \sim (\Omega/c) B \ell_{\rm gap}^2$. These all yield the death line,
\begin{equation}
R \gtrsim 4 \times 10^{9} \left(\frac{\varrho_{c}}{10}\right)^{4/17} P_{3}^{13/17} B_{9}^{-8/17} \quad \rm cm 
\label{eq:minradius}
\end{equation}
where we adopt dimensionless $\varrho_c = \rho_c/R = 10$ for the characteristic field curvature radius. This constraint on $R$ then may be readily expressed in terms of constraint on compactness for a given mass. A more realistic choice for $\varrho_c \gg 10$ commensurate with the expected small polar cap size for a $P\sim 10^3$ s rotator would eliminate any allowed parameter space in Figure~\ref{fig:sdconstraints}.

\section{X-ray Upper-limits}
\label{xray_analysis}

\begin{table}
\footnotesize
\setlength{\tabcolsep}{3pt}
\label{table_obs}
\begin{center}
\begin{tabular}{lcccc}
\hline \hline \hline
 \textbf{Source} & \textbf{Observatory} & \textbf{ObsID} & \textbf{Start Date} & \textbf{Exposure} \\
 & & & \textbf{[MJD]} & \textbf{[ks]}\\
\hline \hline \hline
PSR J0901--4046 & Swift-XRT & 00014019002 & 59245.83 & 3.87 \\
& Swift-XRT & 00014019003 & 59352.04 & 1.35 \\
& Swift-XRT & 00014019004 & 59353.29 & 2.70 \\
 \hline
GLEAM-X J1627 & Chandra &  26228 & 59601.87 & 20.09  \\
& Chandra & 26282 & 59602.25 & 10.11  \\
 \hline
PSR J0250+5854 & XMM-Newton & 0844000201 & 58692.15 & 5.20 \\
& XMM-Newton & 0844000301 & 58696.12 & 3.40 \\ 
& XMM-Newton & 0844000401 & 58694.17 & 2.50 \\
& XMM-Newton & 0844000501 & 58698.12 & 8.00\\
& XMM-Newton & 0844000601 & 58728.04 & 4.90\\
& XMM-Newton & 0844000701 & 58730.03 & 4.50\\
& XMM-Newton & 0844000801 & 58481.59 & 5.30 \\
& XMM-Newton & 0844000901 & 58734.02 & 2.80 \\ 
& XMM-Newton & 0844001001 & 58741.98 & 5.80 \\
& XMM-Newton & 0844001101 & 58743.98 & 5.10 \\
& XMM-Newton & 0844001201 & 58089.03 & 2.70 \\
\hline \hline \hline
\end{tabular}
\end{center}
\caption{Observations of 3 long period NSs. Exposure times for XMM-Newton are calculated after removing instances with high particle background flares.}
\end{table}

PSR~J0901--4046, PSR~J0250+5854, and GLEAM-X~J1627, were observed by Swift-XRT, XMM-Newton (PI:~Tan), and Chandra (PI:~Hurley-Walker), respectively. Table \ref{table_obs} lists a log of all of the observations for each source. The Swift-XRT data were reduced and analyzed using version 6.29 of the HEAsoft software package, while the XMM-Newton data were reduced using SAS version 20.0. For Chandra, we used CIAO version 4.14. We followed the standard procedures for reducing and cleaning each data set. We note that the exposure times for XMM-Newton listed in Table \ref{table_obs} are calculated after removing the times with high particle background flares. Next, we calculated the number of counts at each source's radio position using radii of $30''$, $15''$ and $2''$ for Swift-XRT, XMM-Newton, and Chandra respectively. Then we calculated the net counts for each source by subtracting the total number of expected background counts from large source free regions in each image after correcting for the difference in area of extraction regions. PSR J0901--4046 and GLEAM-X~J1627 both had zero net source counts detected, while PSR J0250+5854 had 15$\pm10$ net source counts detected. We converted these net source counts into $3\sigma$ upper-limits on the count rates in each observatory using Table~1 from \citet{1986ApJ...303..336G} and dividing by the total exposure, while also correcting for the energy containment fraction and exposure map. This lead to $3\sigma$ upper-limit count rates of $10^{-3}$, $9\times10^{-4}$, and $2\times10^{-4}$ cts s$^{-1}$ for PSR~J0901--4046, PSR~J0250+5854, and GLEAM-X~J1627, respectively. These count rates were calculated in the 0.5-10 keV bands for Swift and XMM-Newton, and 0.5-8 keV for Chandra.

We used the count rates to estimate the maximum temperature that each NS could be emitting from its entire surface, yet still remain undetected. We placed each source at its nominal dispersion measure estimated distance to calculate the blackbody model normalizations (i.e., {\tt bbodyrad} in Xspec). The absorbing column density for each source was estimated from the $N_{\rm H}$-DM relation of \cite{2013ApJ...768...64H}. We also checked the 3D extinction maps at the position and inferred DM distances of each source using {\tt mwdust} \citep{Bovy2015} and found good agreement between the $N_{\rm H}$ values derived from both the DM and E(B-V) (using the relationship of \citealt{2009MNRAS.400.2050G}) for PSR J0901 and GLEAM-X J1627. However, we found that the $N_{\rm H}$ value for PSR J0250 from the E(B-V) estimate was about a factor of 2.5 larger than the $N_{\rm H}$ derived from its DM. This larger $N_{\rm H}$ will generally increase the detectable temperature of PSR J0250's surface by $\sim10\%$. Since it is quite common for NSs to only emit from a small fraction of their surface (e.g., from hot spots), we repeated this exercise by shrinking the NS emitting area and recalculating the largest temperature that would remain undetected at the $3\sigma$ level. The curves shown in the left panel of Figure~\ref{fig:thermalage} were calculated adopting this procedure.


\bsp	
\label{lastpage}

\begin{thebibliography}{}
	\makeatletter
	\relax
	\def\mn@urlcharsother{\let\do\@makeother \do\$\do\&\do\#\do\^\do\_\do\%\do\~}
	\def\mn@doi{\begingroup\mn@urlcharsother \@ifnextchar [ {\mn@doi@}
		{\mn@doi@[]}}
	\def\mn@doi@[#1]#2{\def\@tempa{#1}\ifx\@tempa\@empty \href
		{http://dx.doi.org/#2} {doi:#2}\else \href {http://dx.doi.org/#2} {#1}\fi
		\endgroup}
	\def\mn@eprint#1#2{\mn@eprint@#1:#2::\@nil}
	\def\mn@eprint@arXiv#1{\href {http://arxiv.org/abs/#1} {{\tt arXiv:#1}}}
	\def\mn@eprint@dblp#1{\href {http://dblp.uni-trier.de/rec/bibtex/#1.xml}
		{dblp:#1}}
	\def\mn@eprint@#1:#2:#3:#4\@nil{\def\@tempa {#1}\def\@tempb {#2}\def\@tempc
		{#3}\ifx \@tempc \@empty \let \@tempc \@tempb \let \@tempb \@tempa \fi \ifx
		\@tempb \@empty \def\@tempb {arXiv}\fi \@ifundefined
		{mn@eprint@\@tempb}{\@tempb:\@tempc}{\expandafter \expandafter \csname
			mn@eprint@\@tempb\endcsname \expandafter{\@tempc}}}
	
	\bibitem[\protect\citeauthoryear{{Alcock}, {Farhi}  \& {Olinto}}{{Alcock}
		et~al.}{1986}]{1986ApJ...310..261A}
	{Alcock} C.,  {Farhi} E.,   {Olinto} A.,  1986, \mn@doi [\apj]
	{10.1086/164679}, \href
	{https://ui.adsabs.harvard.edu/abs/1986ApJ...310..261A} {310, 261}
	
	\bibitem[\protect\citeauthoryear{{Alexander}, {Hajela}, {Margutti}, {Bright},
		{Eftekhari}, {Kathirgamaraju}  \& {Berger}}{{Alexander}
		et~al.}{2020}]{Alexander2020}
	{Alexander} K.~D.,  {Hajela} A.,  {Margutti} R.,  {Bright} J.,  {Eftekhari} T.,
	{Kathirgamaraju} A.,   {Berger} E.,  2020, GRB Coordinates Network, \href
	{https://ui.adsabs.harvard.edu/abs/2020GCN.29053....1A} {29053, 1}
	
	\bibitem[\protect\citeauthoryear{{Alford}, {Schmitt}, {Rajagopal}  \&
		{Sch{\"a}fer}}{{Alford} et~al.}{2008}]{2008RvMP...80.1455A}
	{Alford} M.~G.,  {Schmitt} A.,  {Rajagopal} K.,   {Sch{\"a}fer} T.,  2008,
	\mn@doi [Reviews of Modern Physics] {10.1103/RevModPhys.80.1455}, \href
	{https://ui.adsabs.harvard.edu/abs/2008RvMP...80.1455A} {80, 1455}
	
	\bibitem[\protect\citeauthoryear{{Archibald} et~al.,}{{Archibald}
		et~al.}{2013}]{Archibald2013}
	{Archibald} R.~F.,  et~al., 2013, \mn@doi [\nat] {10.1038/nature12159}, \href
	{https://ui.adsabs.harvard.edu/abs/2013Natur.497..591A} {497, 591}
	
	\bibitem[\protect\citeauthoryear{{Archibald}, {Scholz}, {Kaspi}, {Tendulkar}
		\& {Beardmore}}{{Archibald} et~al.}{2020}]{Archibald2020}
	{Archibald} R.~F.,  {Scholz} P.,  {Kaspi} V.~M.,  {Tendulkar} S.~P.,
	{Beardmore} A.~P.,  2020, \mn@doi [\apj] {10.3847/1538-4357/ab660c}, \href
	{https://ui.adsabs.harvard.edu/abs/2020ApJ...889..160A} {889, 160}
	
	\bibitem[\protect\citeauthoryear{{Arons} \& {Barnard}}{{Arons} \&
		{Barnard}}{1986}]{1986ApJ...302..120A}
	{Arons} J.,  {Barnard} J.~J.,  1986, \mn@doi [\apj] {10.1086/163978}, \href
	{https://ui.adsabs.harvard.edu/abs/1986ApJ...302..120A} {302, 120}
	
	\bibitem[\protect\citeauthoryear{{Arzoumanian}, {Chernoff}  \&
		{Cordes}}{{Arzoumanian} et~al.}{2002}]{2002ApJ...568..289A}
	{Arzoumanian} Z.,  {Chernoff} D.~F.,   {Cordes} J.~M.,  2002, \mn@doi [\apj]
	{10.1086/338805}, \href
	{https://ui.adsabs.harvard.edu/abs/2002ApJ...568..289A} {568, 289}
	
	\bibitem[\protect\citeauthoryear{{Bachetti} et~al.,}{{Bachetti}
		et~al.}{2014}]{2014Natur.514..202B}
	{Bachetti} M.,  et~al., 2014, \mn@doi [\nat] {10.1038/nature13791}, \href
	{https://ui.adsabs.harvard.edu/abs/2014Natur.514..202B} {514, 202}
	
	\bibitem[\protect\citeauthoryear{{Bachetti} et~al.,}{{Bachetti}
		et~al.}{2022}]{2022ApJ...937..125B}
	{Bachetti} M.,  et~al., 2022, \mn@doi [\apj] {10.3847/1538-4357/ac8d67}, \href
	{https://ui.adsabs.harvard.edu/abs/2022ApJ...937..125B} {937, 125}
	
	\bibitem[\protect\citeauthoryear{{Bahcall} \& {Soneira}}{{Bahcall} \&
		{Soneira}}{1980}]{1980ApJS...44...73B}
	{Bahcall} J.~N.,  {Soneira} R.~M.,  1980, \mn@doi [\apjs] {10.1086/190685},
	\href {https://ui.adsabs.harvard.edu/abs/1980ApJS...44...73B} {44, 73}
	
	\bibitem[\protect\citeauthoryear{{Bailer-Jones}, {Rybizki}, {Fouesneau},
		{Mantelet}  \& {Andrae}}{{Bailer-Jones} et~al.}{2018}]{2018AJ....156...58B}
	{Bailer-Jones} C.~A.~L.,  {Rybizki} J.,  {Fouesneau} M.,  {Mantelet} G.,
	{Andrae} R.,  2018, \mn@doi [\aj] {10.3847/1538-3881/aacb21}, \href
	{https://ui.adsabs.harvard.edu/abs/2018AJ....156...58B} {156, 58}
	
	\bibitem[\protect\citeauthoryear{{Bailer-Jones}, {Rybizki}, {Fouesneau},
		{Demleitner}  \& {Andrae}}{{Bailer-Jones} et~al.}{2021}]{2021AJ....161..147B}
	{Bailer-Jones} C.~A.~L.,  {Rybizki} J.,  {Fouesneau} M.,  {Demleitner} M.,
	{Andrae} R.,  2021, \mn@doi [\aj] {10.3847/1538-3881/abd806}, \href
	{https://ui.adsabs.harvard.edu/abs/2021AJ....161..147B} {161, 147}
	
	\bibitem[\protect\citeauthoryear{{Baring}}{{Baring}}{2018}]{2018NatAs...2..282B}
	{Baring} M.~G.,  2018, \mn@doi [Nature Astronomy] {10.1038/s41550-018-0435-y},
	\href {https://ui.adsabs.harvard.edu/abs/2018NatAs...2..282B} {2, 282}
	
	\bibitem[\protect\citeauthoryear{{Barthelmy} et~al.,}{{Barthelmy}
		et~al.}{2016}]{2016ATel.8831....1B}
	{Barthelmy} S.~D.,  et~al., 2016, The Astronomer's Telegram, \href
	{https://ui.adsabs.harvard.edu/abs/2016ATel.8831....1B} {8831, 1}
	
	\bibitem[\protect\citeauthoryear{{Basu} et~al.,}{{Basu}
		et~al.}{2022}]{2022MNRAS.510.4049B}
	{Basu} A.,  et~al., 2022, \mn@doi [\mnras] {10.1093/mnras/stab3336}, \href
	{https://ui.adsabs.harvard.edu/abs/2022MNRAS.510.4049B} {510, 4049}
	
	\bibitem[\protect\citeauthoryear{{Baym}, {Pethick}  \& {Pines}}{{Baym}
		et~al.}{1969a}]{1969Natur.224..673B}
	{Baym} G.,  {Pethick} C.,   {Pines} D.,  1969a, \mn@doi [\nat]
	{10.1038/224673a0}, \href
	{https://ui.adsabs.harvard.edu/abs/1969Natur.224..673B} {224, 673}
	
	\bibitem[\protect\citeauthoryear{{Baym}, {Pethick}  \& {Pikes}}{{Baym}
		et~al.}{1969b}]{1969Natur.224..674B}
	{Baym} G.,  {Pethick} C.,   {Pikes} D.,  1969b, \mn@doi [\nat]
	{10.1038/224674a0}, \href
	{https://ui.adsabs.harvard.edu/abs/1969Natur.224..674B} {224, 674}
	
	\bibitem[\protect\citeauthoryear{{B{\'e}dard}, {Bergeron}, {Brassard}  \&
		{Fontaine}}{{B{\'e}dard} et~al.}{2020}]{2020ApJ...901...93B}
	{B{\'e}dard} A.,  {Bergeron} P.,  {Brassard} P.,   {Fontaine} G.,  2020,
	\mn@doi [\apj] {10.3847/1538-4357/abafbe}, \href
	{https://ui.adsabs.harvard.edu/abs/2020ApJ...901...93B} {901, 93}
	
	\bibitem[\protect\citeauthoryear{{Beniamini} \& {Piran}}{{Beniamini} \&
		{Piran}}{2019}]{BP2019}
	{Beniamini} P.,  {Piran} T.,  2019, \mn@doi [\mnras] {10.1093/mnras/stz1589},
	\href {https://ui.adsabs.harvard.edu/abs/2019MNRAS.487.4847B} {487, 4847}
	
	\bibitem[\protect\citeauthoryear{{Beniamini}, {Hotokezaka}, {van der Horst}  \&
		{Kouveliotou}}{{Beniamini} et~al.}{2019}]{Beniamini2019}
	{Beniamini} P.,  {Hotokezaka} K.,  {van der Horst} A.,   {Kouveliotou} C.,
	2019, \mn@doi [\mnras] {10.1093/mnras/stz1391}, \href
	{https://ui.adsabs.harvard.edu/abs/2019MNRAS.487.1426B} {487, 1426}
	
	\bibitem[\protect\citeauthoryear{{Beniamini}, {Wadiasingh}  \&
		{Metzger}}{{Beniamini} et~al.}{2020}]{Beniamini+20}
	{Beniamini} P.,  {Wadiasingh} Z.,   {Metzger} B.~D.,  2020, \mn@doi [\mnras]
	{10.1093/mnras/staa1783}, \href
	{https://ui.adsabs.harvard.edu/abs/2020MNRAS.496.3390B} {496, 3390}
	
	\bibitem[\protect\citeauthoryear{{Bethapudi}, {Spitler}, {Main}, {Li}  \&
		{Wharton}}{{Bethapudi} et~al.}{2022}]{2022arXiv220713669B}
	{Bethapudi} S.,  {Spitler} L.~G.,  {Main} R.~A.,  {Li} D.~Z.,   {Wharton}
	R.~S.,  2022, arXiv e-prints, \href
	{https://ui.adsabs.harvard.edu/abs/2022arXiv220713669B} {p. arXiv:2207.13669}
	
	\bibitem[\protect\citeauthoryear{{Bhardwaj} et~al.,}{{Bhardwaj}
		et~al.}{2021}]{2021ApJ...910L..18B}
	{Bhardwaj} M.,  et~al., 2021, \mn@doi [\apjl] {10.3847/2041-8213/abeaa6}, \href
	{https://ui.adsabs.harvard.edu/abs/2021ApJ...910L..18B} {910, L18}
	
	\bibitem[\protect\citeauthoryear{{Bochenek}, {Ravi}, {Belov}, {Hallinan},
		{Kocz}, {Kulkarni}  \& {McKenna}}{{Bochenek} et~al.}{2020}]{STARE2020}
	{Bochenek} C.~D.,  {Ravi} V.,  {Belov} K.~V.,  {Hallinan} G.,  {Kocz} J.,
	{Kulkarni} S.~R.,   {McKenna} D.~L.,  2020, \mn@doi [\nat]
	{10.1038/s41586-020-2872-x}, \href
	{https://ui.adsabs.harvard.edu/abs/2020Natur.587...59B} {587, 59}
	
	\bibitem[\protect\citeauthoryear{{Borghese} et~al.,}{{Borghese}
		et~al.}{2018}]{2018MNRAS.478..741B}
	{Borghese} A.,  et~al., 2018, \mn@doi [\mnras] {10.1093/mnras/sty1119}, \href
	{https://ui.adsabs.harvard.edu/abs/2018MNRAS.478..741B} {478, 741}
	
	\bibitem[\protect\citeauthoryear{{Bovy}}{{Bovy}}{2015}]{Bovy2015}
	{Bovy} J.,  2015, \mn@doi [\apjs] {10.1088/0067-0049/216/2/29}, \href
	{https://ui.adsabs.harvard.edu/abs/2015ApJS..216...29B} {216, 29}
	
	\bibitem[\protect\citeauthoryear{{Bovy}, {Rix}, {Green}, {Schlafly}  \&
		{Finkbeiner}}{{Bovy} et~al.}{2016}]{2016ApJ...818..130B}
	{Bovy} J.,  {Rix} H.-W.,  {Green} G.~M.,  {Schlafly} E.~F.,   {Finkbeiner}
	D.~P.,  2016, \mn@doi [\apj] {10.3847/0004-637X/818/2/130}, \href
	{https://ui.adsabs.harvard.edu/abs/2016ApJ...818..130B} {818, 130}
	
	\bibitem[\protect\citeauthoryear{{Boyles}, {Lorimer}, {Turk}, {Mnatsakanov},
		{Lynch}, {Ransom}, {Freire}  \& {Belczynski}}{{Boyles}
		et~al.}{2011}]{2011ApJ...742...51B}
	{Boyles} J.,  {Lorimer} D.~R.,  {Turk} P.~J.,  {Mnatsakanov} R.,  {Lynch}
	R.~S.,  {Ransom} S.~M.,  {Freire} P.~C.,   {Belczynski} K.,  2011, \mn@doi
	[\apj] {10.1088/0004-637X/742/1/51}, \href
	{https://ui.adsabs.harvard.edu/abs/2011ApJ...742...51B} {742, 51}
	
	\bibitem[\protect\citeauthoryear{{Bozzo}, {Ferrigno}, {Oskinova}  \&
		{Ducci}}{{Bozzo} et~al.}{2022}]{2022MNRAS.510.4645B}
	{Bozzo} E.,  {Ferrigno} C.,  {Oskinova} L.,   {Ducci} L.,  2022, \mn@doi
	[\mnras] {10.1093/mnras/stab3688}, \href
	{https://ui.adsabs.harvard.edu/abs/2022MNRAS.510.4645B} {510, 4645}
	
	\bibitem[\protect\citeauthoryear{{Bransgrove}, {Levin}  \&
		{Beloborodov}}{{Bransgrove} et~al.}{2018}]{2018MNRAS.473.2771B}
	{Bransgrove} A.,  {Levin} Y.,   {Beloborodov} A.,  2018, \mn@doi [\mnras]
	{10.1093/mnras/stx2508}, \href
	{https://ui.adsabs.harvard.edu/abs/2018MNRAS.473.2771B} {473, 2771}
	
	\bibitem[\protect\citeauthoryear{{Braun}, {Safi-Harb}  \& {Fryer}}{{Braun}
		et~al.}{2019}]{2019MNRAS.489.4444B}
	{Braun} C.,  {Safi-Harb} S.,   {Fryer} C.~L.,  2019, \mn@doi [\mnras]
	{10.1093/mnras/stz2437}, \href
	{https://ui.adsabs.harvard.edu/abs/2019MNRAS.489.4444B} {489, 4444}
	
	\bibitem[\protect\citeauthoryear{{Bretz}, {van Eysden}  \& {Link}}{{Bretz}
		et~al.}{2021}]{2021MNRAS.504.5880B}
	{Bretz} J.,  {van Eysden} C.~A.,   {Link} B.,  2021, \mn@doi [\mnras]
	{10.1093/mnras/stab1220}, \href
	{https://ui.adsabs.harvard.edu/abs/2021MNRAS.504.5880B} {504, 5880}
	
	\bibitem[\protect\citeauthoryear{{Brightman} et~al.,}{{Brightman}
		et~al.}{2018}]{2018NatAs...2..312B}
	{Brightman} M.,  et~al., 2018, \mn@doi [Nature Astronomy]
	{10.1038/s41550-018-0391-6}, \href
	{https://ui.adsabs.harvard.edu/abs/2018NatAs...2..312B} {2, 312}
	
	\bibitem[\protect\citeauthoryear{{Buckley}, {Meintjes}, {Potter}, {Marsh}  \&
		{G{\"a}nsicke}}{{Buckley} et~al.}{2017}]{2017NatAs...1E..29B}
	{Buckley} D.~A.~H.,  {Meintjes} P.~J.,  {Potter} S.~B.,  {Marsh} T.~R.,
	{G{\"a}nsicke} B.~T.,  2017, \mn@doi [Nature Astronomy]
	{10.1038/s41550-016-0029}, \href
	{https://ui.adsabs.harvard.edu/abs/2017NatAs...1E..29B} {1, 0029}
	
	\bibitem[\protect\citeauthoryear{{Burgay} et~al.,}{{Burgay}
		et~al.}{2006}]{2006MNRAS.368..283B}
	{Burgay} M.,  et~al., 2006, \mn@doi [\mnras]
	{10.1111/j.1365-2966.2006.10100.x}, \href
	{https://ui.adsabs.harvard.edu/abs/2006MNRAS.368..283B} {368, 283}
	
	\bibitem[\protect\citeauthoryear{{Caleb} et~al.,}{{Caleb}
		et~al.}{2022}]{caleb2022}
	{Caleb} M.,  et~al., 2022, \mn@doi [Nature Astronomy]
	{10.1038/s41550-022-01688-x}, \href
	{https://ui.adsabs.harvard.edu/abs/2022NatAs.tmp..123C} {}
	
	\bibitem[\protect\citeauthoryear{{Castro-Tirado} et~al.,}{{Castro-Tirado}
		et~al.}{2021}]{2021Natur.600..621C}
	{Castro-Tirado} A.~J.,  et~al., 2021, \mn@doi [\nat]
	{10.1038/s41586-021-04101-1}, \href
	{https://ui.adsabs.harvard.edu/abs/2021Natur.600..621C} {600, 621}
	
	\bibitem[\protect\citeauthoryear{{Chau}, {Cheng}  \& {Ding}}{{Chau}
		et~al.}{1992}]{1992ApJ...399..213C}
	{Chau} H.~F.,  {Cheng} K.~S.,   {Ding} K.~Y.,  1992, \mn@doi [\apj]
	{10.1086/171917}, \href
	{https://ui.adsabs.harvard.edu/abs/1992ApJ...399..213C} {399, 213}
	
	\bibitem[\protect\citeauthoryear{{Cheng}, {Epstein}, {Guyer}  \&
		{Young}}{{Cheng} et~al.}{1996}]{1996Natur.382..518C}
	{Cheng} B.,  {Epstein} R.~I.,  {Guyer} R.~A.,   {Young} A.~C.,  1996, \mn@doi
	[\nat] {10.1038/382518a0}, \href
	{https://ui.adsabs.harvard.edu/abs/1996Natur.382..518C} {382, 518}
	
	\bibitem[\protect\citeauthoryear{{Chime/Frb Collaboration} et~al.,}{{Chime/Frb
			Collaboration} et~al.}{2020}]{CHIMEperiodicity}
	{Chime/Frb Collaboration} et~al., 2020, \mn@doi [\nat]
	{10.1038/s41586-020-2398-2}, \href
	{https://ui.adsabs.harvard.edu/abs/2020Natur.582..351C} {582, 351}
	
	\bibitem[\protect\citeauthoryear{{Chrimes} et~al.,}{{Chrimes}
		et~al.}{2022}]{2022MNRAS.513.3550C}
	{Chrimes} A.~A.,  et~al., 2022, \mn@doi [\mnras] {10.1093/mnras/stac1090},
	\href {https://ui.adsabs.harvard.edu/abs/2022MNRAS.513.3550C} {513, 3550}
	
	\bibitem[\protect\citeauthoryear{{Colaiuda} \& {Kokkotas}}{{Colaiuda} \&
		{Kokkotas}}{2011}]{2011MNRAS.414.3014C}
	{Colaiuda} A.,  {Kokkotas} K.~D.,  2011, \mn@doi [\mnras]
	{10.1111/j.1365-2966.2011.18602.x}, \href
	{https://ui.adsabs.harvard.edu/abs/2011MNRAS.414.3014C} {414, 3014}
	
	\bibitem[\protect\citeauthoryear{{Colaiuda} \& {Kokkotas}}{{Colaiuda} \&
		{Kokkotas}}{2012}]{2012MNRAS.423..811C}
	{Colaiuda} A.,  {Kokkotas} K.~D.,  2012, \mn@doi [\mnras]
	{10.1111/j.1365-2966.2012.20919.x}, \href
	{https://ui.adsabs.harvard.edu/abs/2012MNRAS.423..811C} {423, 811}
	
	\bibitem[\protect\citeauthoryear{{Colpi}, {Geppert}  \& {Page}}{{Colpi}
		et~al.}{2000}]{Colpi2000}
	{Colpi} M.,  {Geppert} U.,   {Page} D.,  2000, \mn@doi [\apjl]
	{10.1086/312448}, \href
	{https://ui.adsabs.harvard.edu/abs/2000ApJ...529L..29C} {529, L29}
	
	\bibitem[\protect\citeauthoryear{{Cooper}, {Gupta}, {Wadiasingh}, {Wijers},
		{Boersma}, {Andreoni}, {Rowlinson}  \& {Gourdji}}{{Cooper}
		et~al.}{2022}]{2022arXiv221017205C}
	{Cooper} A.~J.,  {Gupta} O.,  {Wadiasingh} Z.,  {Wijers} R.~A.~M.~J.,
	{Boersma} O.~M.,  {Andreoni} I.,  {Rowlinson} A.,   {Gourdji} K.,  2022,
	arXiv e-prints, \href {https://ui.adsabs.harvard.edu/abs/2022arXiv221017205C}
	{p. arXiv:2210.17205}
	
	\bibitem[\protect\citeauthoryear{{Corbet}}{{Corbet}}{1984}]{1984A&A...141...91C}
	{Corbet} R.~H.~D.,  1984, \aap, \href
	{https://ui.adsabs.harvard.edu/abs/1984A&A...141...91C} {141, 91}
	
	\bibitem[\protect\citeauthoryear{{Corbet}}{{Corbet}}{1986}]{1986MNRAS.220.1047C}
	{Corbet} R.~H.~D.,  1986, \mn@doi [\mnras] {10.1093/mnras/220.4.1047}, \href
	{https://ui.adsabs.harvard.edu/abs/1986MNRAS.220.1047C} {220, 1047}
	
	\bibitem[\protect\citeauthoryear{{Cordes}}{{Cordes}}{1978}]{1978ApJ...222.1006C}
	{Cordes} J.~M.,  1978, \mn@doi [\apj] {10.1086/156218}, \href
	{https://ui.adsabs.harvard.edu/abs/1978ApJ...222.1006C} {222, 1006}
	
	\bibitem[\protect\citeauthoryear{{Cordes} \& {McLaughlin}}{{Cordes} \&
		{McLaughlin}}{2003}]{2003ApJ...596.1142C}
	{Cordes} J.~M.,  {McLaughlin} M.~A.,  2003, \mn@doi [\apj] {10.1086/378231},
	\href {https://ui.adsabs.harvard.edu/abs/2003ApJ...596.1142C} {596, 1142}
	
	\bibitem[\protect\citeauthoryear{{Cruces} et~al.,}{{Cruces}
		et~al.}{2021}]{2021MNRAS.500..448C}
	{Cruces} M.,  et~al., 2021, \mn@doi [\mnras] {10.1093/mnras/staa3223}, \href
	{https://ui.adsabs.harvard.edu/abs/2021MNRAS.500..448C} {500, 448}
	
	\bibitem[\protect\citeauthoryear{{Cumming}}{{Cumming}}{2002}]{2002MNRAS.333..589C}
	{Cumming} A.,  2002, \mn@doi [\mnras] {10.1046/j.1365-8711.2002.05434.x}, \href
	{https://ui.adsabs.harvard.edu/abs/2002MNRAS.333..589C} {333, 589}
	
	\bibitem[\protect\citeauthoryear{{Cumming}, {Arras}  \& {Zweibel}}{{Cumming}
		et~al.}{2004}]{2004ApJ...609..999C}
	{Cumming} A.,  {Arras} P.,   {Zweibel} E.,  2004, \mn@doi [\apj]
	{10.1086/421324}, \href
	{https://ui.adsabs.harvard.edu/abs/2004ApJ...609..999C} {609, 999}
	
	\bibitem[\protect\citeauthoryear{{D'A{\`\i}} et~al.,}{{D'A{\`\i}}
		et~al.}{2016}]{2016MNRAS.463.2394D}
	{D'A{\`\i}} A.,  et~al., 2016, \mn@doi [\mnras] {10.1093/mnras/stw2023}, \href
	{https://ui.adsabs.harvard.edu/abs/2016MNRAS.463.2394D} {463, 2394}
	
	\bibitem[\protect\citeauthoryear{{Dall'Osso}, {Granot}  \& {Piran}}{{Dall'Osso}
		et~al.}{2012}]{Dall'Osso2012D}
	{Dall'Osso} S.,  {Granot} J.,   {Piran} T.,  2012, \mn@doi [\mnras]
	{10.1111/j.1365-2966.2012.20612.x}, \href
	{https://ui.adsabs.harvard.edu/abs/2012MNRAS.422.2878D} {422, 2878}
	
	\bibitem[\protect\citeauthoryear{{De Luca}, {Caraveo}, {Mereghetti}, {Tiengo}
		\& {Bignami}}{{De Luca} et~al.}{2006}]{DeLuca2006}
	{De Luca} A.,  {Caraveo} P.~A.,  {Mereghetti} S.,  {Tiengo} A.,   {Bignami}
	G.~F.,  2006, \mn@doi [Science] {10.1126/science.1129185}, \href
	{https://ui.adsabs.harvard.edu/abs/2006Sci...313..814D} {313, 814}
	
	\bibitem[\protect\citeauthoryear{{De Luca}, {Mignani}, {Zaggia}, {Beccari},
		{Mereghetti}, {Caraveo}  \& {Bignami}}{{De Luca}
		et~al.}{2008}]{2008ApJ...682.1185D}
	{De Luca} A.,  {Mignani} R.~P.,  {Zaggia} S.,  {Beccari} G.,  {Mereghetti} S.,
	{Caraveo} P.~A.,   {Bignami} G.~F.,  2008, \mn@doi [\apj] {10.1086/588600},
	\href {https://ui.adsabs.harvard.edu/abs/2008ApJ...682.1185D} {682, 1185}
	
	\bibitem[\protect\citeauthoryear{{De Luca} et~al.,}{{De Luca}
		et~al.}{2021}]{deluca21AA}
	{De Luca} A.,  et~al., 2021, \mn@doi [\aap] {10.1051/0004-6361/202039783},
	\href {https://ui.adsabs.harvard.edu/abs/2021A&A...650A.167D} {650, A167}
	
	\bibitem[\protect\citeauthoryear{Dewdney, Braun  \& Turner}{Dewdney
		et~al.}{2017}]{dewdney2017}
	Dewdney P.~E.,  Braun R.,   Turner W.,  2017, in 2017 XXXIInd General Assembly
	and Scientific Symposium of the International Union of Radio Science (URSI
	GASS). pp~1--4, \mn@doi{10.23919/URSIGASS.2017.8105425}
	
	\bibitem[\protect\citeauthoryear{{Dib} \& {Kaspi}}{{Dib} \&
		{Kaspi}}{2014}]{dib14ApJ}
	{Dib} R.,  {Kaspi} V.~M.,  2014, \mn@doi [\apj] {10.1088/0004-637X/784/1/37},
	\href {https://ui.adsabs.harvard.edu/abs/2014ApJ...784...37D} {784, 37}
	
	\bibitem[\protect\citeauthoryear{{Dommes} \& {Gusakov}}{{Dommes} \&
		{Gusakov}}{2017}]{2017MNRAS.467L.115D}
	{Dommes} V.~A.,  {Gusakov} M.~E.,  2017, \mn@doi [\mnras]
	{10.1093/mnrasl/slx011}, \href
	{https://ui.adsabs.harvard.edu/abs/2017MNRAS.467L.115D} {467, L115}
	
	\bibitem[\protect\citeauthoryear{{Doroshenko}, {Santangelo}, {Tsygankov}  \&
		{Ji}}{{Doroshenko} et~al.}{2021}]{2021A&A...647A.165D}
	{Doroshenko} V.,  {Santangelo} A.,  {Tsygankov} S.~S.,   {Ji} L.,  2021,
	\mn@doi [\aap] {10.1051/0004-6361/202039785}, \href
	{https://ui.adsabs.harvard.edu/abs/2021A&A...647A.165D} {647, A165}
	
	\bibitem[\protect\citeauthoryear{{Duncan} \& {Thompson}}{{Duncan} \&
		{Thompson}}{1992}]{1992ApJ...392L...9D}
	{Duncan} R.~C.,  {Thompson} C.,  1992, \mn@doi [\apjl] {10.1086/186413}, \href
	{https://ui.adsabs.harvard.edu/abs/1992ApJ...392L...9D} {392, L9}
	
	\bibitem[\protect\citeauthoryear{{Ek{\c{s}}i} \&
		{{\c{S}}a{\c{s}}maz}}{{Ek{\c{s}}i} \&
		{{\c{S}}a{\c{s}}maz}}{2022}]{2022arXiv220205160E}
	{Ek{\c{s}}i} K.~Y.,  {{\c{S}}a{\c{s}}maz} S.,  2022, arXiv e-prints, \href
	{https://ui.adsabs.harvard.edu/abs/2022arXiv220205160E} {p. arXiv:2202.05160}
	
	\bibitem[\protect\citeauthoryear{{Elfritz}, {Pons}, {Rea}, {Glampedakis}  \&
		{Vigan{\`o}}}{{Elfritz} et~al.}{2016}]{2016MNRAS.456.4461E}
	{Elfritz} J.~G.,  {Pons} J.~A.,  {Rea} N.,  {Glampedakis} K.,   {Vigan{\`o}}
	D.,  2016, \mn@doi [\mnras] {10.1093/mnras/stv2963}, \href
	{https://ui.adsabs.harvard.edu/abs/2016MNRAS.456.4461E} {456, 4461}
	
	\bibitem[\protect\citeauthoryear{{Enoto} et~al.,}{{Enoto}
		et~al.}{2014}]{2014ApJ...786..127E}
	{Enoto} T.,  et~al., 2014, \mn@doi [\apj] {10.1088/0004-637X/786/2/127}, \href
	{https://ui.adsabs.harvard.edu/abs/2014ApJ...786..127E} {786, 127}
	
	\bibitem[\protect\citeauthoryear{{Enoto} et~al.,}{{Enoto}
		et~al.}{2017}]{2017ApJS..231....8E}
	{Enoto} T.,  et~al., 2017, \mn@doi [\apjs] {10.3847/1538-4365/aa6f0a}, \href
	{https://ui.adsabs.harvard.edu/abs/2017ApJS..231....8E} {231, 8}
	
	\bibitem[\protect\citeauthoryear{{Erber}}{{Erber}}{1966}]{1966RvMP...38..626E}
	{Erber} T.,  1966, \mn@doi [Reviews of Modern Physics]
	{10.1103/RevModPhys.38.626}, \href
	{https://ui.adsabs.harvard.edu/abs/1966RvMP...38..626E} {38, 626}
	
	\bibitem[\protect\citeauthoryear{{Erkut}}{{Erkut}}{2022}]{2022MNRAS.514L..41E}
	{Erkut} M.~H.,  2022, \mn@doi [\mnras] {10.1093/mnrasl/slac057}, \href
	{https://ui.adsabs.harvard.edu/abs/2022MNRAS.514L..41E} {514, L41}
	
	\bibitem[\protect\citeauthoryear{{Espinoza}, {Lyne}, {Stappers}  \&
		{Kramer}}{{Espinoza} et~al.}{2011}]{2011MNRAS.414.1679E}
	{Espinoza} C.~M.,  {Lyne} A.~G.,  {Stappers} B.~W.,   {Kramer} M.,  2011,
	\mn@doi [\mnras] {10.1111/j.1365-2966.2011.18503.x}, \href
	{https://ui.adsabs.harvard.edu/abs/2011MNRAS.414.1679E} {414, 1679}
	
	\bibitem[\protect\citeauthoryear{{Esposito}, {Turolla}, {de Luca}, {Israel},
		{Possenti}  \& {Burrows}}{{Esposito} et~al.}{2011}]{2011MNRAS.418..170E}
	{Esposito} P.,  {Turolla} R.,  {de Luca} A.,  {Israel} G.~L.,  {Possenti} A.,
	{Burrows} D.~N.,  2011, \mn@doi [\mnras] {10.1111/j.1365-2966.2011.19473.x},
	\href {https://ui.adsabs.harvard.edu/abs/2011MNRAS.418..170E} {418, 170}
	
	\bibitem[\protect\citeauthoryear{{Faucher-Gigu{\`e}re} \&
		{Kaspi}}{{Faucher-Gigu{\`e}re} \& {Kaspi}}{2006}]{GiguereKaspi2006}
	{Faucher-Gigu{\`e}re} C.-A.,  {Kaspi} V.~M.,  2006, \mn@doi [\apj]
	{10.1086/501516}, \href
	{https://ui.adsabs.harvard.edu/abs/2006ApJ...643..332F} {643, 332}
	
	\bibitem[\protect\citeauthoryear{{Ferrario}, {de Martino}  \&
		{G{\"a}nsicke}}{{Ferrario} et~al.}{2015}]{2015SSRv..191..111F}
	{Ferrario} L.,  {de Martino} D.,   {G{\"a}nsicke} B.~T.,  2015, \mn@doi [\ssr]
	{10.1007/s11214-015-0152-0}, \href
	{https://ui.adsabs.harvard.edu/abs/2015SSRv..191..111F} {191, 111}
	
	\bibitem[\protect\citeauthoryear{{Ferrario}, {Wickramasinghe}  \&
		{Kawka}}{{Ferrario} et~al.}{2020}]{2020AdSpR..66.1025F}
	{Ferrario} L.,  {Wickramasinghe} D.,   {Kawka} A.,  2020, \mn@doi [Advances in
	Space Research] {10.1016/j.asr.2019.11.012}, \href
	{https://ui.adsabs.harvard.edu/abs/2020AdSpR..66.1025F} {66, 1025}
	
	\bibitem[\protect\citeauthoryear{{Fitzpatrick}}{{Fitzpatrick}}{1999}]{1999PASP..111...63F}
	{Fitzpatrick} E.~L.,  1999, \mn@doi [\pasp] {10.1086/316293}, \href
	{https://ui.adsabs.harvard.edu/abs/1999PASP..111...63F} {111, 63}
	
	\bibitem[\protect\citeauthoryear{{Fuentes}, {Espinoza}, {Reisenegger}, {Shaw},
		{Stappers}  \& {Lyne}}{{Fuentes} et~al.}{2017}]{2017A&A...608A.131F}
	{Fuentes} J.~R.,  {Espinoza} C.~M.,  {Reisenegger} A.,  {Shaw} B.,  {Stappers}
	B.~W.,   {Lyne} A.~G.,  2017, \mn@doi [\aap] {10.1051/0004-6361/201731519},
	\href {https://ui.adsabs.harvard.edu/abs/2017A&A...608A.131F} {608, A131}
	
	\bibitem[\protect\citeauthoryear{{F{\"u}rst} et~al.,}{{F{\"u}rst}
		et~al.}{2016}]{2016ApJ...831L..14F}
	{F{\"u}rst} F.,  et~al., 2016, \mn@doi [\apjl] {10.3847/2041-8205/831/2/L14},
	\href {https://ui.adsabs.harvard.edu/abs/2016ApJ...831L..14F} {831, L14}
	
	\bibitem[\protect\citeauthoryear{{Gabler}, {Cerd{\'a}-Dur{\'a}n},
		{Stergioulas}, {Font}  \& {M{\"u}ller}}{{Gabler}
		et~al.}{2012}]{2012MNRAS.421.2054G}
	{Gabler} M.,  {Cerd{\'a}-Dur{\'a}n} P.,  {Stergioulas} N.,  {Font} J.~A.,
	{M{\"u}ller} E.,  2012, \mn@doi [\mnras] {10.1111/j.1365-2966.2012.20454.x},
	\href {https://ui.adsabs.harvard.edu/abs/2012MNRAS.421.2054G} {421, 2054}
	
	\bibitem[\protect\citeauthoryear{{Gehrels}}{{Gehrels}}{1986}]{1986ApJ...303..336G}
	{Gehrels} N.,  1986, \mn@doi [\apj] {10.1086/164079}, \href
	{https://ui.adsabs.harvard.edu/abs/1986ApJ...303..336G} {303, 336}
	
	\bibitem[\protect\citeauthoryear{{Glampedakis}, {Andersson}  \&
		{Samuelsson}}{{Glampedakis} et~al.}{2011a}]{2011MNRAS.410..805G}
	{Glampedakis} K.,  {Andersson} N.,   {Samuelsson} L.,  2011a, \mn@doi [\mnras]
	{10.1111/j.1365-2966.2010.17484.x}, \href
	{https://ui.adsabs.harvard.edu/abs/2011MNRAS.410..805G} {410, 805}
	
	\bibitem[\protect\citeauthoryear{{Glampedakis}, {Jones}  \&
		{Samuelsson}}{{Glampedakis} et~al.}{2011b}]{2011MNRAS.413.2021G}
	{Glampedakis} K.,  {Jones} D.~I.,   {Samuelsson} L.,  2011b, \mn@doi [\mnras]
	{10.1111/j.1365-2966.2011.18278.x}, \href
	{https://ui.adsabs.harvard.edu/abs/2011MNRAS.413.2021G} {413, 2021}
	
	\bibitem[\protect\citeauthoryear{{Goldreich} \& {Reisenegger}}{{Goldreich} \&
		{Reisenegger}}{1992}]{1992ApJ...395..250G}
	{Goldreich} P.,  {Reisenegger} A.,  1992, \mn@doi [\apj] {10.1086/171646},
	\href {https://ui.adsabs.harvard.edu/abs/1992ApJ...395..250G} {395, 250}
	
	\bibitem[\protect\citeauthoryear{{Gourgouliatos} \& {Cumming}}{{Gourgouliatos}
		\& {Cumming}}{2014a}]{2014PhRvL.112q1101G}
	{Gourgouliatos} K.~N.,  {Cumming} A.,  2014a, \mn@doi [\prl]
	{10.1103/PhysRevLett.112.171101}, \href
	{https://ui.adsabs.harvard.edu/abs/2014PhRvL.112q1101G} {112, 171101}
	
	\bibitem[\protect\citeauthoryear{{Gourgouliatos} \& {Cumming}}{{Gourgouliatos}
		\& {Cumming}}{2014b}]{2014MNRAS.438.1618G}
	{Gourgouliatos} K.~N.,  {Cumming} A.,  2014b, \mn@doi [\mnras]
	{10.1093/mnras/stt2300}, \href
	{https://ui.adsabs.harvard.edu/abs/2014MNRAS.438.1618G} {438, 1618}
	
	\bibitem[\protect\citeauthoryear{{Gourgouliatos}, {Cumming}, {Reisenegger},
		{Armaza}, {Lyutikov}  \& {Valdivia}}{{Gourgouliatos}
		et~al.}{2013}]{2013MNRAS.434.2480G}
	{Gourgouliatos} K.~N.,  {Cumming} A.,  {Reisenegger} A.,  {Armaza} C.,
	{Lyutikov} M.,   {Valdivia} J.~A.,  2013, \mn@doi [\mnras]
	{10.1093/mnras/stt1195}, \href
	{https://ui.adsabs.harvard.edu/abs/2013MNRAS.434.2480G} {434, 2480}
	
	\bibitem[\protect\citeauthoryear{{Gourgouliatos}, {Wood}  \&
		{Hollerbach}}{{Gourgouliatos} et~al.}{2016}]{2016PNAS..113.3944G}
	{Gourgouliatos} K.~N.,  {Wood} T.~S.,   {Hollerbach} R.,  2016, \mn@doi
	[Proceedings of the National Academy of Science] {10.1073/pnas.1522363113},
	\href {https://ui.adsabs.harvard.edu/abs/2016PNAS..113.3944G} {113, 3944}
	
	\bibitem[\protect\citeauthoryear{{Gourgouliatos}, {De Grandis}  \&
		{Igoshev}}{{Gourgouliatos} et~al.}{2022}]{2022Symm...14..130G}
	{Gourgouliatos} K.~N.,  {De Grandis} D.,   {Igoshev} A.,  2022, \mn@doi
	[Symmetry] {10.3390/sym14010130}, \href
	{https://ui.adsabs.harvard.edu/abs/2022Symm...14..130G} {14, 130}
	
	\bibitem[\protect\citeauthoryear{{Graber}}{{Graber}}{2017}]{2017AN....338.1090G}
	{Graber} V.,  2017, \mn@doi [Astronomische Nachrichten]
	{10.1002/asna.201713441}, \href
	{https://ui.adsabs.harvard.edu/abs/2017AN....338.1090G} {338, 1090}
	
	\bibitem[\protect\citeauthoryear{{Graber}, {Andersson}, {Glampedakis}  \&
		{Lander}}{{Graber} et~al.}{2015}]{2015MNRAS.453..671G}
	{Graber} V.,  {Andersson} N.,  {Glampedakis} K.,   {Lander} S.~K.,  2015,
	\mn@doi [\mnras] {10.1093/mnras/stv1648}, \href
	{https://ui.adsabs.harvard.edu/abs/2015MNRAS.453..671G} {453, 671}
	
	\bibitem[\protect\citeauthoryear{{Green}}{{Green}}{2019}]{2019JApA...40...36G}
	{Green} D.~A.,  2019, \mn@doi [Journal of Astrophysics and Astronomy]
	{10.1007/s12036-019-9601-6}, \href
	{https://ui.adsabs.harvard.edu/abs/2019JApA...40...36G} {40, 36}
	
	\bibitem[\protect\citeauthoryear{{Gusakov}}{{Gusakov}}{2016}]{2016PhRvD..93f4033G}
	{Gusakov} M.~E.,  2016, \mn@doi [\prd] {10.1103/PhysRevD.93.064033}, \href
	{https://ui.adsabs.harvard.edu/abs/2016PhRvD..93f4033G} {93, 064033}
	
	\bibitem[\protect\citeauthoryear{{Gusakov} \& {Dommes}}{{Gusakov} \&
		{Dommes}}{2016}]{2016PhRvD..94h3006G}
	{Gusakov} M.~E.,  {Dommes} V.~A.,  2016, \mn@doi [\prd]
	{10.1103/PhysRevD.94.083006}, \href
	{https://ui.adsabs.harvard.edu/abs/2016PhRvD..94h3006G} {94, 083006}
	
	\bibitem[\protect\citeauthoryear{{G{\"u}ver} \& {{\"O}zel}}{{G{\"u}ver} \&
		{{\"O}zel}}{2009}]{2009MNRAS.400.2050G}
	{G{\"u}ver} T.,  {{\"O}zel} F.,  2009, \mn@doi [\mnras]
	{10.1111/j.1365-2966.2009.15598.x}, \href
	{https://ui.adsabs.harvard.edu/abs/2009MNRAS.400.2050G} {400, 2050}
	
	\bibitem[\protect\citeauthoryear{{Haberl} \& {Sturm}}{{Haberl} \&
		{Sturm}}{2016}]{2016A&A...586A..81H}
	{Haberl} F.,  {Sturm} R.,  2016, \mn@doi [\aap] {10.1051/0004-6361/201527326},
	\href {https://ui.adsabs.harvard.edu/abs/2016A&A...586A..81H} {586, A81}
	
	\bibitem[\protect\citeauthoryear{{Haberl}, {Sturm}, {Filipovi{\'c}}, {Pietsch}
		\& {Crawford}}{{Haberl} et~al.}{2012}]{2012A&A...537L...1H}
	{Haberl} F.,  {Sturm} R.,  {Filipovi{\'c}} M.~D.,  {Pietsch} W.,   {Crawford}
	E.~J.,  2012, \mn@doi [\aap] {10.1051/0004-6361/201118369}, \href
	{https://ui.adsabs.harvard.edu/abs/2012A&A...537L...1H} {537, L1}
	
	\bibitem[\protect\citeauthoryear{{Hallinan} et~al.,}{{Hallinan}
		et~al.}{2019}]{Hallinan2019}
	{Hallinan} G.,  et~al., 2019, in Bulletin of the American Astronomical Society.
	p.~255 (\mn@eprint {arXiv} {1907.07648})
	
	\bibitem[\protect\citeauthoryear{{Harding}, {Contopoulos}  \&
		{Kazanas}}{{Harding} et~al.}{1999}]{Harding1999}
	{Harding} A.~K.,  {Contopoulos} I.,   {Kazanas} D.,  1999, \mn@doi [\apjl]
	{10.1086/312339}, \href
	{https://ui.adsabs.harvard.edu/abs/1999ApJ...525L.125H} {525, L125}
	
	\bibitem[\protect\citeauthoryear{{Hare}, {Kargaltsev}  \& {Rangelov}}{{Hare}
		et~al.}{2018}]{2018ApJ...865...33H}
	{Hare} J.,  {Kargaltsev} O.,   {Rangelov} B.,  2018, \mn@doi [\apj]
	{10.3847/1538-4357/aad90d}, \href
	{https://ui.adsabs.harvard.edu/abs/2018ApJ...865...33H} {865, 33}
	
	\bibitem[\protect\citeauthoryear{{Harrison}, {Lynch}  \& {NRAO Green Bank
			Telescope}}{{Harrison} et~al.}{2017}]{2017AAS...22943104H}
	{Harrison} A.,  {Lynch} R.,   {NRAO Green Bank Telescope} 2017, in American
	Astronomical Society Meeting Abstracts \#229. p. 431.04
	
	\bibitem[\protect\citeauthoryear{{Haskell} \& {Sedrakian}}{{Haskell} \&
		{Sedrakian}}{2018}]{2018ASSL..457..401H}
	{Haskell} B.,  {Sedrakian} A.,  2018, in {Rezzolla} L.,  {Pizzochero} P.,
	{Jones} D.~I.,  {Rea} N.,   {Vida{\~n}a} I.,  eds,  Astrophysics and Space
	Science Library Vol. 457, Astrophysics and Space Science Library. p.~401
	(\mn@eprint {arXiv} {1709.10340}), \mn@doi{10.1007/978-3-319-97616-7_8}
	
	\bibitem[\protect\citeauthoryear{{He}, {Ng}  \& {Kaspi}}{{He}
		et~al.}{2013}]{2013ApJ...768...64H}
	{He} C.,  {Ng} C.~Y.,   {Kaspi} V.~M.,  2013, \mn@doi [\apj]
	{10.1088/0004-637X/768/1/64}, \href
	{https://ui.adsabs.harvard.edu/abs/2013ApJ...768...64H} {768, 64}
	
	\bibitem[\protect\citeauthoryear{{Heber}}{{Heber}}{2009}]{Heber2009}
	{Heber} U.,  2009, \mn@doi [\araa] {10.1146/annurev-astro-082708-101836}, \href
	{https://ui.adsabs.harvard.edu/abs/2009ARA&A..47..211H} {47, 211}
	
	\bibitem[\protect\citeauthoryear{{Henriksson} \& {Wasserman}}{{Henriksson} \&
		{Wasserman}}{2013}]{2013MNRAS.431.2986H}
	{Henriksson} K.~T.,  {Wasserman} I.,  2013, \mn@doi [\mnras]
	{10.1093/mnras/stt338}, \href
	{https://ui.adsabs.harvard.edu/abs/2013MNRAS.431.2986H} {431, 2986}
	
	\bibitem[\protect\citeauthoryear{{Hewitt} et~al.,}{{Hewitt}
		et~al.}{2022}]{2022MNRAS.515.3577H}
	{Hewitt} D.~M.,  et~al., 2022, \mn@doi [\mnras] {10.1093/mnras/stac1960}, \href
	{https://ui.adsabs.harvard.edu/abs/2022MNRAS.515.3577H} {515, 3577}
	
	\bibitem[\protect\citeauthoryear{{Hills}}{{Hills}}{1978}]{Hills1978}
	{Hills} J.~G.,  1978, \mn@doi [\apj] {10.1086/155813}, \href
	{https://ui.adsabs.harvard.edu/abs/1978ApJ...219..550H} {219, 550}
	
	\bibitem[\protect\citeauthoryear{{Hinkle}, {Lebzelter}, {Fekel}, {Straniero},
		{Joyce}, {Prato}, {Karnath}  \& {Habel}}{{Hinkle}
		et~al.}{2020}]{2020ApJ...904..143H}
	{Hinkle} K.~H.,  {Lebzelter} T.,  {Fekel} F.~C.,  {Straniero} O.,  {Joyce}
	R.~R.,  {Prato} L.,  {Karnath} N.,   {Habel} N.,  2020, \mn@doi [\apj]
	{10.3847/1538-4357/abbe01}, \href
	{https://ui.adsabs.harvard.edu/abs/2020ApJ...904..143H} {904, 143}
	
	\bibitem[\protect\citeauthoryear{{Ho} \& {Andersson}}{{Ho} \&
		{Andersson}}{2017}]{2017MNRAS.464L..65H}
	{Ho} W. C.~G.,  {Andersson} N.,  2017, \mn@doi [\mnras]
	{10.1093/mnrasl/slw186}, \href
	{https://ui.adsabs.harvard.edu/abs/2017MNRAS.464L..65H} {464, L65}
	
	\bibitem[\protect\citeauthoryear{{Ho}, {Glampedakis}  \& {Andersson}}{{Ho}
		et~al.}{2012}]{2012MNRAS.422.2632H}
	{Ho} W. C.~G.,  {Glampedakis} K.,   {Andersson} N.,  2012, \mn@doi [\mnras]
	{10.1111/j.1365-2966.2012.20826.x}, \href
	{https://ui.adsabs.harvard.edu/abs/2012MNRAS.422.2632H} {422, 2632}
	
	\bibitem[\protect\citeauthoryear{{Ho}, {Andersson}  \& {Graber}}{{Ho}
		et~al.}{2017}]{2017PhRvC..96f5801H}
	{Ho} W. C.~G.,  {Andersson} N.,   {Graber} V.,  2017, \mn@doi [\prc]
	{10.1103/PhysRevC.96.065801}, \href
	{https://ui.adsabs.harvard.edu/abs/2017PhRvC..96f5801H} {96, 065801}
	
	\bibitem[\protect\citeauthoryear{{Hu} et~al.,}{{Hu} et~al.}{2020}]{Hu2020}
	{Hu} C.-P.,  et~al., 2020, \mn@doi [\apj] {10.3847/1538-4357/abb3c9}, \href
	{https://ui.adsabs.harvard.edu/abs/2020ApJ...902....1H} {902, 1}
	
	\bibitem[\protect\citeauthoryear{Hurley-Walker et~al.,}{Hurley-Walker
		et~al.}{2022}]{Hurley-Walker2022}
	Hurley-Walker N.,  et~al., 2022, \mn@doi [Nature] {10.1038/s41586-021-04272-x},
	601, 526
	
	\bibitem[\protect\citeauthoryear{{Hyman}, {Lazio}, {Kassim}, {Ray}, {Markwardt}
		\& {Yusef-Zadeh}}{{Hyman} et~al.}{2005}]{2005Natur.434...50H}
	{Hyman} S.~D.,  {Lazio} T. J.~W.,  {Kassim} N.~E.,  {Ray} P.~S.,  {Markwardt}
	C.~B.,   {Yusef-Zadeh} F.,  2005, \mn@doi [\nat] {10.1038/nature03400}, \href
	{https://ui.adsabs.harvard.edu/abs/2005Natur.434...50H} {434, 50}
	
	\bibitem[\protect\citeauthoryear{{Hyman}, {Roy}, {Pal}, {Lazio}, {Ray},
		{Kassim}  \& {Bhatnagar}}{{Hyman} et~al.}{2007}]{2007ApJ...660L.121H}
	{Hyman} S.~D.,  {Roy} S.,  {Pal} S.,  {Lazio} T. J.~W.,  {Ray} P.~S.,  {Kassim}
	N.~E.,   {Bhatnagar} S.,  2007, \mn@doi [\apjl] {10.1086/518245}, \href
	{https://ui.adsabs.harvard.edu/abs/2007ApJ...660L.121H} {660, L121}
	
	\bibitem[\protect\citeauthoryear{{Igoshev}}{{Igoshev}}{2020}]{2020MNRAS.494.3663I}
	{Igoshev} A.~P.,  2020, \mn@doi [\mnras] {10.1093/mnras/staa958}, \href
	{https://ui.adsabs.harvard.edu/abs/2020MNRAS.494.3663I} {494, 3663}
	
	\bibitem[\protect\citeauthoryear{{Ikhsanov} \& {Beskrovnaya}}{{Ikhsanov} \&
		{Beskrovnaya}}{2010}]{2010Ap.....53..237I}
	{Ikhsanov} N.~R.,  {Beskrovnaya} N.~G.,  2010, \mn@doi [Astrophysics]
	{10.1007/s10511-010-9115-z}, \href
	{https://ui.adsabs.harvard.edu/abs/2010Ap.....53..237I} {53, 237}
	
	\bibitem[\protect\citeauthoryear{{Ikhsanov} \& {Beskrovnaya}}{{Ikhsanov} \&
		{Beskrovnaya}}{2013}]{2013ARep...57..287I}
	{Ikhsanov} N.~R.,  {Beskrovnaya} N.~G.,  2013, \mn@doi [Astronomy Reports]
	{10.1134/S1063772913030013}, \href
	{https://ui.adsabs.harvard.edu/abs/2013ARep...57..287I} {57, 287}
	
	\bibitem[\protect\citeauthoryear{{Ioka} \& {Zhang}}{{Ioka} \&
		{Zhang}}{2020}]{IokaZhang20}
	{Ioka} K.,  {Zhang} B.,  2020, \mn@doi [\apjl] {10.3847/2041-8213/ab83fb},
	\href {https://ui.adsabs.harvard.edu/abs/2020ApJ...893L..26I} {893, L26}
	
	\bibitem[\protect\citeauthoryear{{Israel}, {Esposito}, {Rodr{\'\i}guez
			Castillo}  \& {Sidoli}}{{Israel} et~al.}{2016}]{israel16MNRAS}
	{Israel} G.~L.,  {Esposito} P.,  {Rodr{\'\i}guez Castillo} G.~A.,   {Sidoli}
	L.,  2016, \mn@doi [\mnras] {10.1093/mnras/stw1897}, \href
	{https://ui.adsabs.harvard.edu/abs/2016MNRAS.462.4371I} {462, 4371}
	
	\bibitem[\protect\citeauthoryear{{Israel} et~al.,}{{Israel}
		et~al.}{2017a}]{2017Sci...355..817I}
	{Israel} G.~L.,  et~al., 2017a, \mn@doi [Science] {10.1126/science.aai8635},
	\href {https://ui.adsabs.harvard.edu/abs/2017Sci...355..817I} {355, 817}
	
	\bibitem[\protect\citeauthoryear{{Israel} et~al.,}{{Israel}
		et~al.}{2017b}]{2017MNRAS.466L..48I}
	{Israel} G.~L.,  et~al., 2017b, \mn@doi [\mnras] {10.1093/mnrasl/slw218}, \href
	{https://ui.adsabs.harvard.edu/abs/2017MNRAS.466L..48I} {466, L48}
	
	\bibitem[\protect\citeauthoryear{{Jawor} \& {Tauris}}{{Jawor} \&
		{Tauris}}{2022}]{Jawor2022}
	{Jawor} J.~A.,  {Tauris} T.~M.,  2022, \mn@doi [\mnras]
	{10.1093/mnras/stab2677}, \href
	{https://ui.adsabs.harvard.edu/abs/2022MNRAS.509..634J} {509, 634}
	
	\bibitem[\protect\citeauthoryear{{Jones}}{{Jones}}{2006}]{2006MNRAS.365..339J}
	{Jones} P.~B.,  2006, \mn@doi [\mnras] {10.1111/j.1365-2966.2005.09724.x},
	\href {https://ui.adsabs.harvard.edu/abs/2006MNRAS.365..339J} {365, 339}
	
	\bibitem[\protect\citeauthoryear{{Kaplan}}{{Kaplan}}{2008}]{2008AIPC..983..331K}
	{Kaplan} D.~L.,  2008, in {Bassa} C.,  {Wang} Z.,  {Cumming} A.,   {Kaspi}
	V.~M.,  eds,  American Institute of Physics Conference Series Vol. 983, 40
	Years of Pulsars: Millisecond Pulsars, Magnetars and More. pp 331--339,
	\mn@doi{10.1063/1.2900177}
	
	\bibitem[\protect\citeauthoryear{{Kaplan} \& {van Kerkwijk}}{{Kaplan} \& {van
			Kerkwijk}}{2009}]{2009ApJ...705..798K}
	{Kaplan} D.~L.,  {van Kerkwijk} M.~H.,  2009, \mn@doi [\apj]
	{10.1088/0004-637X/705/1/798}, \href
	{https://ui.adsabs.harvard.edu/abs/2009ApJ...705..798K} {705, 798}
	
	\bibitem[\protect\citeauthoryear{{Kaplan}, {Hyman}, {Roy}, {Bandyopadhyay},
		{Chakrabarty}, {Kassim}, {Lazio}  \& {Ray}}{{Kaplan}
		et~al.}{2008}]{2008ApJ...687..262K}
	{Kaplan} D.~L.,  {Hyman} S.~D.,  {Roy} S.,  {Bandyopadhyay} R.~M.,
	{Chakrabarty} D.,  {Kassim} N.~E.,  {Lazio} T.~J.~W.,   {Ray} P.~S.,  2008,
	\mn@doi [\apj] {10.1086/591436}, \href
	{https://ui.adsabs.harvard.edu/abs/2008ApJ...687..262K} {687, 262}
	
	\bibitem[\protect\citeauthoryear{{Kaspi} \& {Beloborodov}}{{Kaspi} \&
		{Beloborodov}}{2017}]{KaspiBeloborodov2017}
	{Kaspi} V.~M.,  {Beloborodov} A.~M.,  2017, \mn@doi [\araa]
	{10.1146/annurev-astro-081915-023329}, \href
	{https://ui.adsabs.harvard.edu/abs/2017ARA&A..55..261K} {55, 261}
	
	\bibitem[\protect\citeauthoryear{{Kaspi}, {Gavriil}, {Woods}, {Jensen},
		{Roberts}  \& {Chakrabarty}}{{Kaspi} et~al.}{2003}]{2003ApJ...588L..93K}
	{Kaspi} V.~M.,  {Gavriil} F.~P.,  {Woods} P.~M.,  {Jensen} J.~B.,  {Roberts}
	M.~S.~E.,   {Chakrabarty} D.,  2003, \mn@doi [\apjl] {10.1086/375683}, \href
	{https://ui.adsabs.harvard.edu/abs/2003ApJ...588L..93K} {588, L93}
	
	\bibitem[\protect\citeauthoryear{{Katz}}{{Katz}}{2022}]{2022arXiv220308112K}
	{Katz} J.~I.,  2022, arXiv e-prints, \href
	{https://ui.adsabs.harvard.edu/abs/2022arXiv220308112K} {p. arXiv:2203.08112}
	
	\bibitem[\protect\citeauthoryear{{Keane} et~al.,}{{Keane}
		et~al.}{2015}]{keane2015}
	{Keane} E.,  et~al., 2015, in Advancing Astrophysics with the Square Kilometre
	Array (AASKA14). p.~40 (\mn@eprint {arXiv} {1501.00056})
	
	\bibitem[\protect\citeauthoryear{{Kharchenko}, {Piskunov}, {Schilbach},
		{R{\"o}ser}  \& {Scholz}}{{Kharchenko} et~al.}{2013}]{2013A&A...558A..53K}
	{Kharchenko} N.~V.,  {Piskunov} A.~E.,  {Schilbach} E.,  {R{\"o}ser} S.,
	{Scholz} R.~D.,  2013, \mn@doi [\aap] {10.1051/0004-6361/201322302}, \href
	{https://ui.adsabs.harvard.edu/abs/2013A&A...558A..53K} {558, A53}
	
	\bibitem[\protect\citeauthoryear{{Kijak} \& {Gil}}{{Kijak} \&
		{Gil}}{1998}]{1998MNRAS.299..855K}
	{Kijak} J.,  {Gil} J.,  1998, \mn@doi [\mnras]
	{10.1046/j.1365-8711.1998.01832.x}, \href
	{https://ui.adsabs.harvard.edu/abs/1998MNRAS.299..855K} {299, 855}
	
	\bibitem[\protect\citeauthoryear{{Kijak} \& {Gil}}{{Kijak} \&
		{Gil}}{2003}]{2003A&A...397..969K}
	{Kijak} J.,  {Gil} J.,  2003, \mn@doi [\aap] {10.1051/0004-6361:20021583},
	\href {https://ui.adsabs.harvard.edu/abs/2003A&A...397..969K} {397, 969}
	
	\bibitem[\protect\citeauthoryear{{Kiman}, {Schmidt}, {Angus}, {Cruz}, {Faherty}
		\& {Rice}}{{Kiman} et~al.}{2019}]{2019AJ....157..231K}
	{Kiman} R.,  {Schmidt} S.~J.,  {Angus} R.,  {Cruz} K.~L.,  {Faherty} J.~K.,
	{Rice} E.,  2019, \mn@doi [\aj] {10.3847/1538-3881/ab1753}, \href
	{https://ui.adsabs.harvard.edu/abs/2019AJ....157..231K} {157, 231}
	
	\bibitem[\protect\citeauthoryear{{Kirsten} et~al.,}{{Kirsten}
		et~al.}{2022}]{2022Natur.602..585K}
	{Kirsten} F.,  et~al., 2022, \mn@doi [\nat] {10.1038/s41586-021-04354-w}, \href
	{https://ui.adsabs.harvard.edu/abs/2022Natur.602..585K} {602, 585}
	
	\bibitem[\protect\citeauthoryear{{Kojima}}{{Kojima}}{2022}]{2022ApJ...938...91K}
	{Kojima} Y.,  2022, \mn@doi [\apj] {10.3847/1538-4357/ac9184}, \href
	{https://ui.adsabs.harvard.edu/abs/2022ApJ...938...91K} {938, 91}
	
	\bibitem[\protect\citeauthoryear{{Kouveliotou} et~al.,}{{Kouveliotou}
		et~al.}{1998}]{Kouveliotou98}
	{Kouveliotou} C.,  et~al., 1998, \mn@doi [\nat] {10.1038/30410}, \href
	{https://ui.adsabs.harvard.edu/abs/1998Natur.393..235K} {393, 235}
	
	\bibitem[\protect\citeauthoryear{{Kouveliotou}, {Patel}, {Tennant}, {Woods},
		{Finger}  \& {Wachter}}{{Kouveliotou} et~al.}{2003}]{2003IAUC.8109....2K}
	{Kouveliotou} C.,  {Patel} S.,  {Tennant} A.,  {Woods} P.,  {Finger} M.,
	{Wachter} S.,  2003, \iaucirc, \href
	{https://ui.adsabs.harvard.edu/abs/2003IAUC.8109....2K} {8109, 2}
	
	\bibitem[\protect\citeauthoryear{{Kremer}, {Piro}  \& {Li}}{{Kremer}
		et~al.}{2021}]{Kremer2021}
	{Kremer} K.,  {Piro} A.~L.,   {Li} D.,  2021, \mn@doi [\apjl]
	{10.3847/2041-8213/ac13a0}, \href
	{https://ui.adsabs.harvard.edu/abs/2021ApJ...917L..11K} {917, L11}
	
	\bibitem[\protect\citeauthoryear{{Kretschmar} et~al.,}{{Kretschmar}
		et~al.}{2019}]{2019NewAR..8601546K}
	{Kretschmar} P.,  et~al., 2019, \mn@doi [\nar] {10.1016/j.newar.2020.101546},
	\href {https://ui.adsabs.harvard.edu/abs/2019NewAR..8601546K} {86, 101546}
	
	\bibitem[\protect\citeauthoryear{{Laha} et~al.,}{{Laha}
		et~al.}{2022a}]{2022ApJ...929..173L}
	{Laha} S.,  et~al., 2022a, \mn@doi [\apj] {10.3847/1538-4357/ac5f3c}, \href
	{https://ui.adsabs.harvard.edu/abs/2022ApJ...929..173L} {929, 173}
	
	\bibitem[\protect\citeauthoryear{{Laha} et~al.,}{{Laha}
		et~al.}{2022b}]{2022ApJ...930..172L}
	{Laha} S.,  et~al., 2022b, \mn@doi [\apj] {10.3847/1538-4357/ac63a8}, \href
	{https://ui.adsabs.harvard.edu/abs/2022ApJ...930..172L} {930, 172}
	
	\bibitem[\protect\citeauthoryear{{Lander}}{{Lander}}{2013}]{2013PhRvL.110g1101L}
	{Lander} S.~K.,  2013, \mn@doi [\prl] {10.1103/PhysRevLett.110.071101}, \href
	{https://ui.adsabs.harvard.edu/abs/2013PhRvL.110g1101L} {110, 071101}
	
	\bibitem[\protect\citeauthoryear{{Lander}}{{Lander}}{2014}]{2014MNRAS.437..424L}
	{Lander} S.~K.,  2014, \mn@doi [\mnras] {10.1093/mnras/stt1894}, \href
	{https://ui.adsabs.harvard.edu/abs/2014MNRAS.437..424L} {437, 424}
	
	\bibitem[\protect\citeauthoryear{{Lander}}{{Lander}}{2022}]{2022arXiv220908598L}
	{Lander} S.~K.,  2022, arXiv e-prints, \href
	{https://ui.adsabs.harvard.edu/abs/2022arXiv220908598L} {p. arXiv:2209.08598}
	
	\bibitem[\protect\citeauthoryear{{Lander}, {Andersson}, {Antonopoulou}  \&
		{Watts}}{{Lander} et~al.}{2015}]{2015MNRAS.449.2047L}
	{Lander} S.~K.,  {Andersson} N.,  {Antonopoulou} D.,   {Watts} A.~L.,  2015,
	\mn@doi [\mnras] {10.1093/mnras/stv432}, \href
	{https://ui.adsabs.harvard.edu/abs/2015MNRAS.449.2047L} {449, 2047}
	
	\bibitem[\protect\citeauthoryear{{Levin}}{{Levin}}{2006}]{2006MNRAS.368L..35L}
	{Levin} Y.,  2006, \mn@doi [\mnras] {10.1111/j.1745-3933.2006.00155.x}, \href
	{https://ui.adsabs.harvard.edu/abs/2006MNRAS.368L..35L} {368, L35}
	
	\bibitem[\protect\citeauthoryear{{Levin}, {Beloborodov}  \&
		{Bransgrove}}{{Levin} et~al.}{2020}]{Levin+20}
	{Levin} Y.,  {Beloborodov} A.~M.,   {Bransgrove} A.,  2020, arXiv e-prints,
	\href {https://ui.adsabs.harvard.edu/abs/2020arXiv200204595L} {p.
		arXiv:2002.04595}
	
	\bibitem[\protect\citeauthoryear{{Li} \& {van den Heuvel}}{{Li} \& {van den
			Heuvel}}{1999}]{1999ApJ...513L..45L}
	{Li} X.~D.,  {van den Heuvel} E.~P.~J.,  1999, \mn@doi [\apjl]
	{10.1086/311904}, \href
	{https://ui.adsabs.harvard.edu/abs/1999ApJ...513L..45L} {513, L45}
	
	\bibitem[\protect\citeauthoryear{{Li} et~al.,}{{Li} et~al.}{2020}]{Li2020}
	{Li} C.~K.,  et~al., 2020, arXiv e-prints, \href
	{https://ui.adsabs.harvard.edu/abs/2020arXiv200511071L} {p. arXiv:2005.11071}
	
	\bibitem[\protect\citeauthoryear{{Li} et~al.,}{{Li}
		et~al.}{2021}]{2021Natur.598..267L}
	{Li} D.,  et~al., 2021, \mn@doi [\nat] {10.1038/s41586-021-03878-5}, \href
	{https://ui.adsabs.harvard.edu/abs/2021Natur.598..267L} {598, 267}
	
	\bibitem[\protect\citeauthoryear{{Li} et~al.,}{{Li}
		et~al.}{2022}]{2022ApJ...931...56L}
	{Li} X.,  et~al., 2022, \mn@doi [\apj] {10.3847/1538-4357/ac6587}, \href
	{https://ui.adsabs.harvard.edu/abs/2022ApJ...931...56L} {931, 56}
	
	\bibitem[\protect\citeauthoryear{{Link}}{{Link}}{2012a}]{2012MNRAS.421.2682L}
	{Link} B.,  2012a, \mn@doi [\mnras] {10.1111/j.1365-2966.2012.20498.x}, \href
	{https://ui.adsabs.harvard.edu/abs/2012MNRAS.421.2682L} {421, 2682}
	
	\bibitem[\protect\citeauthoryear{{Link}}{{Link}}{2012b}]{2012MNRAS.422.1640L}
	{Link} B.,  2012b, \mn@doi [\mnras] {10.1111/j.1365-2966.2012.20740.x}, \href
	{https://ui.adsabs.harvard.edu/abs/2012MNRAS.422.1640L} {422, 1640}
	
	\bibitem[\protect\citeauthoryear{{Linscott} \& {Erkes}}{{Linscott} \&
		{Erkes}}{1980}]{Linscott1980}
	{Linscott} I.~R.,  {Erkes} J.~W.,  1980, \mn@doi [\apjl] {10.1086/183209},
	\href {https://ui.adsabs.harvard.edu/abs/1980ApJ...236L.109L} {236, L109}
	
	\bibitem[\protect\citeauthoryear{{Lipunov} \& {Panchenko}}{{Lipunov} \&
		{Panchenko}}{1996}]{1996A&A...312..937L}
	{Lipunov} V.~M.,  {Panchenko} I.~E.,  1996, \aap, \href
	{https://ui.adsabs.harvard.edu/abs/1996A&A...312..937L} {312, 937}
	
	\bibitem[\protect\citeauthoryear{{Liu}, {van Paradijs}  \& {van den
			Heuvel}}{{Liu} et~al.}{2006}]{2006A&A...455.1165L}
	{Liu} Q.~Z.,  {van Paradijs} J.,   {van den Heuvel} E.~P.~J.,  2006, \mn@doi
	[\aap] {10.1051/0004-6361:20064987}, \href
	{https://ui.adsabs.harvard.edu/abs/2006A&A...455.1165L} {455, 1165}
	
	\bibitem[\protect\citeauthoryear{{Loeb} \& {Maoz}}{{Loeb} \&
		{Maoz}}{2022}]{2022RNAAS...6...27L}
	{Loeb} A.,  {Maoz} D.,  2022, \mn@doi [Research Notes of the American
	Astronomical Society] {10.3847/2515-5172/ac52f1}, \href
	{https://ui.adsabs.harvard.edu/abs/2022RNAAS...6...27L} {6, 27}
	
	\bibitem[\protect\citeauthoryear{{Lorimer}, {Bailes}, {McLaughlin}, {Narkevic}
		\& {Crawford}}{{Lorimer} et~al.}{2007}]{Lorimer+07}
	{Lorimer} D.~R.,  {Bailes} M.,  {McLaughlin} M.~A.,  {Narkevic} D.~J.,
	{Crawford} F.,  2007, \mn@doi [Science] {10.1126/science.1147532}, \href
	{https://ui.adsabs.harvard.edu/abs/2007Sci...318..777L} {318, 777}
	
	\bibitem[\protect\citeauthoryear{{Lower} et~al.,}{{Lower}
		et~al.}{2020}]{lower2020}
	{Lower} M.~E.,  et~al., 2020, \mn@doi [\mnras] {10.1093/mnras/staa615}, \href
	{https://ui.adsabs.harvard.edu/abs/2020MNRAS.494..228L} {494, 228}
	
	\bibitem[\protect\citeauthoryear{{Lu}, {Kumar}  \& {Zhang}}{{Lu}
		et~al.}{2020}]{LKZ2020}
	{Lu} W.,  {Kumar} P.,   {Zhang} B.,  2020, \mn@doi [\mnras]
	{10.1093/mnras/staa2450}, \href
	{https://ui.adsabs.harvard.edu/abs/2020MNRAS.498.1397L} {498, 1397}
	
	\bibitem[\protect\citeauthoryear{{Lu}, {Beniamini}  \& {Kumar}}{{Lu}
		et~al.}{2021}]{LBK2021}
	{Lu} W.,  {Beniamini} P.,   {Kumar} P.,  2021, arXiv e-prints, \href
	{https://ui.adsabs.harvard.edu/abs/2021arXiv210704059L} {p. arXiv:2107.04059}
	
	\bibitem[\protect\citeauthoryear{{Lynch}, {Lorimer}, {Ransom}  \&
		{Boyles}}{{Lynch} et~al.}{2012}]{2012ApJ...756...78L}
	{Lynch} R.~S.,  {Lorimer} D.~R.,  {Ransom} S.~M.,   {Boyles} J.,  2012, \mn@doi
	[\apj] {10.1088/0004-637X/756/1/78}, \href
	{https://ui.adsabs.harvard.edu/abs/2012ApJ...756...78L} {756, 78}
	
	\bibitem[\protect\citeauthoryear{{Lyne}, {Manchester}  \& {D'Amico}}{{Lyne}
		et~al.}{1996}]{1996ApJ...460L..41L}
	{Lyne} A.~G.,  {Manchester} R.~N.,   {D'Amico} N.,  1996, \mn@doi [\apjl]
	{10.1086/309972}, \href
	{https://ui.adsabs.harvard.edu/abs/1996ApJ...460L..41L} {460, L41}
	
	\bibitem[\protect\citeauthoryear{{Lyutikov}}{{Lyutikov}}{2017}]{2017ApJ...838L..13L}
	{Lyutikov} M.,  2017, \mn@doi [\apjl] {10.3847/2041-8213/aa62fa}, \href
	{https://ui.adsabs.harvard.edu/abs/2017ApJ...838L..13L} {838, L13}
	
	\bibitem[\protect\citeauthoryear{{Lyutikov}}{{Lyutikov}}{2019}]{Lyutikov2019}
	{Lyutikov} M.,  2019, \mn@doi [\mnras] {10.1093/mnras/sty3303}, \href
	{https://ui.adsabs.harvard.edu/abs/2019MNRAS.483.2766L} {483, 2766}
	
	\bibitem[\protect\citeauthoryear{{Lyutikov}, {Barkov}  \&
		{Giannios}}{{Lyutikov} et~al.}{2020}]{Lyutikov+20}
	{Lyutikov} M.,  {Barkov} M.~V.,   {Giannios} D.,  2020, \mn@doi [\apjl]
	{10.3847/2041-8213/ab87a4}, \href
	{https://ui.adsabs.harvard.edu/abs/2020ApJ...893L..39L} {893, L39}
	
	\bibitem[\protect\citeauthoryear{{Majid} et~al.,}{{Majid}
		et~al.}{2021}]{2021ApJ...919L...6M}
	{Majid} W.~A.,  et~al., 2021, \mn@doi [\apjl] {10.3847/2041-8213/ac1921}, \href
	{https://ui.adsabs.harvard.edu/abs/2021ApJ...919L...6M} {919, L6}
	
	\bibitem[\protect\citeauthoryear{{Malacaria}, {Jenke}, {Roberts},
		{Wilson-Hodge}, {Cleveland}, {Mailyan}  \& {GBM Accreting Pulsars Program
			Team}}{{Malacaria} et~al.}{2020}]{2020ApJ...896...90M}
	{Malacaria} C.,  {Jenke} P.,  {Roberts} O.~J.,  {Wilson-Hodge} C.~A.,
	{Cleveland} W.~H.,  {Mailyan} B.,   {GBM Accreting Pulsars Program Team}
	2020, \mn@doi [\apj] {10.3847/1538-4357/ab855c}, \href
	{https://ui.adsabs.harvard.edu/abs/2020ApJ...896...90M} {896, 90}
	
	\bibitem[\protect\citeauthoryear{{Manchester} et~al.,}{{Manchester}
		et~al.}{2001}]{manchester2001}
	{Manchester} R.~N.,  et~al., 2001, \mn@doi [\mnras]
	{10.1046/j.1365-8711.2001.04751.x}, \href
	{https://ui.adsabs.harvard.edu/abs/2001MNRAS.328...17M} {328, 17}
	
	\bibitem[\protect\citeauthoryear{{Manchester}, {Hobbs}, {Teoh}  \&
		{Hobbs}}{{Manchester} et~al.}{2005}]{2005AJ....129.1993M}
	{Manchester} R.~N.,  {Hobbs} G.~B.,  {Teoh} A.,   {Hobbs} M.,  2005, \mn@doi
	[\aj] {10.1086/428488}, \href
	{https://ui.adsabs.harvard.edu/abs/2005AJ....129.1993M} {129, 1993}
	
	\bibitem[\protect\citeauthoryear{{Mannarelli}, {Pagliaroli}, {Parisi}, {Pilo}
		\& {Tonelli}}{{Mannarelli} et~al.}{2015}]{2015ApJ...815...81M}
	{Mannarelli} M.,  {Pagliaroli} G.,  {Parisi} A.,  {Pilo} L.,   {Tonelli} F.,
	2015, \mn@doi [\apj] {10.1088/0004-637X/815/2/81}, \href
	{https://ui.adsabs.harvard.edu/abs/2015ApJ...815...81M} {815, 81}
	
	\bibitem[\protect\citeauthoryear{{Margalit}, {Beniamini}, {Sridhar}  \&
		{Metzger}}{{Margalit} et~al.}{2020}]{MBSM2020}
	{Margalit} B.,  {Beniamini} P.,  {Sridhar} N.,   {Metzger} B.~D.,  2020,
	\mn@doi [\apjl] {10.3847/2041-8213/abac57}, \href
	{https://ui.adsabs.harvard.edu/abs/2020ApJ...899L..27M} {899, L27}
	
	\bibitem[\protect\citeauthoryear{{Marsh} et~al.,}{{Marsh}
		et~al.}{2016}]{2016Natur.537..374M}
	{Marsh} T.~R.,  et~al., 2016, \mn@doi [\nat] {10.1038/nature18620}, \href
	{https://ui.adsabs.harvard.edu/abs/2016Natur.537..374M} {537, 374}
	
	\bibitem[\protect\citeauthoryear{{Mckinven} et~al.,}{{Mckinven}
		et~al.}{2022}]{2022arXiv220509221M}
	{Mckinven} R.,  et~al., 2022, arXiv e-prints, \href
	{https://ui.adsabs.harvard.edu/abs/2022arXiv220509221M} {p. arXiv:2205.09221}
	
	\bibitem[\protect\citeauthoryear{{Melrose}}{{Melrose}}{1979}]{1979AuJPh..32...61M}
	{Melrose} D.~B.,  1979, \mn@doi [Australian Journal of Physics]
	{10.1071/PH790061}, \href
	{https://ui.adsabs.harvard.edu/abs/1979AuJPh..32...61M} {32, 61}
	
	\bibitem[\protect\citeauthoryear{{Mereghetti} et~al.,}{{Mereghetti}
		et~al.}{2020}]{Mereghetti+20}
	{Mereghetti} S.,  et~al., 2020, \mn@doi [\apjl] {10.3847/2041-8213/aba2cf},
	\href {https://ui.adsabs.harvard.edu/abs/2020ApJ...898L..29M} {898, L29}
	
	\bibitem[\protect\citeauthoryear{{Mereghetti}, {Topinka}, {Rigoselli}  \&
		{G{\"o}tz}}{{Mereghetti} et~al.}{2021}]{2021ApJ...921L...3M}
	{Mereghetti} S.,  {Topinka} M.,  {Rigoselli} M.,   {G{\"o}tz} D.,  2021,
	\mn@doi [\apjl] {10.3847/2041-8213/ac2ee7}, \href
	{https://ui.adsabs.harvard.edu/abs/2021ApJ...921L...3M} {921, L3}
	
	\bibitem[\protect\citeauthoryear{{Michel}}{{Michel}}{1993}]{1993MNRAS.265..449M}
	{Michel} F.~C.,  1993, \mn@doi [\mnras] {10.1093/mnras/265.2.449}, \href
	{https://ui.adsabs.harvard.edu/abs/1993MNRAS.265..449M} {265, 449}
	
	\bibitem[\protect\citeauthoryear{{Morello} et~al.,}{{Morello}
		et~al.}{2020}]{2020MNRAS.493.1165M}
	{Morello} V.,  et~al., 2020, \mn@doi [\mnras] {10.1093/mnras/staa321}, \href
	{https://ui.adsabs.harvard.edu/abs/2020MNRAS.493.1165M} {493, 1165}
	
	\bibitem[\protect\citeauthoryear{{Morello}, {Rajwade}  \& {Stappers}}{{Morello}
		et~al.}{2022}]{2022MNRAS.510.1393M}
	{Morello} V.,  {Rajwade} K.~M.,   {Stappers} B.~W.,  2022, \mn@doi [\mnras]
	{10.1093/mnras/stab3493}, \href
	{https://ui.adsabs.harvard.edu/abs/2022MNRAS.510.1393M} {510, 1393}
	
	\bibitem[\protect\citeauthoryear{{Morozova}, {Ahmedov}  \&
		{Zanotti}}{{Morozova} et~al.}{2010}]{2010MNRAS.408..490M}
	{Morozova} V.~S.,  {Ahmedov} B.~J.,   {Zanotti} O.,  2010, \mn@doi [\mnras]
	{10.1111/j.1365-2966.2010.17131.x}, \href
	{https://ui.adsabs.harvard.edu/abs/2010MNRAS.408..490M} {408, 490}
	
	\bibitem[\protect\citeauthoryear{{Muslimov} \& {Tsygan}}{{Muslimov} \&
		{Tsygan}}{1985}]{1985Ap&SS.115...43M}
	{Muslimov} A.~G.,  {Tsygan} A.~I.,  1985, \mn@doi [\apss] {10.1007/BF00653825},
	\href {https://ui.adsabs.harvard.edu/abs/1985Ap&SS.115...43M} {115, 43}
	
	\bibitem[\protect\citeauthoryear{{Niebergal}, {Ouyed}  \& {Leahy}}{{Niebergal}
		et~al.}{2006}]{2006ApJ...646L..17N}
	{Niebergal} B.,  {Ouyed} R.,   {Leahy} D.,  2006, \mn@doi [\apjl]
	{10.1086/506521}, \href
	{https://ui.adsabs.harvard.edu/abs/2006ApJ...646L..17N} {646, L17}
	
	\bibitem[\protect\citeauthoryear{{Nimmo} et~al.,}{{Nimmo}
		et~al.}{2022}]{2022arXiv220603759N}
	{Nimmo} K.,  et~al., 2022, arXiv e-prints, \href
	{https://ui.adsabs.harvard.edu/abs/2022arXiv220603759N} {p. arXiv:2206.03759}
	
	\bibitem[\protect\citeauthoryear{{Paczynski}}{{Paczynski}}{1992}]{Paczynski1992}
	{Paczynski} B.,  1992, \actaa, \href
	{https://ui.adsabs.harvard.edu/abs/1992AcA....42..145P} {42, 145}
	
	\bibitem[\protect\citeauthoryear{{Pakull} \& {Gris{\'e}}}{{Pakull} \&
		{Gris{\'e}}}{2008}]{2008AIPC.1010..303P}
	{Pakull} M.~W.,  {Gris{\'e}} F.,  2008, in {Bandyopadhyay} R.~M.,  {Wachter}
	S.,  {Gelino} D.,   {Gelino} C.~R.,  eds,  American Institute of Physics
	Conference Series Vol. 1010, A Population Explosion: The Nature \& Evolution
	of X-ray Binaries in Diverse Environments. pp 303--307 (\mn@eprint {arXiv}
	{0803.4345}), \mn@doi{10.1063/1.2945062}
	
	\bibitem[\protect\citeauthoryear{{Pakull} \& {Mirioni}}{{Pakull} \&
		{Mirioni}}{2002}]{2002astro.ph..2488P}
	{Pakull} M.~W.,  {Mirioni} L.,  2002, arXiv e-prints, \href
	{https://ui.adsabs.harvard.edu/abs/2002astro.ph..2488P} {pp
		astro--ph/0202488}
	
	\bibitem[\protect\citeauthoryear{{Palapanidis}, {Stergioulas}  \&
		{Lander}}{{Palapanidis} et~al.}{2015}]{2015MNRAS.452.3246P}
	{Palapanidis} K.,  {Stergioulas} N.,   {Lander} S.~K.,  2015, \mn@doi [\mnras]
	{10.1093/mnras/stv1536}, \href
	{https://ui.adsabs.harvard.edu/abs/2015MNRAS.452.3246P} {452, 3246}
	
	\bibitem[\protect\citeauthoryear{{Passamonti}, {Akg{\"u}n}, {Pons}  \&
		{Miralles}}{{Passamonti} et~al.}{2017a}]{2017MNRAS.465.3416P}
	{Passamonti} A.,  {Akg{\"u}n} T.,  {Pons} J.~A.,   {Miralles} J.~A.,  2017a,
	\mn@doi [\mnras] {10.1093/mnras/stw2936}, \href
	{https://ui.adsabs.harvard.edu/abs/2017MNRAS.465.3416P} {465, 3416}
	
	\bibitem[\protect\citeauthoryear{{Passamonti}, {Akg{\"u}n}, {Pons}  \&
		{Miralles}}{{Passamonti} et~al.}{2017b}]{2017MNRAS.469.4979P}
	{Passamonti} A.,  {Akg{\"u}n} T.,  {Pons} J.~A.,   {Miralles} J.~A.,  2017b,
	\mn@doi [\mnras] {10.1093/mnras/stx1192}, \href
	{https://ui.adsabs.harvard.edu/abs/2017MNRAS.469.4979P} {469, 4979}
	
	\bibitem[\protect\citeauthoryear{{Pastor-Marazuela} et~al.,}{{Pastor-Marazuela}
		et~al.}{2020}]{Pastor-Marazuela2020}
	{Pastor-Marazuela} I.,  et~al., 2020, arXiv e-prints, \href
	{https://ui.adsabs.harvard.edu/abs/2020arXiv201208348P} {p. arXiv:2012.08348}
	
	\bibitem[\protect\citeauthoryear{{Patel} et~al.,}{{Patel}
		et~al.}{2004}]{2004ApJ...602L..45P}
	{Patel} S.~K.,  et~al., 2004, \mn@doi [\apjl] {10.1086/382210}, \href
	{https://ui.adsabs.harvard.edu/abs/2004ApJ...602L..45P} {602, L45}
	
	\bibitem[\protect\citeauthoryear{{Patel} et~al.,}{{Patel}
		et~al.}{2007}]{2007ApJ...657..994P}
	{Patel} S.~K.,  et~al., 2007, \mn@doi [\apj] {10.1086/510374}, \href
	{https://ui.adsabs.harvard.edu/abs/2007ApJ...657..994P} {657, 994}
	
	\bibitem[\protect\citeauthoryear{{Pavan}, {Bozzo}, {Ferrigno}, {Ricci},
		{Manousakis}, {Walter}  \& {Stella}}{{Pavan}
		et~al.}{2011}]{2011A&A...526A.122P}
	{Pavan} L.,  {Bozzo} E.,  {Ferrigno} C.,  {Ricci} C.,  {Manousakis} A.,
	{Walter} R.,   {Stella} L.,  2011, \mn@doi [\aap]
	{10.1051/0004-6361/201015561}, \href
	{https://ui.adsabs.harvard.edu/abs/2011A&A...526A.122P} {526, A122}
	
	\bibitem[\protect\citeauthoryear{{Peacock}, {Zepf}  \& {Maccarone}}{{Peacock}
		et~al.}{2012}]{2012ApJ...752...90P}
	{Peacock} M.~B.,  {Zepf} S.~E.,   {Maccarone} T.~J.,  2012, \mn@doi [\apj]
	{10.1088/0004-637X/752/2/90}, \href
	{https://ui.adsabs.harvard.edu/abs/2012ApJ...752...90P} {752, 90}
	
	\bibitem[\protect\citeauthoryear{{Pelisoli} et~al.,}{{Pelisoli}
		et~al.}{2022}]{2022MNRAS.516.5052P}
	{Pelisoli} I.,  et~al., 2022, \mn@doi [\mnras] {10.1093/mnras/stac2391}, \href
	{https://ui.adsabs.harvard.edu/abs/2022MNRAS.516.5052P} {516, 5052}
	
	\bibitem[\protect\citeauthoryear{{Philippov}, {Timokhin}  \&
		{Spitkovsky}}{{Philippov} et~al.}{2020}]{2020PhRvL.124x5101P}
	{Philippov} A.,  {Timokhin} A.,   {Spitkovsky} A.,  2020, \mn@doi [\prl]
	{10.1103/PhysRevLett.124.245101}, \href
	{https://ui.adsabs.harvard.edu/abs/2020PhRvL.124x5101P} {124, 245101}
	
	\bibitem[\protect\citeauthoryear{{Pietka}, {Fender}  \& {Keane}}{{Pietka}
		et~al.}{2015}]{2015MNRAS.446.3687P}
	{Pietka} M.,  {Fender} R.~P.,   {Keane} E.~F.,  2015, \mn@doi [\mnras]
	{10.1093/mnras/stu2335}, \href
	{https://ui.adsabs.harvard.edu/abs/2015MNRAS.446.3687P} {446, 3687}
	
	\bibitem[\protect\citeauthoryear{{Pires}, {Schwope}  \& {Motch}}{{Pires}
		et~al.}{2017}]{2017AN....338..213P}
	{Pires} A.~M.,  {Schwope} A.~D.,   {Motch} C.,  2017, \mn@doi [Astronomische
	Nachrichten] {10.1002/asna.201713333}, \href
	{https://ui.adsabs.harvard.edu/abs/2017AN....338..213P} {338, 213}
	
	\bibitem[\protect\citeauthoryear{{Piro} et~al.,}{{Piro}
		et~al.}{2021}]{2021A&A...656L..15P}
	{Piro} L.,  et~al., 2021, \mn@doi [\aap] {10.1051/0004-6361/202141903}, \href
	{https://ui.adsabs.harvard.edu/abs/2021A&A...656L..15P} {656, L15}
	
	\bibitem[\protect\citeauthoryear{{Pleunis} et~al.,}{{Pleunis}
		et~al.}{2021a}]{Pleunis2021}
	{Pleunis} Z.,  et~al., 2021a, \mn@doi [\apjl] {10.3847/2041-8213/abec72}, \href
	{https://ui.adsabs.harvard.edu/abs/2021ApJ...911L...3P} {911, L3}
	
	\bibitem[\protect\citeauthoryear{{Pleunis} et~al.,}{{Pleunis}
		et~al.}{2021b}]{2021ApJ...923....1P}
	{Pleunis} Z.,  et~al., 2021b, \mn@doi [\apj] {10.3847/1538-4357/ac33ac}, \href
	{https://ui.adsabs.harvard.edu/abs/2021ApJ...923....1P} {923, 1}
	
	\bibitem[\protect\citeauthoryear{{Pons} \& {Vigan{\`o}}}{{Pons} \&
		{Vigan{\`o}}}{2019}]{2019LRCA....5....3P}
	{Pons} J.~A.,  {Vigan{\`o}} D.,  2019, \mn@doi [Living Reviews in Computational
	Astrophysics] {10.1007/s41115-019-0006-7}, \href
	{https://ui.adsabs.harvard.edu/abs/2019LRCA....5....3P} {5, 3}
	
	\bibitem[\protect\citeauthoryear{{Popov}}{{Popov}}{2022}]{2022arXiv220107507P}
	{Popov} S.~B.,  2022, arXiv e-prints, \href
	{https://ui.adsabs.harvard.edu/abs/2022arXiv220107507P} {p. arXiv:2201.07507}
	
	\bibitem[\protect\citeauthoryear{{Popov}, {Igoshev}, {Taverna}  \&
		{Turolla}}{{Popov} et~al.}{2017}]{2017JPhCS.932a2048P}
	{Popov} S.~B.,  {Igoshev} A.~P.,  {Taverna} R.,   {Turolla} R.,  2017, in
	Journal of Physics Conference Series. p. 012048 (\mn@eprint {arXiv}
	{1710.09190}), \mn@doi{10.1088/1742-6596/932/1/012048}
	
	\bibitem[\protect\citeauthoryear{{Potekhin}, {Zyuzin}, {Yakovlev}, {Beznogov}
		\& {Shibanov}}{{Potekhin} et~al.}{2020}]{2020MNRAS.496.5052P}
	{Potekhin} A.~Y.,  {Zyuzin} D.~A.,  {Yakovlev} D.~G.,  {Beznogov} M.~V.,
	{Shibanov} Y.~A.,  2020, \mn@doi [\mnras] {10.1093/mnras/staa1871}, \href
	{https://ui.adsabs.harvard.edu/abs/2020MNRAS.496.5052P} {496, 5052}
	
	\bibitem[\protect\citeauthoryear{{Predehl} et~al.,}{{Predehl}
		et~al.}{2021}]{erosita21AA}
	{Predehl} P.,  et~al., 2021, \mn@doi [\aap] {10.1051/0004-6361/202039313},
	\href {https://ui.adsabs.harvard.edu/abs/2021A&A...647A...1P} {647, A1}
	
	\bibitem[\protect\citeauthoryear{{Rafat}, {Melrose}  \& {Mastrano}}{{Rafat}
		et~al.}{2019}]{2019JPlPh..85c9011R}
	{Rafat} M.~Z.,  {Melrose} D.~B.,   {Mastrano} A.,  2019, \mn@doi [Journal of
	Plasma Physics] {10.1017/S0022377819000448}, \href
	{https://ui.adsabs.harvard.edu/abs/2019JPlPh..85c9011R} {85, 905850311}
	
	\bibitem[\protect\citeauthoryear{{Rajwade}, {Chennamangalam}, {Lorimer}  \&
		{Karastergiou}}{{Rajwade} et~al.}{2018}]{Rajwade2018}
	{Rajwade} K.,  {Chennamangalam} J.,  {Lorimer} D.,   {Karastergiou} A.,  2018,
	\mn@doi [\mnras] {10.1093/mnras/sty1695}, \href
	{https://ui.adsabs.harvard.edu/abs/2018MNRAS.479.3094R} {479, 3094}
	
	\bibitem[\protect\citeauthoryear{{Rajwade} et~al.,}{{Rajwade}
		et~al.}{2020a}]{Rajwade2020}
	{Rajwade} K.~M.,  et~al., 2020a, \mn@doi [\mnras] {10.1093/mnras/staa1237},
	\href {https://ui.adsabs.harvard.edu/abs/2020MNRAS.495.3551R} {495, 3551}
	
	\bibitem[\protect\citeauthoryear{{Rajwade} et~al.,}{{Rajwade}
		et~al.}{2020b}]{2020SPIE11447E..0JR}
	{Rajwade} K.,  et~al., 2020b, in Society of Photo-Optical Instrumentation
	Engineers (SPIE) Conference Series. p. 114470J, \mn@doi{10.1117/12.2559937}
	
	\bibitem[\protect\citeauthoryear{{Rankin}}{{Rankin}}{1983}]{1983ApJ...274..333R}
	{Rankin} J.~M.,  1983, \mn@doi [\apj] {10.1086/161450}, \href
	{https://ui.adsabs.harvard.edu/abs/1983ApJ...274..333R} {274, 333}
	
	\bibitem[\protect\citeauthoryear{{Rankin}}{{Rankin}}{1990}]{1990ApJ...352..247R}
	{Rankin} J.~M.,  1990, \mn@doi [\apj] {10.1086/168530}, \href
	{https://ui.adsabs.harvard.edu/abs/1990ApJ...352..247R} {352, 247}
	
	\bibitem[\protect\citeauthoryear{{Rankin}}{{Rankin}}{1993}]{1993ApJ...405..285R}
	{Rankin} J.~M.,  1993, \mn@doi [\apj] {10.1086/172361}, \href
	{https://ui.adsabs.harvard.edu/abs/1993ApJ...405..285R} {405, 285}
	
	\bibitem[\protect\citeauthoryear{{Rau} \& {Wasserman}}{{Rau} \&
		{Wasserman}}{2021}]{2021MNRAS.506.4632R}
	{Rau} P.~B.,  {Wasserman} I.,  2021, \mn@doi [\mnras] {10.1093/mnras/stab1538},
	\href {https://ui.adsabs.harvard.edu/abs/2021MNRAS.506.4632R} {506, 4632}
	
	\bibitem[\protect\citeauthoryear{{Ravi}}{{Ravi}}{2019}]{Ravi2019b}
	{Ravi} V.,  2019, \mn@doi [Nature Astronomy] {10.1038/s41550-019-0831-y}, \href
	{https://ui.adsabs.harvard.edu/abs/2019NatAs...3..928R} {3, 928}
	
	\bibitem[\protect\citeauthoryear{{Rea} et~al.,}{{Rea}
		et~al.}{2013}]{2013ApJ...770...65R}
	{Rea} N.,  et~al., 2013, \mn@doi [\apj] {10.1088/0004-637X/770/1/65}, \href
	{https://ui.adsabs.harvard.edu/abs/2013ApJ...770...65R} {770, 65}
	
	\bibitem[\protect\citeauthoryear{{Rea}, {Borghese}, {Esposito}, {Coti Zelati},
		{Bachetti}, {Israel}  \& {De Luca}}{{Rea} et~al.}{2016}]{2016ApJ...828L..13R}
	{Rea} N.,  {Borghese} A.,  {Esposito} P.,  {Coti Zelati} F.,  {Bachetti} M.,
	{Israel} G.~L.,   {De Luca} A.,  2016, \mn@doi [\apjl]
	{10.3847/2041-8205/828/1/L13}, \href
	{https://ui.adsabs.harvard.edu/abs/2016ApJ...828L..13R} {828, L13}
	
	\bibitem[\protect\citeauthoryear{{Revnivtsev}, {Tuerler}, {Del Santo},
		{Westergaard}, {Gehrels}  \& {Winkler}}{{Revnivtsev}
		et~al.}{2003}]{2003IAUC.8097....2R}
	{Revnivtsev} M.,  {Tuerler} M.,  {Del Santo} M.,  {Westergaard} N.~J.,
	{Gehrels} N.,   {Winkler} C.,  2003, \iaucirc, \href
	{https://ui.adsabs.harvard.edu/abs/2003IAUC.8097....2R} {8097, 2}
	
	\bibitem[\protect\citeauthoryear{{Ridnaia} et~al.,}{{Ridnaia}
		et~al.}{2020}]{Ridania2020}
	{Ridnaia} A.,  et~al., 2020, arXiv e-prints, \href
	{https://ui.adsabs.harvard.edu/abs/2020arXiv200511178R} {p. arXiv:2005.11178}
	
	\bibitem[\protect\citeauthoryear{{Rigoselli}, {Mereghetti}, {Suleimanov},
		{Potekhin}, {Turolla}, {Taverna}  \& {Pintore}}{{Rigoselli}
		et~al.}{2019}]{2019A&A...627A..69R}
	{Rigoselli} M.,  {Mereghetti} S.,  {Suleimanov} V.,  {Potekhin} A.~Y.,
	{Turolla} R.,  {Taverna} R.,   {Pintore} F.,  2019, \mn@doi [\aap]
	{10.1051/0004-6361/201935485}, \href
	{https://ui.adsabs.harvard.edu/abs/2019A&A...627A..69R} {627, A69}
	
	\bibitem[\protect\citeauthoryear{{Rodes-Roca}, {Torrej{\'o}n},
		{Mart{\'\i}nez-N{\'u}{\~n}ez}, {Bernab{\'e}u}  \& {Magazz{\'u}}}{{Rodes-Roca}
		et~al.}{2013}]{2013A&A...555A.115R}
	{Rodes-Roca} J.~J.,  {Torrej{\'o}n} J.~M.,  {Mart{\'\i}nez-N{\'u}{\~n}ez} S.,
	{Bernab{\'e}u} G.,   {Magazz{\'u}} A.,  2013, \mn@doi [\aap]
	{10.1051/0004-6361/201321923}, \href
	{https://ui.adsabs.harvard.edu/abs/2013A&A...555A.115R} {555, A115}
	
	\bibitem[\protect\citeauthoryear{{Ruderman}}{{Ruderman}}{1991a}]{1991ApJ...382..576R}
	{Ruderman} R.,  1991a, \mn@doi [\apj] {10.1086/170744}, \href
	{https://ui.adsabs.harvard.edu/abs/1991ApJ...382..576R} {382, 576}
	
	\bibitem[\protect\citeauthoryear{{Ruderman}}{{Ruderman}}{1991b}]{1991ApJ...382..587R}
	{Ruderman} M.,  1991b, \mn@doi [\apj] {10.1086/170745}, \href
	{https://ui.adsabs.harvard.edu/abs/1991ApJ...382..587R} {382, 587}
	
	\bibitem[\protect\citeauthoryear{{Sanjurjo-Ferrr{\'\i}n}, {Torrej{\'o}n},
		{Postnov}, {Oskinova}, {Rodes-Roca}  \& {Bernabeu}}{{Sanjurjo-Ferrr{\'\i}n}
		et~al.}{2017}]{2017A&A...606A.145S}
	{Sanjurjo-Ferrr{\'\i}n} G.,  {Torrej{\'o}n} J.~M.,  {Postnov} K.,  {Oskinova}
	L.,  {Rodes-Roca} J.~J.,   {Bernabeu} G.,  2017, \mn@doi [\aap]
	{10.1051/0004-6361/201630119}, \href
	{https://ui.adsabs.harvard.edu/abs/2017A&A...606A.145S} {606, A145}
	
	\bibitem[\protect\citeauthoryear{{Schlafly} \& {Finkbeiner}}{{Schlafly} \&
		{Finkbeiner}}{2011}]{2011ApJ...737..103S}
	{Schlafly} E.~F.,  {Finkbeiner} D.~P.,  2011, \mn@doi [\apj]
	{10.1088/0004-637X/737/2/103}, \href
	{https://ui.adsabs.harvard.edu/abs/2011ApJ...737..103S} {737, 103}
	
	\bibitem[\protect\citeauthoryear{{Schlafly} et~al.,}{{Schlafly}
		et~al.}{2018}]{2018ApJS..234...39S}
	{Schlafly} E.~F.,  et~al., 2018, \mn@doi [\apjs] {10.3847/1538-4365/aaa3e2},
	\href {https://ui.adsabs.harvard.edu/abs/2018ApJS..234...39S} {234, 39}
	
	\bibitem[\protect\citeauthoryear{{Shaham}}{{Shaham}}{1977}]{1977ApJ...214..251S}
	{Shaham} J.,  1977, \mn@doi [\apj] {10.1086/155249}, \href
	{https://ui.adsabs.harvard.edu/abs/1977ApJ...214..251S} {214, 251}
	
	\bibitem[\protect\citeauthoryear{{Shapiro} \& {Teukolsky}}{{Shapiro} \&
		{Teukolsky}}{1983}]{1983bhwd.book.....S}
	{Shapiro} S.~L.,  {Teukolsky} S.~A.,  1983, {Black holes, white dwarfs, and
		neutron stars : the physics of compact objects}
	
	\bibitem[\protect\citeauthoryear{{Sidoli}, {Israel}, {Esposito},
		{Rodr{\'\i}guez Castillo}  \& {Postnov}}{{Sidoli}
		et~al.}{2017}]{2017MNRAS.469.3056S}
	{Sidoli} L.,  {Israel} G.~L.,  {Esposito} P.,  {Rodr{\'\i}guez Castillo} G.~A.,
	{Postnov} K.,  2017, \mn@doi [\mnras] {10.1093/mnras/stx1105}, \href
	{https://ui.adsabs.harvard.edu/abs/2017MNRAS.469.3056S} {469, 3056}
	
	\bibitem[\protect\citeauthoryear{{Sinha} \& {Sedrakian}}{{Sinha} \&
		{Sedrakian}}{2015}]{2015PhRvC..91c5805S}
	{Sinha} M.,  {Sedrakian} A.,  2015, \mn@doi [\prc]
	{10.1103/PhysRevC.91.035805}, \href
	{https://ui.adsabs.harvard.edu/abs/2015PhRvC..91c5805S} {91, 035805}
	
	\bibitem[\protect\citeauthoryear{{Sotani} \& {Kokkotas}}{{Sotani} \&
		{Kokkotas}}{2009}]{2009MNRAS.395.1163S}
	{Sotani} H.,  {Kokkotas} K.~D.,  2009, \mn@doi [\mnras]
	{10.1111/j.1365-2966.2009.14631.x}, \href
	{https://ui.adsabs.harvard.edu/abs/2009MNRAS.395.1163S} {395, 1163}
	
	\bibitem[\protect\citeauthoryear{{Speagle}, {Kaplan}  \& {van
			Kerkwijk}}{{Speagle} et~al.}{2011}]{2011ApJ...743..183S}
	{Speagle} J.~S.,  {Kaplan} D.~L.,   {van Kerkwijk} M.~H.,  2011, \mn@doi [\apj]
	{10.1088/0004-637X/743/2/183}, \href
	{https://ui.adsabs.harvard.edu/abs/2011ApJ...743..183S} {743, 183}
	
	\bibitem[\protect\citeauthoryear{{Spitler} et~al.,}{{Spitler}
		et~al.}{2016}]{Spitler+16}
	{Spitler} L.~G.,  et~al., 2016, \mn@doi [\nat] {10.1038/nature17168}, \href
	{https://ui.adsabs.harvard.edu/abs/2016Natur.531..202S} {531, 202}
	
	\bibitem[\protect\citeauthoryear{{Spreeuw}, {Scheers}, {Braun}, {Wijers},
		{Miller-Jones}, {Stappers}  \& {Fender}}{{Spreeuw}
		et~al.}{2009}]{2009A&A...502..549S}
	{Spreeuw} H.,  {Scheers} B.,  {Braun} R.,  {Wijers} R.~A.~M.~J.,
	{Miller-Jones} J.~C.~A.,  {Stappers} B.~W.,   {Fender} R.~P.,  2009, \mn@doi
	[\aap] {10.1051/0004-6361/200810449}, \href
	{https://ui.adsabs.harvard.edu/abs/2009A&A...502..549S} {502, 549}
	
	\bibitem[\protect\citeauthoryear{{Stiller}, {Littlefield}, {Garnavich}, {Wood},
		{Hambsch}  \& {Myers}}{{Stiller} et~al.}{2018}]{2018AJ....156..150S}
	{Stiller} R.~A.,  {Littlefield} C.,  {Garnavich} P.,  {Wood} C.,  {Hambsch}
	F.-J.,   {Myers} G.,  2018, \mn@doi [\aj] {10.3847/1538-3881/aad5dd}, \href
	{https://ui.adsabs.harvard.edu/abs/2018AJ....156..150S} {156, 150}
	
	\bibitem[\protect\citeauthoryear{{Sur} \& {Haskell}}{{Sur} \&
		{Haskell}}{2021}]{2021PASA...38...43S}
	{Sur} A.,  {Haskell} B.,  2021, \mn@doi [\pasa] {10.1017/pasa.2021.39}, \href
	{https://ui.adsabs.harvard.edu/abs/2021PASA...38...43S} {38, e043}
	
	\bibitem[\protect\citeauthoryear{{Tan} et~al.,}{{Tan}
		et~al.}{2018}]{2018ApJ...866...54T}
	{Tan} C.~M.,  et~al., 2018, \mn@doi [\apj] {10.3847/1538-4357/aade88}, \href
	{https://ui.adsabs.harvard.edu/abs/2018ApJ...866...54T} {866, 54}
	
	\bibitem[\protect\citeauthoryear{{Tavani} et~al.,}{{Tavani}
		et~al.}{2020}]{Tavani2020}
	{Tavani} M.,  et~al., 2020, arXiv e-prints, \href
	{https://ui.adsabs.harvard.edu/abs/2020arXiv200512164T} {p. arXiv:2005.12164}
	
	\bibitem[\protect\citeauthoryear{{Tendulkar}, {Cameron}  \&
		{Kulkarni}}{{Tendulkar} et~al.}{2012}]{2012ApJ...761...76T}
	{Tendulkar} S.~P.,  {Cameron} P.~B.,   {Kulkarni} S.~R.,  2012, \mn@doi [\apj]
	{10.1088/0004-637X/761/1/76}, \href
	{https://ui.adsabs.harvard.edu/abs/2012ApJ...761...76T} {761, 76}
	
	\bibitem[\protect\citeauthoryear{{Tendulkar}, {Kaspi}, {Archibald}  \&
		{Scholz}}{{Tendulkar} et~al.}{2017}]{2017ApJ...841...11T}
	{Tendulkar} S.~P.,  {Kaspi} V.~M.,  {Archibald} R.~F.,   {Scholz} P.,  2017,
	\mn@doi [\apj] {10.3847/1538-4357/aa6d0c}, \href
	{https://ui.adsabs.harvard.edu/abs/2017ApJ...841...11T} {841, 11}
	
	\bibitem[\protect\citeauthoryear{{Tendulkar} et~al.,}{{Tendulkar}
		et~al.}{2021}]{2021ApJ...908L..12T}
	{Tendulkar} S.~P.,  et~al., 2021, \mn@doi [\apjl] {10.3847/2041-8213/abdb38},
	\href {https://ui.adsabs.harvard.edu/abs/2021ApJ...908L..12T} {908, L12}
	
	\bibitem[\protect\citeauthoryear{{Tetzlaff}, {Eisenbeiss}, {Neuh{\"a}user}  \&
		{Hohle}}{{Tetzlaff} et~al.}{2011}]{2011MNRAS.417..617T}
	{Tetzlaff} N.,  {Eisenbeiss} T.,  {Neuh{\"a}user} R.,   {Hohle} M.~M.,  2011,
	\mn@doi [\mnras] {10.1111/j.1365-2966.2011.19302.x}, \href
	{https://ui.adsabs.harvard.edu/abs/2011MNRAS.417..617T} {417, 617}
	
	\bibitem[\protect\citeauthoryear{{The Chime/Frb Collaboration} Andersen
		et~al.,}{{The Chime/Frb Collaboration} et~al.}{2020}]{CHIME2020}
	{The Chime/Frb Collaboration} Andersen B.~C.,  et~al., 2020, \mn@doi [\nat]
	{10.1038/s41586-020-2863-y}, \href
	{https://ui.adsabs.harvard.edu/abs/2020Natur.587...54T} {587, 54}
	
	\bibitem[\protect\citeauthoryear{{Thompson} \& {Blaes}}{{Thompson} \&
		{Blaes}}{1998}]{BT1998}
	{Thompson} C.,  {Blaes} O.,  1998, \mn@doi [\prd] {10.1103/PhysRevD.57.3219},
	\href {https://ui.adsabs.harvard.edu/abs/1998PhRvD..57.3219T} {57, 3219}
	
	\bibitem[\protect\citeauthoryear{{Thompson} \& {Duncan}}{{Thompson} \&
		{Duncan}}{1995}]{1995MNRAS.275..255T}
	{Thompson} C.,  {Duncan} R.~C.,  1995, \mn@doi [\mnras]
	{10.1093/mnras/275.2.255}, \href
	{https://ui.adsabs.harvard.edu/abs/1995MNRAS.275..255T} {275, 255}
	
	\bibitem[\protect\citeauthoryear{{Thompson} \& {Duncan}}{{Thompson} \&
		{Duncan}}{1996}]{1996ApJ...473..322T}
	{Thompson} C.,  {Duncan} R.~C.,  1996, \mn@doi [\apj] {10.1086/178147}, \href
	{https://ui.adsabs.harvard.edu/abs/1996ApJ...473..322T} {473, 322}
	
	\bibitem[\protect\citeauthoryear{{Tiengo} et~al.,}{{Tiengo}
		et~al.}{2013}]{2013Natur.500..312T}
	{Tiengo} A.,  et~al., 2013, \mn@doi [\nat] {10.1038/nature12386}, \href
	{https://ui.adsabs.harvard.edu/abs/2013Natur.500..312T} {500, 312}
	
	\bibitem[\protect\citeauthoryear{{Timokhin}}{{Timokhin}}{2010}]{2010MNRAS.408.2092T}
	{Timokhin} A.~N.,  2010, \mn@doi [\mnras] {10.1111/j.1365-2966.2010.17286.x},
	\href {https://ui.adsabs.harvard.edu/abs/2010MNRAS.408.2092T} {408, 2092}
	
	\bibitem[\protect\citeauthoryear{{Timokhin} \& {Arons}}{{Timokhin} \&
		{Arons}}{2013}]{2013MNRAS.429...20T}
	{Timokhin} A.~N.,  {Arons} J.,  2013, \mn@doi [\mnras] {10.1093/mnras/sts298},
	\href {https://ui.adsabs.harvard.edu/abs/2013MNRAS.429...20T} {429, 20}
	
	\bibitem[\protect\citeauthoryear{{Timokhin} \& {Harding}}{{Timokhin} \&
		{Harding}}{2015}]{2015ApJ...810..144T}
	{Timokhin} A.~N.,  {Harding} A.~K.,  2015, \mn@doi [\apj]
	{10.1088/0004-637X/810/2/144}, \href
	{https://ui.adsabs.harvard.edu/abs/2015ApJ...810..144T} {810, 144}
	
	\bibitem[\protect\citeauthoryear{{Timokhin}, {Bisnovatyi-Kogan}  \&
		{Spruit}}{{Timokhin} et~al.}{2000}]{2000MNRAS.316..734T}
	{Timokhin} A.~N.,  {Bisnovatyi-Kogan} G.~S.,   {Spruit} H.~C.,  2000, \mn@doi
	[\mnras] {10.1046/j.1365-8711.2000.03535.x}, \href
	{https://ui.adsabs.harvard.edu/abs/2000MNRAS.316..734T} {316, 734}
	
	\bibitem[\protect\citeauthoryear{{Tolman}, {Philippov}  \& {Timokhin}}{{Tolman}
		et~al.}{2022}]{2022ApJ...933L..37T}
	{Tolman} E.~A.,  {Philippov} A.~A.,   {Timokhin} A.~N.,  2022, \mn@doi [\apjl]
	{10.3847/2041-8213/ac7c71}, \href
	{https://ui.adsabs.harvard.edu/abs/2022ApJ...933L..37T} {933, L37}
	
	\bibitem[\protect\citeauthoryear{{Tong}, {Wang}, {Liu}  \& {Xu}}{{Tong}
		et~al.}{2016}]{2016ApJ...833..265T}
	{Tong} H.,  {Wang} W.,  {Liu} X.~W.,   {Xu} R.~X.,  2016, \mn@doi [\apj]
	{10.3847/1538-4357/833/2/265}, \href
	{https://ui.adsabs.harvard.edu/abs/2016ApJ...833..265T} {833, 265}
	
	\bibitem[\protect\citeauthoryear{{Torrej{\'o}n}, {Reig}, {F{\"u}rst},
		{Martinez-Chicharro}, {Postnov}  \& {Oskinova}}{{Torrej{\'o}n}
		et~al.}{2018}]{2018MNRAS.479.3366T}
	{Torrej{\'o}n} J.~M.,  {Reig} P.,  {F{\"u}rst} F.,  {Martinez-Chicharro} M.,
	{Postnov} K.,   {Oskinova} L.,  2018, \mn@doi [\mnras]
	{10.1093/mnras/sty1628}, \href
	{https://ui.adsabs.harvard.edu/abs/2018MNRAS.479.3366T} {479, 3366}
	
	\bibitem[\protect\citeauthoryear{{Troja} et~al.,}{{Troja}
		et~al.}{2020}]{Troja2020}
	{Troja} E.,  et~al., 2020, \mn@doi [\mnras] {10.1093/mnras/staa2626}, \href
	{https://ui.adsabs.harvard.edu/abs/2020MNRAS.498.5643T} {498, 5643}
	
	\bibitem[\protect\citeauthoryear{{Usov}}{{Usov}}{1984}]{1984Ap&SS.107..191U}
	{Usov} V.~V.,  1984, \mn@doi [\apss] {10.1007/BF00649624}, \href
	{https://ui.adsabs.harvard.edu/abs/1984Ap&SS.107..191U} {107, 191}
	
	\bibitem[\protect\citeauthoryear{{Usov}}{{Usov}}{2004}]{2004PhRvD..70f7301U}
	{Usov} V.~V.,  2004, \mn@doi [\prd] {10.1103/PhysRevD.70.067301}, \href
	{https://ui.adsabs.harvard.edu/abs/2004PhRvD..70f7301U} {70, 067301}
	
	\bibitem[\protect\citeauthoryear{{Vasilopoulos}, {Maitra}, {Haberl},
		{Hatzidimitriou}  \& {Petropoulou}}{{Vasilopoulos}
		et~al.}{2018}]{2018MNRAS.475..220V}
	{Vasilopoulos} G.,  {Maitra} C.,  {Haberl} F.,  {Hatzidimitriou} D.,
	{Petropoulou} M.,  2018, \mn@doi [\mnras] {10.1093/mnras/stx3139}, \href
	{https://ui.adsabs.harvard.edu/abs/2018MNRAS.475..220V} {475, 220}
	
	\bibitem[\protect\citeauthoryear{{Vigan{\`o}}, {Rea}, {Pons}, {Perna},
		{Aguilera}  \& {Miralles}}{{Vigan{\`o}} et~al.}{2013}]{Vigano2013}
	{Vigan{\`o}} D.,  {Rea} N.,  {Pons} J.~A.,  {Perna} R.,  {Aguilera} D.~N.,
	{Miralles} J.~A.,  2013, \mn@doi [\mnras] {10.1093/mnras/stt1008}, \href
	{https://ui.adsabs.harvard.edu/abs/2013MNRAS.434..123V} {434, 123}
	
	\bibitem[\protect\citeauthoryear{{Wadiasingh} \& {Chirenti}}{{Wadiasingh} \&
		{Chirenti}}{2020}]{2020ApJ...903L..38W}
	{Wadiasingh} Z.,  {Chirenti} C.,  2020, \mn@doi [\apjl]
	{10.3847/2041-8213/abc562}, \href
	{https://ui.adsabs.harvard.edu/abs/2020ApJ...903L..38W} {903, L38}
	
	\bibitem[\protect\citeauthoryear{{Wadiasingh} \& {Timokhin}}{{Wadiasingh} \&
		{Timokhin}}{2019}]{Wadiasingh2019}
	{Wadiasingh} Z.,  {Timokhin} A.,  2019, \mn@doi [\apj]
	{10.3847/1538-4357/ab2240}, \href
	{https://ui.adsabs.harvard.edu/abs/2019ApJ...879....4W} {879, 4}
	
	\bibitem[\protect\citeauthoryear{{Wadiasingh}, {Beniamini}, {Timokhin},
		{Baring}, {van der Horst}, {Harding}  \& {Kazanas}}{{Wadiasingh}
		et~al.}{2020}]{Wadiasingh2020}
	{Wadiasingh} Z.,  {Beniamini} P.,  {Timokhin} A.,  {Baring} M.~G.,  {van der
		Horst} A.~J.,  {Harding} A.~K.,   {Kazanas} D.,  2020, \mn@doi [\apj]
	{10.3847/1538-4357/ab6d69}, \href
	{https://ui.adsabs.harvard.edu/abs/2020ApJ...891...82W} {891, 82}
	
	\bibitem[\protect\citeauthoryear{{Walton} et~al.,}{{Walton}
		et~al.}{2018}]{2018ApJ...857L...3W}
	{Walton} D.~J.,  et~al., 2018, \mn@doi [\apjl] {10.3847/2041-8213/aabadc},
	\href {https://ui.adsabs.harvard.edu/abs/2018ApJ...857L...3W} {857, L3}
	
	\bibitem[\protect\citeauthoryear{{Wang} et~al.,}{{Wang}
		et~al.}{2021}]{2021ApJ...920...45W}
	{Wang} Z.,  et~al., 2021, \mn@doi [\apj] {10.3847/1538-4357/ac2360}, \href
	{https://ui.adsabs.harvard.edu/abs/2021ApJ...920...45W} {920, 45}
	
	\bibitem[\protect\citeauthoryear{{Wang} et~al.,}{{Wang}
		et~al.}{2022}]{2022arXiv220902352W}
	{Wang} Z.,  et~al., 2022, arXiv e-prints, \href
	{https://ui.adsabs.harvard.edu/abs/2022arXiv220902352W} {p. arXiv:2209.02352}
	
	\bibitem[\protect\citeauthoryear{{Weber}}{{Weber}}{2005}]{2005PrPNP..54..193W}
	{Weber} F.,  2005, \mn@doi [Progress in Particle and Nuclear Physics]
	{10.1016/j.ppnp.2004.07.001}, \href
	{https://ui.adsabs.harvard.edu/abs/2005PrPNP..54..193W} {54, 193}
	
	\bibitem[\protect\citeauthoryear{{Witten}}{{Witten}}{1984}]{1984PhRvD..30..272W}
	{Witten} E.,  1984, \mn@doi [\prd] {10.1103/PhysRevD.30.272}, \href
	{https://ui.adsabs.harvard.edu/abs/1984PhRvD..30..272W} {30, 272}
	
	\bibitem[\protect\citeauthoryear{{Wood} \& {Graber}}{{Wood} \&
		{Graber}}{2022}]{2022Univ....8..228W}
	{Wood} T.~S.,  {Graber} V.,  2022, \mn@doi [Universe]
	{10.3390/universe8040228}, \href
	{https://ui.adsabs.harvard.edu/abs/2022Univ....8..228W} {8, 228}
	
	\bibitem[\protect\citeauthoryear{{Wood} \& {Hollerbach}}{{Wood} \&
		{Hollerbach}}{2015}]{2015PhRvL.114s1101W}
	{Wood} T.~S.,  {Hollerbach} R.,  2015, \mn@doi [\prl]
	{10.1103/PhysRevLett.114.191101}, \href
	{https://ui.adsabs.harvard.edu/abs/2015PhRvL.114s1101W} {114, 191101}
	
	\bibitem[\protect\citeauthoryear{{Woods}, {Kouveliotou}, {Finger},
		{G{\"o}{\v{g}}{\"u}{\c{s}}}, {Wilson}, {Patel}, {Hurley}  \& {Swank}}{{Woods}
		et~al.}{2007}]{Woods2007}
	{Woods} P.~M.,  {Kouveliotou} C.,  {Finger} M.~H.,  {G{\"o}{\v{g}}{\"u}{\c{s}}}
	E.,  {Wilson} C.~A.,  {Patel} S.~K.,  {Hurley} K.,   {Swank} J.~H.,  2007,
	\mn@doi [\apj] {10.1086/507459}, \href
	{https://ui.adsabs.harvard.edu/abs/2007ApJ...654..470W} {654, 470}
	
	\bibitem[\protect\citeauthoryear{{Yakovlev} \& {Pethick}}{{Yakovlev} \&
		{Pethick}}{2004}]{2004ARA&A..42..169Y}
	{Yakovlev} D.~G.,  {Pethick} C.~J.,  2004, \mn@doi [\araa]
	{10.1146/annurev.astro.42.053102.134013}, \href
	{https://ui.adsabs.harvard.edu/abs/2004ARA&A..42..169Y} {42, 169}
	
	\bibitem[\protect\citeauthoryear{{Yoneyama}, {Hayashida}, {Nakajima}  \&
		{Matsumoto}}{{Yoneyama} et~al.}{2019}]{2019PASJ...71...17Y}
	{Yoneyama} T.,  {Hayashida} K.,  {Nakajima} H.,   {Matsumoto} H.,  2019,
	\mn@doi [\pasj] {10.1093/pasj/psy135}, \href
	{https://ui.adsabs.harvard.edu/abs/2019PASJ...71...17Y} {71, 17}
	
	\bibitem[\protect\citeauthoryear{{Younes}, {Kouveliotou}  \& {Kaspi}}{{Younes}
		et~al.}{2015}]{Younes2015}
	{Younes} G.,  {Kouveliotou} C.,   {Kaspi} V.~M.,  2015, \mn@doi [\apj]
	{10.1088/0004-637X/809/2/165}, \href
	{https://ui.adsabs.harvard.edu/abs/2015ApJ...809..165Y} {809, 165}
	
	\bibitem[\protect\citeauthoryear{{Younes}, {Ray}, {Baring}, {Kouveliotou},
		{Fletcher}, {Wadiasingh}, {Harding}  \& {Goldstein}}{{Younes}
		et~al.}{2020a}]{younes20ApJ:2259}
	{Younes} G.,  {Ray} P.~S.,  {Baring} M.~G.,  {Kouveliotou} C.,  {Fletcher} C.,
	{Wadiasingh} Z.,  {Harding} A.~K.,   {Goldstein} A.,  2020a, \mn@doi [\apjl]
	{10.3847/2041-8213/ab9a48}, \href
	{https://ui.adsabs.harvard.edu/abs/2020ApJ...896L..42Y} {896, L42}
	
	\bibitem[\protect\citeauthoryear{{Younes} et~al.,}{{Younes}
		et~al.}{2020b}]{2020ApJ...904L..21Y}
	{Younes} G.,  et~al., 2020b, \mn@doi [\apjl] {10.3847/2041-8213/abc94c}, \href
	{https://ui.adsabs.harvard.edu/abs/2020ApJ...904L..21Y} {904, L21}
	
	\bibitem[\protect\citeauthoryear{{Younes} et~al.,}{{Younes}
		et~al.}{2021}]{2021NatAs...5..408Y}
	{Younes} G.,  et~al., 2021, \mn@doi [Nature Astronomy]
	{10.1038/s41550-020-01292-x}, \href
	{https://ui.adsabs.harvard.edu/abs/2021NatAs...5..408Y} {5, 408}
	
	\bibitem[\protect\citeauthoryear{{Younes} et~al.,}{{Younes}
		et~al.}{2022}]{younes2022:1935}
	{Younes} G.,  et~al., 2022, arXiv e-prints, \href
	{https://ui.adsabs.harvard.edu/abs/2022arXiv221011518Y} {p. arXiv:2210.11518}
	
	\bibitem[\protect\citeauthoryear{{Zanazzi} \& {Lai}}{{Zanazzi} \&
		{Lai}}{2020}]{Zanazzi&Lai20}
	{Zanazzi} J.~J.,  {Lai} D.,  2020, \mn@doi [\apjl] {10.3847/2041-8213/ab7cdd},
	\href {https://ui.adsabs.harvard.edu/abs/2020ApJ...892L..15Z} {892, L15}
	
	\bibitem[\protect\citeauthoryear{{Zhang} et~al.,}{{Zhang}
		et~al.}{2022}]{2022HEAD...1910845Z}
	{Zhang} W.,  et~al., 2022, in AAS/High Energy Astrophysics Division. p. 108.45
	
	\bibitem[\protect\citeauthoryear{{du Plessis}, {Wadiasingh}, {Venter}  \&
		{Harding}}{{du Plessis} et~al.}{2019}]{2019ApJ...887...44D}
	{du Plessis} L.,  {Wadiasingh} Z.,  {Venter} C.,   {Harding} A.~K.,  2019,
	\mn@doi [\apj] {10.3847/1538-4357/ab4e19}, \href
	{https://ui.adsabs.harvard.edu/abs/2019ApJ...887...44D} {887, 44}
	
	\makeatother
\end{thebibliography}
\end{document}